\documentclass[aps,prc,twocolumn,floatfix,superscriptaddress,amsmath,amssymb,nofootinbib]{revtex4-2}

\usepackage{dcolumn}
\usepackage{float}
\usepackage{graphicx}
\usepackage{xspace}
\usepackage{xcolor}
\usepackage[super]{nth}
\usepackage{physics}
\usepackage[colorlinks=true,linkcolor=blue,citecolor=blue,urlcolor=blue]{hyperref}
\usepackage{nameref}
\usepackage{orcidlink}
\usepackage[linewidth=1pt]{mdframed}

\newcommand{\nuclide}[2]{\ensuremath{^{#2}\rm{#1}}}

\begin{document}

\title{A nuclear mass model rooted in chiral effective field theory}

\author{C.~Mishra\orcidlink{0000-0003-1400-7898}}
\affiliation{Department of Physics and Astronomy, University of
  Tennessee, Knoxville, Tennessee 37996, USA}
\affiliation{Lawrence Livermore National Laboratory, Livermore, California 94551, USA}
\email{mishra10@llnl.gov}

\author{A.~Ekstr\"om}
\affiliation{Department of Physics, Chalmers University of Technology, SE-412 96 G\"oteborg, Sweden}

\author{G.~Hagen}
\affiliation{Physics Division, Oak Ridge National Laboratory, Oak
  Ridge, Tennessee 37831, USA}
\affiliation{Department of Physics and Astronomy, University of
  Tennessee, Knoxville, Tennessee 37996, USA}

\author {M. Menickelly\orcidlink{0000-0002-2023-0837}}
\affiliation{Mathematics and Computer Science Division, Argonne National Laboratory, Lemont, IL, 60439, USA}

\author{T.~Papenbrock\orcidlink{0000-0001-8733-2849}}
\affiliation{Department of Physics and Astronomy, University of
  Tennessee, Knoxville, Tennessee 37996, USA}
\affiliation{Physics Division, Oak Ridge National Laboratory, Oak
  Ridge, Tennessee 37831, USA}

\author{S. M. Wild\orcidlink{0000-0002-6099-2772}}
\affiliation{Applied Math \& Computational Research Division, Lawrence Berkeley National Laboratory, Berkeley, CA 94720, USA}
    
\begin{abstract}
We develop a nuclear mass model that is based on chiral effective field theory\xspace at next-to-next-to leading order. Nuclear binding energies are computed via the Hartree-Fock method using a Hamiltonian from delta-full chiral effective field theory. We employ Hartree-Fock emulators to adjust $11$ \xspace {low-energy constants}\xspace in the chiral interaction to binding energies of $18$ even-even nuclei.  When applied to 
$107$ even-even nuclei with mass numbers $16\leq A\leq 56$ 
the chiral mass model exhibits an overall root-mean-square deviation of $3.5$~MeV.
\end{abstract}
\maketitle

\section{Introduction} 

\label{sec:mass-model-intro}

About 3000 nuclei have been discovered so far~\cite{thoennessen2016}, while about 7000 with atomic numbers between 2 and 120 are predicted to exist~\cite{erler2012}.
Experiments have generated precise measurements for many of their binding energies, spectra, electric and magnetic multipole moments, decay modes, etc.
Our primary concern in this work is the nuclear binding energy.
Precise knowledge of binding energies is relevant for
predictions of the neutron dripline, 
nucleosynthesis, and
fundamental symmetries~\cite{blaum2013}.
In particular, understanding the origins of heavy elements produced via rapid neutron capture (the r-process) in stellar environments, which account for the generation of heavy elements with mass numbers $A > 56$, requires accurate binding energies~\cite{Horowitz:2018ndv,sprouse2020,Holmbeck:2023bjs}.

Theoretical models  for nuclear binding energies have been constructed and calibrated to known nuclear data since the early \nth{20} century.
Bethe \& Weizs\"acker's semi-empirical mass formula 
\cite{Weizsacker1935, Bethe1937}
models the nucleus as a liquid drop (plus pairing effects) and captures binding energies within a root-mean-square (RMS) error of about 3 to 4 MeV. A statistical analysis~\cite{Kejzlar:2020vla} shows that the liquid drop model effectively depends on only two parameters. 
Today's state-of-the-art mass models are more accurate.
Examples include 
the microscopic-macroscopic Finite-Range Liquid Drop Model \cite{moller2016}, with a RMS error of $\sigma = 0.56$~MeV, 
as well as the microscopic models of Refs. \cite{Kortelainen2010} and \cite{goriely2013} based on the Hartree-Fock-Bogoliubov framework and Skyrme forces.
The RMS deviations in these microscopic approaches were $0.97$~MeV and $0.50$~MeV, respectively;
see Refs.~\cite{Lunney2003, Pearson2006, Stone2007, Kortelainen2010, Grasso2019, Robledo2019} for further discussions.
The parameters appearing in mass models are fixed using global fits.
For example,
Ref.~\cite{Duflo1995} (Ref.~\cite{moller2016}) tuned  $28$ ($10$) parameters to $1751$ (2149) binding energies, while the authors of Ref.~\cite{Kortelainen2010} tuned their $12$ parameters to the bulk properties of a much smaller set of $72$ nuclei.

It is not obvious how to make nuclear mass models more accurate~\cite{carlsson2008,carlsson2010,Raimondi2014,Sobiczewski2014, Kejzlar:2023tlm} and this poses a challenge because half-an-MeV uncertainties in binding energies can generate order-of-magnitude changes in the abundance patterns emerging from r-process simulations \cite{Mumpower2015,Mumpower:2015ova,Saito:2023seh}. We note that neutron separation energies, i.e. the differences in binding energies between neighboring nuclei that differ by a single neutron, enter such simulations. One might consider  \emph{ab initio} computations, i.e. computations based on Hamiltonians from effective field theories of quantum chromodynamics that only use controlled approximations~\cite{Ekstrom:2022yea} to compute nuclear binding energies~\cite{stroberg2021}. However, such computations are often less accurate than more phenomenological models for bulk properties such as binding energies and charge radii, which can only be computed to about 2 and 3\% of accuracy, respectively.  In contrast, differences of observables in neighboring nuclei, e.g., separation energies and differential radii are often reproduced more accurately~\cite{scalesi2024,koszorus2021,kortelainen2021}.       

This motivates us to look at constructions of  nuclear mass models that are rooted in effective field theories of quantum chromodynamics. There have been several of such efforts within
nuclear energy density functionals~\cite{
Stoitsov2010,gebremariam2011,
NavarroPerez2018, Salvioni2020,
Zurek2021,Duguet2023a, 
Zurek2024,Zurek:2025rvt}.
These functionals employ the density-matrix expansion of Refs.~\cite{negele1972,negele1975,Bogner:2008kj}. 
\textcite{Zurek2024} included long-range physics from pion-exchange  in chiral effective field theory\xspace and short-range physics from Skyrme forces~\cite{skyrme1956,skyrme1958}.  
They found an improved RMS deviation of $1.5$~MeV at next-to-next-to leading order (NNLO)  (including three-nucleon forces and $\Delta$-excitations), compared to $2.1$~MeV from a density functional without pion contributions. 

In this work, we follow a somewhat different path and use a Hamiltonian from chiral effective field theory instead of a density functional. However, the Hamiltonan will only be solved via the Hartree-Fock method and the idea is to renormalize the short-range contributions of the Hamiltonian appropriately such that a Hartree-Fock computation yields accurate results for nuclear masses. This idea is based on the demonstration~\cite{Sun2022b} that three-particle--three-hole correlations in coupled-cluster computations of nuclei and infinite nucleonic matter can approximately be removed (i.e. integrated out) via a renormalization of the three-body contact that enters at NNLO. Therefore, removing two-particle--two-hole correlations in beyond mean-field computations, such as coupled-cluster with singles-and-doubles excitations~\cite{hagen2014}, could then presumably be accomplished via renormalization of two-body contacts. Thus, a Hamiltonian from chiral effective field theory with properly renormalized contact terms -- when solved at the Hartree-Fock approximation -- could yield an accurate nuclear mass model.

Hartree-Fock computations of nuclei based on Hamiltonians from chiral effective field theory are still computationally expensive. This is because the computation involves three-nucleon forces and large model spaces. This makes the envisioned renormalization impractical. To meet this challenge, we develop projection-based emulators of \emph{ab initio}\xspace Hartree-Fock computation following Ref.~\cite{sun2024}. 

The article is organized as follows.
We discuss the Hamiltonians from chiral effective field  theory and the construction of emulators in Sec.~\ref{sec:background}. 
In Sec.~\ref{sec:methods}, we present the calibration of the mass model. 
In Sec.~\ref{sec:mass-model-results}, we report the results from our mass model, and in Sec.~\ref{sec:discussion}  summarize and discuss our results.

\section{Theoretical Background} 
\label{sec:background}

\subsection{The Nuclear Hamiltonian} 

The nuclear interaction derived from chiral effective field theory\xspace consists of two-, three- and higher-body potentials organized via a power counting into a hierarchy of decreasing significance~\cite{vankolck1994,epelbaum2009, Hammer2020, machleidt2011}. At third order in Weinberg power counting (NNLO), the Hamiltonian for the $A$-nucleon system is linear in the low-energy constants that govern the strengths of the two- and three-nucleon forces, and can be expressed as
\begin{align}
  H(\mathbf{x}) = h_0 + \sum_{i=1}^n x_i h_i \,,
  \label{eq:h-matrices}
\end{align}
where $h_i$ (with $i=0,\ldots,n$) are operators and $x_i$ (with $i=1,\ldots,n$) denote \xspace {low-energy constants}\xspace. Here we singled out $h_0$, which consists of operators that do not contain short-range physics. These are, for instance, the kinetic energy and interaction terms that exclusively contain pion physics. As they contain the (long-ranged) chiral dynamics, they will be kept fixed. The terms $h_i$ with $i=1,\ldots,n$ all contain short-range (contact) interactions. These are the $s$ and $p$-wave contacts in the nucleon-nucleon interaction and the short-ranged three nucleon potentials with low-energy constants $c_D$ and $c_E$~\cite{epelbaum2002}.  The corresponding low-energy constants  $x_i$ with $i=1,\ldots,n$ will be renormalized in what follows.

In this work, we use the momentum-space  NNLO delta-full chiral interaction of Refs.~\cite{ekstrom2018,jiang2020}, 
with the cutoffs $\Lambda=394$~MeV in the non-local regulator and $\Lambda_{\textrm{SFR}} = 700$~MeV in the spectral function regulators.
This interaction, \xspace$\Delta$NNLO$_{\rm GO}(394)$\xspace, has a total of 17 parameters. However, the four pion-nucleon couplings $c_1,c_2,c_3,c_4$ are kept as in Ref.~\cite{jiang2020}, because they are not associated with short-range physics. Furthermore, we neglect charge dependence beyond static Coulomb and pion-mass splitting in the one-pion exchange interaction, and therefore only use a single leading-order $^1S_0$ contact $\tilde{C}_{^1S_0}=\tilde{C}_{^1S_0}^{(nn)}=\tilde{C}_{^1S_0}^{(np)}=\tilde{C}_{^1S_0}^{(pp)}$ instead of three separate ones corresponding to neutron-neutron, neutron-proton and proton-proton channels. 
We consider the size of charge independence breaking in nuclei small enough to not warrant a more detailed treatment. The differences between the charge dependent contact values is much smaller than their value, see Refs.~\cite{machleidt2011,jiang2020}. This implies that $n=11$ in the Hamiltonian~(\ref{eq:h-matrices}). 
These \xspace {low-energy constants}\xspace are listed in Table~\ref{table:lecs-by-order} below, along with the order within chiral effective field theory\xspace at which they appear.
  
\renewcommand{\arraystretch}{1.2}
\begin{table}[htp]
  \centering
  \caption{Short range contacts from the lowest three orders in the delta-full chiral effective field theory\xspace being renormalized in the present work.}
  \label{table:lecs-by-order}
    \begin{tabular}{|>
    {\centering\arraybackslash}m{0.09\textwidth}|>
    {\centering\arraybackslash}m{0.375\textwidth}|}
    \hline
    EFT Order & LECs \\
    \hline
    LO      &  $\tilde{C}_{^1{\rm{S}_0}}$, $\tilde{C}_{^3{\rm{S}_1}}$ \\
    NLO     &  
    $C_{^1{\rm{S}_0}}$, 
    $C_{^3{\rm{P}_0}}$, 
    $C_{^1{\rm{P}_1}}$, 
    $C_{^3{\rm{P}_1}}$, 
    $C_{^3{\rm{S}_1}}$, 
    $C_{^3{\rm{S}_1}-^3{\rm{D}_1}}$, 
    $C_{^3{\rm{P}_2}}$\\
    NNLO &  $c_D$, $c_E$\\
    \hline
    \end{tabular}
    \end{table}

\subsection{The Hartree-Fock Mass Model} 
In what follows we compute the ground-state energy with the Hartree-Fock method. Our computations start from the spherical harmonic-oscillator basis. The Hartree-Fock method yields the unitary transformation to a product state with variationally-minimal energy expectation value for the Hamiltonian~\eqref{eq:h-matrices}. 
In terms of the Hartree-Fock Hamiltonian $H^{\rm HF}$ obtained from the Hamiltonian~\eqref{eq:h-matrices}, one solves the eigenvalue problem
\begin{align}
  H^{\rm{HF}}(\mathbf{x}) \ket{\psi(\mathbf{x})} = E(\mathbf{x}) \ket{\psi(\mathbf{x})}\,.
  \label{eq:hf-hamil}
\end{align}
Here we combined the low-energy constants in a vector $\mathbf{x}=(x_1,\ldots,x_n)$ and denote the Hartree-Fock energy as $E(\mathbf{x})$ and the Hartree-Fock state as $\ket{\psi(\mathbf{x})}$.
Note that while the Hamiltonian~\eqref{eq:h-matrices} depends linearly on the \xspace {low-energy constants}\xspace $\mathbf{x}$, the Hartree-Fock energy and state exhibit a nonlinear dependence. We elaborate on this in Sec.~\ref{sec:calibration}, where we detail the construction of Hartree-Fock emulators.

Our Hartree-Fock mass model should lead to ground-state energies that are close to the (negative) experimental binding energies for nuclei in (a region of) the nuclear chart. Thus, 
we propose to calibrate the mass model by solving the nonlinear least-squares problem 
\begin{equation}
  \begin{aligned}\label{eq:least-squares}
    \displaystyle\min_{\mathbf{x}} &  \sum_{a=1}^{n_\text{learned}} (\delta B_a(\mathbf{x}))^2, \quad 
    \delta B_a(\mathbf{x}) := E_a(\mathbf{x}) + B_a\,.
  \end{aligned}
\end{equation}
Here $B_a$ (intrinsically positive) is the binding energy of the $a^\textrm{th}$ nucleus, and $E_a(\mathbf{x})$ is the Hartree-Fock energy obtained from our mass model for the parameterization $\mathbf{x}$, and $n_\text{learned}$ is the number of nuclear binding energies included as input in the calibration process. 

We optimize the parameters $\mathbf{x}$ in Eq.~\eqref{eq:least-squares} to obtain \xspace {low-energy constants}\xspace $\mathbf{x}_\textrm{opt}$; we refer to this set of parameters as a renormalization of the Hamiltonian~(\ref{eq:h-matrices}), which specifies the mass model. 
One can then use the mass model to compute  ground-state energies  for nuclei not included in the calibration stage simply with the Hartree-Fock method. 

In this work, we restricted ourselves  to $107$ even-even nuclei with mass numbers in the range  $16 \leq A \leq 56$. The lower bound is motivated by the fact that the emergence of a mean-field is doubtful for too light nuclei.
The upper bound $A=56$ is chosen because Hartree-Fock computations become much more cumbersome for heavier open-shell nuclei~\cite{hu2024,hu2024b}. 
Of these 107 nuclei, we choose $n_\text{learned} = 18$ spherical and deformed nuclei to calibrate our mass model. They are listed in Table~\ref{table:learned-nuclides} 

\begin{table}[htp]
  \centering
  \caption{Nuclei included in the objective function. $Z$ and $N$ are the proton- and neutron-numbers, $B$ is the total binding energy and $A \equiv Z + N$ is the mass number. The ratio $E_{4^+}/E_{2^+}$ of the excitation energies of the lowest states with spin parity $J^\pi=2^+$ and $4^+$ is a measure of the deformation of the nucleus -- a value close to $10/3\approx 3.33$ implies rigid deformation. 
  Data taken from {\rm{NuDat 3.0}}.}
    \begin{tabular}{|c|c|c|c|c|}
    \hline
    Nucleus & $Z$ & $N$ & $B/A$ (MeV) & $E_{4^+}/E_{2^+}$\\
    \hline
    \nuclide{ O }{ 16 }  & $ 8 $  & $ 8 $  &  $7.98$ & $1.50$ \\
    \nuclide{ Ne }{ 20 } & $ 10 $ & $ 10 $ &  $8.03$ & $2.60$ \\
    \nuclide{ O }{ 22 }  & $ 8 $  & $ 14 $ &  $7.36$ & $2.17$ \\
    \nuclide{ Mg }{ 24 } & $ 12 $ & $ 12 $ &  $8.26$ & $3.01$ \\
    \nuclide{ Ne }{ 32 } & $ 10 $ & $ 22 $ &  $6.67$ & $ -  $ \\
    \nuclide{ Mg }{ 34 } & $ 12 $ & $ 22 $ &  $7.55$ & $3.21$ \\
    \nuclide{ Mg }{ 36 } & $ 12 $ & $ 24 $ &  $7.24$ & $3.03$ \\
    \nuclide{ Ca }{ 40 } & $ 20 $ & $ 20 $ &  $8.55$ & $1.35$ \\
    \nuclide{ S }{ 42 }  & $ 16 $ & $ 26 $ &  $8.19$ & $3.02$ \\
    \nuclide{ Ti }{ 44 } & $ 22 $ & $ 22 $ &  $8.53$ & $2.27$ \\
    \nuclide{ Ti }{ 46 } & $ 22 $ & $ 24 $ &  $8.66$ & $2.26$ \\
    \nuclide{ Ca }{ 48 } & $ 20 $ & $ 28 $ &  $8.67$ & $1.18$ \\
    \nuclide{ Cr }{ 48 } & $ 24 $ & $ 24 $ &  $8.57$ & $2.47$ \\
    \nuclide{ Ca }{ 52 } & $ 20 $ & $ 32 $ &  $8.43$ & $ -  $ \\
    \nuclide{ Cr }{ 52 } & $ 24 $ & $ 28 $ &  $8.78$ & $1.65$ \\
    \nuclide{ Fe }{ 52 } & $ 26 $ & $ 26 $ &  $8.61$ & $2.81$ \\
    \nuclide{ Ti }{ 54 } & $ 22 $ & $ 32 $ &  $8.60$ & $1.67$ \\
    \nuclide{ Ni }{ 56 } & $ 28 $ & $ 28 $ &  $8.64$ & $1.45$ \\
    \hline
  \end{tabular}
  \label{table:learned-nuclides}
\end{table}

A look at Eq.~\eqref{eq:least-squares} shows that for a given test point (i.e., for a given set of $11$ \xspace {low-energy constants}\xspace $\mathbf{x}$), the evaluation of the objective function requires  $n_\text{learned} = 18$ Hartree-Fock calculations.
The construction of fast Hartree-Fock emulators makes the calibration practical, and this is discussed in the next section.
\section{Mass Model Calibration via Fast Hartree-Fock Emulation}\label{sec:calibration} 
\label{sec:methods}
The three-body potential of the \xspace$\Delta$NNLO$_{\rm GO}(394)$\xspace interaction is included in the normal-ordered two-body approximation~\cite{hagen2007a,roth2012}, and we  follow \textcite{Frosini:2021tuj} for open-shell nuclei. Thus, for a fixed parameter set $\mathbf{x}$ one first computes the normal-ordered two-body Hamiltonian by filling open harmonic oscillator orbitals with fractional and equal occupation numbers. Then the normal-ordering approximation is performed and residual three-nucleon forces are neglected in what follows. In a second step one performs a symmetry-breaking Hartree-Fock computation using the one-body and two-body matrix elements of the normal-ordered interaction. For all deformed nuclei we chose an axially symmetric prolate shape. 

The evaluation of the objective function~\eqref{eq:least-squares} is still expensive because 
(i) the three-body interaction has to be read-in once per-nucleus to perform the normal ordered two-body approximation,
(ii) the resulting zero-, one- and two-body matrix elements have to be stored to disk, and
(iii) they need to be loaded into memory for each nucleus and for each Hartree-Fock calculation.
As a result, a single Hartree-Fock calculation takes on the order of tens of minutes on current supercomputing hardware.  
    
To reduce this computational burden, we constructed fast Hartree-Fock emulators following~\cite{sun2024}, using a reduced basis method~\cite{Benner2017} known within the nuclear physics community as eigenvector continuation
\cite{frame2018,ekstrom2019,Konig:2019adq,
Drischler2021a,
Drischler2022,
Yapa2022,
Duguet2023b}.

\subsection{Emulator Construction} 

\label{sec:emulator-construction}
We are interested in emulating the Hartree-Fock energy for a given nucleus at a target point $\mathbf{x}$ in the vicinity of a ``central'' point $\mathbf{x}_0$ and follow Ref.~\cite{sun2024}. The main idea consists of computing $d_{\mathrm{RB}}$ exact Hartree-Fock wavefunctions (or ``snapshots'') $\ket{\psi(\mathbf{x}_\alpha)}$ corresponding to  \xspace {low-energy constants}\xspace  $\mathbf{x}_\alpha$, with $\alpha = 1, 2, \ldots,  d_{\mathrm{RB}}$. 
Here, $d_{\mathrm{RB}}$ is the dimension of the reduced basis model.

Next, the snapshots are utilized to construct an $d_{\mathrm{RB}}\times d_{\mathrm{RB}}$ matrix model  where the Hamiltonian and norm matrices have elements
\begin{equation}
  \begin{aligned}
    H^{\rm{RB}}_{\alpha\beta}(\mathbf{x}) &= \bra{\psi(\mathbf{x}_\alpha)} H^{\rm{HF}}(\mathbf{x}) \ket{\psi(\mathbf{x}_\beta)}\,,\\
    N^{\rm{RB}}_{\alpha\beta} &= \braket{\psi(\mathbf{x}_\alpha)}{\psi(\mathbf{x}_\beta)}\,.
  \end{aligned}
  \label{eq:evc-projection-ideal}
\end{equation}
Here, $\alpha$ and $\beta$ are indices running from 1 to $d_\text{RB}$. 

The estimated Hartree-Fock energy obtained from the reduced basis approach $E^{\textrm{RB}}$ can be obtained by solving the generalized eigenvalue problem 
\begin{equation}
    H^{\rm{RB}}(\mathbf{x}) \ket{\psi^{\text{RB}}} = E^{\textrm{RB}} N^{\rm{RB}} \ket{\psi^{\text{RB}}}\,.
    \label{eq:emulated-hf-energy}
\end{equation}
An emulator is said to be accurate when  $E^{\textrm{RB}} \approx E^{\textrm{HF}}$ holds with high accuracy. 
Note that the reduced basis Hamiltonian, along with its eigenstate and corresponding eigenvalue, depends explicitly on the \xspace {low-energy constants}\xspace $\mathbf{x}$.

While the evaluation of the Hamiltonian and norm kernels in Eqs.~\eqref{eq:evc-projection-ideal} is straightforward, we must address a complication relating to the Hartree-Fock Hamiltonian therein.
Unlike the case of Hamiltonian~\eqref{eq:h-matrices}, the Hartree-Fock matrices corresponding to the operators appearing on the right hand side of  Eq.~\eqref{eq:h-matrices} depend also on the \xspace {low-energy constants}\xspace $\mathbf{x}$. In Eqs.~(\ref{eq:evc-projection-ideal}) we therefore use 
\begin{equation}
  \begin{aligned}
    H^{\rm{RB}}(\mathbf{x})\ket{\psi(\mathbf{x}_\beta)} &\equiv h^{\rm{RB}}_0(\mathbf{x}_\beta)\ket{\psi(\mathbf{x}_\beta)}\\
    &+ \sum_{i=1}^{11} x_i h^{\rm{RB}}_i(\mathbf{x}_\beta)\ket{\psi(\mathbf{x}_\beta)}\,.
  \end{aligned}
  \label{eq:reduced-hf-hamil-actual}
\end{equation}
Here, $h^{\textrm{RB}}_0$ and $h_{i}^{{\rm{RB}}}$ are the Hartree-Fock matrices corresponding to the respective terms in the full Hamiltonian~\eqref{eq:h-matrices}, obtained using the \xspace {low-energy constants}\xspace $\mathbf{x}_\beta$ appearing on the right (ket) side in Eq.~\eqref{eq:evc-projection-ideal}.
This approach ensures that product states (and not a superposition of product states) are exact solutions of the emulator at the training points $\mathbf{x}_\beta$. It inherently linearizes the nonlinear Hartree Fock emulation problem. Assuming that the set of training points $\{\mathbf{x}_\alpha\}$ is sufficiently close to the point $\mathbf{x}$ of interest, the Hartree-Fock emulator is expected to be accurate. 

We note that the generalized eigenvalue problem~(\ref{eq:emulated-hf-energy}) is non-symmetric.
We have used the right eigenvalues in the present work (and checked, of course, that right and left eigenvalues agreed). We constructed a Hartree-Fock emulator for each nucleus entering the objective function~(\ref{eq:least-squares}), listed in  Table~\ref{table:learned-nuclides}.  

We have to decide about the region in the parameter space of low-energy constants for which to construct the emulators. Starting from the parameterization $\mathbf{x}_0$ of the \xspace$\Delta$NNLO$_{\rm GO}(394)$\xspace interaction, we choose a $\pm 10$\% region around $\mathbf{x}_0$
for all low-energy constants except $c_D$ and $c_E$. 
The three-body contact strengths $c_D$ and $c_E$ were unnaturally small in the original \xspace$\Delta$NNLO$_{\rm GO}(394)$\xspace interaction, being $+0.081$ and $-0.002$, respectively, and we have allowed these to span the natural ranges of about $[-0.7, 0.9]$ and $[-0.7, 0.7]$,
respectively.

In this neighborhood of $\mathbf{x}_0$, we choose $d_{\rm{RB}} = 68$ points using the latin hypercube sampling scheme to capture the Hartree-Fock snapshots.
Using those, we construct and store to disk the $68\times 68$ subspace matrices $h^{\textrm{RB}}_0$, $h_{i}^{{\rm{RB}}}$ and $N^{\textrm{RB}}$.
Then, given any set of \xspace {low-energy constants}\xspace $\mathbf{x}$ in the target region, the reduced basis Hartree-Fock Hamiltonian can be quickly computed using the expansion appearing in  Eq.~\eqref{eq:reduced-hf-hamil-actual} to solve for the Hartree-Fock energy. 
This completes construction of the fast Hartree-Fock emulator for a given nucleus.

\subsection{Emulator Validation} 
\label{sec:emulator-validation}

For a set of \xspace {low-energy constants}\xspace, the emulator generates an approximation for the Hartree-Fock energy for the nucleus to which it corresponds.
To check the emulator's accuracy, we compare results from exact Hartree-Fock calculations at randomly chosen validation points $\mathbf{x}_a$ in the neighborhood of $\mathbf{x}_0$ with the corresponding emulator results. 

\begin{figure*}[htb]
  \centering
  \includegraphics[width=0.32\textwidth]{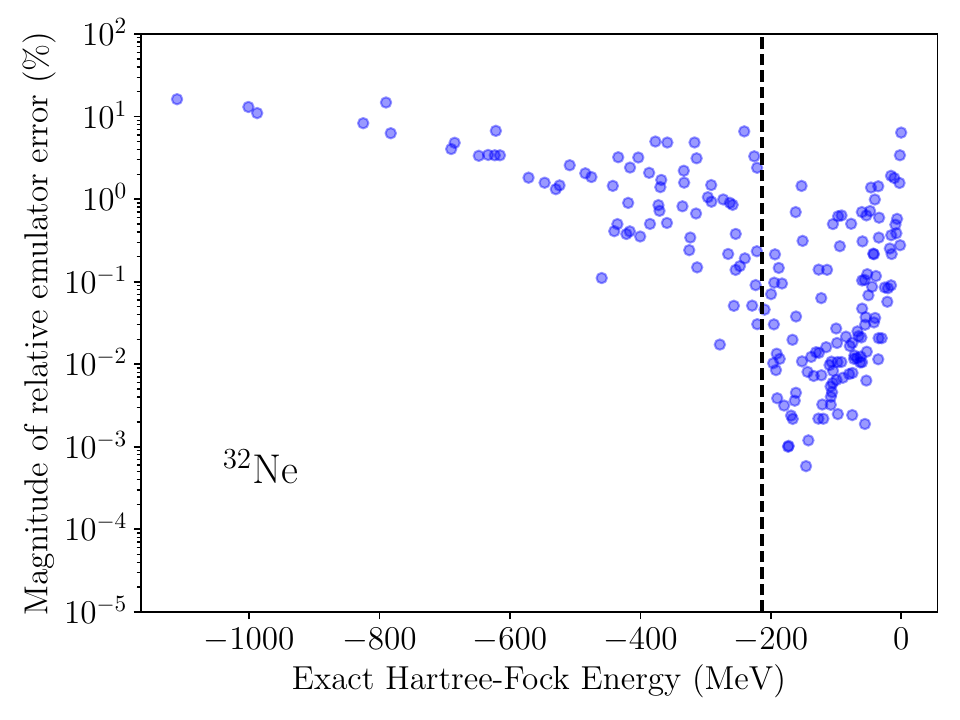}
  \includegraphics[width=0.32\textwidth]{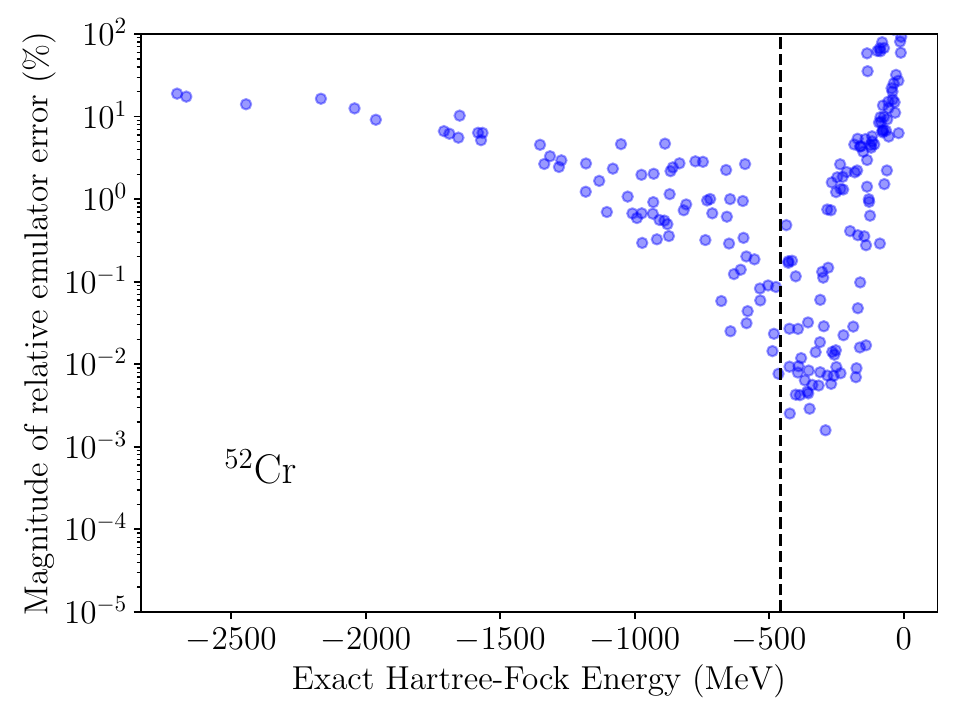}
  \includegraphics[width=0.32\textwidth]{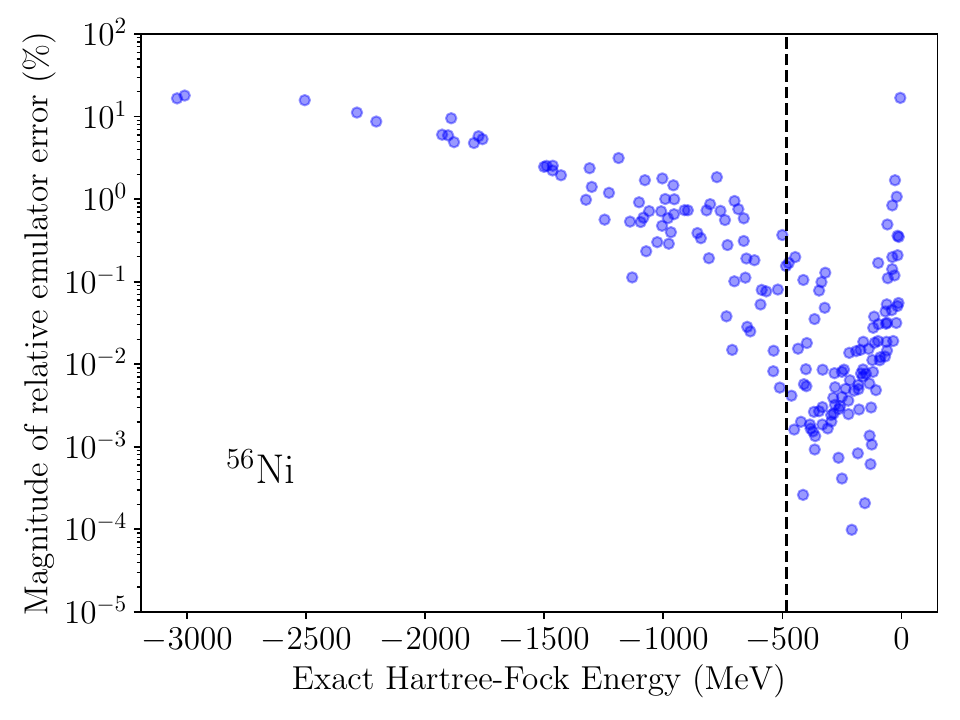}  
  \caption{
  Magnitude of the relative error (in \%) between the emulated and exact Hartree-Fock energy for $220$ pseudo-randomly sampled \xspace {low-energy constants}\xspace (shown as blue points) in the nuclei 
  \nuclide{Ne}{32},
  \nuclide{Cr}{52} and
  \nuclide{Ni}{56}.
  Dashed vertical lines mark the experimental ground state energies.}
  \label{fig:emulator-validation}
\end{figure*}

Figure~\ref{fig:emulator-validation} shows the validation results for 
\nuclide{Ne}{32},
\nuclide{Cr}{52}, and
\nuclide{Ni}{56}.
They are based on about 220 validation points $\mathbf{x}_a$ in the ``10\% neighborhood'' of $\mathbf{x}_0$, with $c_D$ and $c_E$ being allowed to span the range $[-2,2]$.
The magnitude of relative errors in percentage are plotted against the exact Hartree-Fock energies and vertical dashed lines mark the experimental ground-state energies. The Hartree-Fock emulator typically reproduces ground-state energies with an absolute percentage error below 10\%. However, the emulator is frequently accurate to better than $1\%$ in the vicinity of the experimentally relevant energy region.

This will allow us to credibly go forward with the optimization using emulators.  
The Hartree-Fock energies at the central \xspace$\Delta$NNLO$_{\rm GO}(394)$\xspace point $\mathbf{x}_0$ yields binding energies of about 50\% of the experimental value. Thus, a significant renormalization will be required. However, our chosen neighborhood of parameter values seems to be sufficiently large to accomplish the desired renormalization. 

\begin{figure}[hbt!]
  \centering
  \includegraphics[width=1.0\columnwidth]{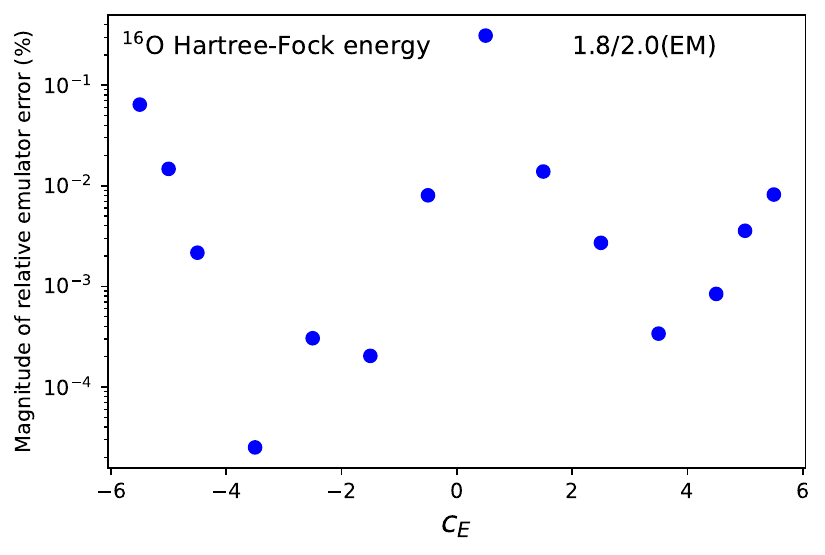}
  \caption{
    Magnitude of the relative emulator error (in \%) for the Hartree-Fock energy of $^{16}$O computed with the interaction $1.8/2.0$(EM). We omit showing the training points which have a relative error of $\sim 10^{-10}$ \%. 
  }
  \label{fig:O16-bench}
\end{figure}

To put the emulator accuracy into perspective, we also show results for a Hartree-Fock emulator for the nucleus $^{16}$O based on the interaction $1.8/2.0$(EM) of Ref.~\cite{hebeler2011}. Here, we varied only a single low-energy constant and chose the three-body contact $c_E$. Figure~\ref{fig:O16-bench} shows the results. The training points were the nine points of the set $[-4, -3, -2,\ldots 4]$. We see that the emulator is of similar quality as the ones depicted in Fig.~\ref{fig:emulator-validation}, taking into account that many low-energy constants were varied for the latter.

\subsection{Mass Model Calibration} 

\label{sec:mass-model-calibration}
We repeat the procedure described above to build emulators for the $18$ nuclei listed in Table~\ref{table:learned-nuclides}.
This allows for fast evaluation of the objective function in Eq.~\eqref{eq:least-squares} at a given test point $\mathbf{x}$, enabling local optimization methods to identify an approximate solution $\mathbf{x}_\text{opt}$ to Eq.~\eqref{eq:least-squares} in the neighborhood of an initial parameterization $\mathbf{x}_0$.

For the optimization we use the Levenberg-Marquardt method included in scipy \cite{virtanen2020scipy}, which is identical to the implementation employed in MINPACK \cite{more1980user}.
The residuals $\delta B_a$ permit a closed-form  computation of their Jacobians, which were directly supplied to the Levenberg-Marquardt method. 
We note that the computation of the Hartree-Fock mass model binding energy entails solving a generalized eigenvalue problem, viz.  Eq.~\eqref{eq:emulated-hf-energy}.
As such, the residuals $\delta B_a$ may not be strictly differentiable as a function of $\mathbf{x}$ when the least generalized eigenvalue exhibits multiplicity. 
However, 
in our tests, we never encountered this situation to within numerical precision.  
Thus, while the Levenberg-Marquardt method is not generally applicable to a nondifferentiable objective function, which \eqref{eq:least-squares} technically features, we did not encounter situations where nondifferentiability of the objective function caused an issue. 

\begin{figure}[htb]
  \centering
  \includegraphics[width=1.0\columnwidth]{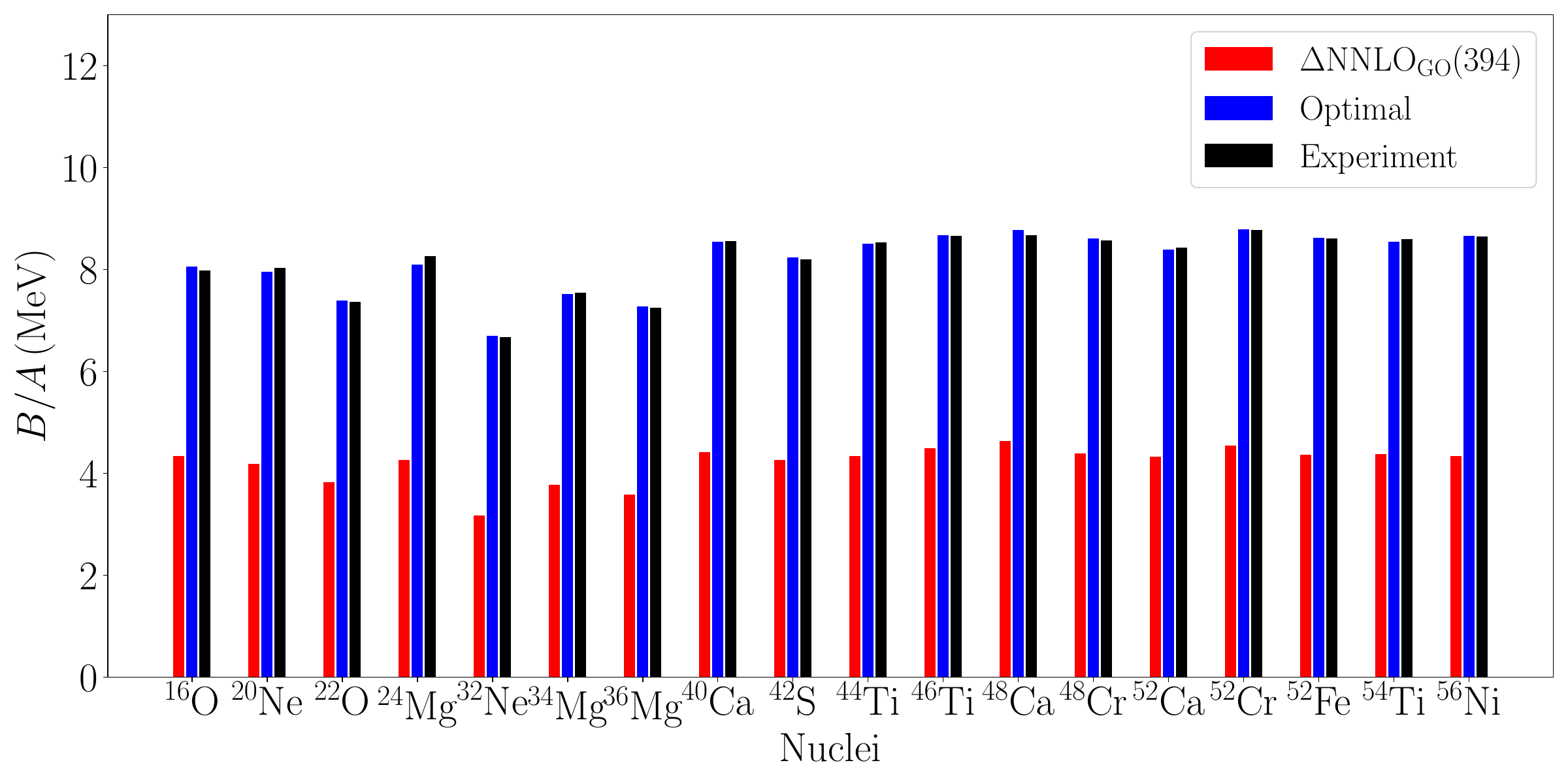}
  \caption{
    Hartree-Fock energy per nucleon before (red) and after (blue) renormalization and compared to experimental results (black) for the nuclei used in the calibration.
  }
  \label{fig:binding-bar-chart}
\end{figure}

Our optimization yields a RMS deviation of $2.17$~MeV for the 18 nuclei used in the calibration. Figure~\ref{fig:binding-bar-chart}
shows the binding energies from Hartree-Fock before and after the renormalization and compares them to experiment. The renormalization clearly is effective.  

The renormalized \xspace {low-energy constants}\xspace $\mathbf{x}_{\rm opt}$ that resulted from the calibration of the mass  model are listed in Table~\ref{table:lecs}. For comparison,  we also show the original values $\mathbf{x}_0$ of the \xspace$\Delta$NNLO$_{\rm GO}(394)$\xspace interaction. With the exception of the three-body contacts, the renormalized couplings are close to the original ones. We note that $c_D=-2$ is at the boundary allowed in the optimization.

\begin{table}[htp]
  \centering
  \caption{
    Low-energy constants of the short-range contact terms of the delta-full chiral interaction, ordered in three groups of leading-order, next-to-leading-order and three-body contacts (from top to bottom). 
    The values $\mathbf{x}_0$ of the \xspace$\Delta$NNLO$_{\rm GO}(394)$\xspace interaction are from Ref.~\cite{jiang2020} at the regulator cutoff of $\Lambda = 394$~MeV.
    The last column shows the optimal \xspace {low-energy constants}\xspace $\mathbf{x}_\mathrm{opt}$ from the renormalization.
    The leading-order and next-to-leading order low-energy constants are in units of
    $10^4$~GeV$^{-2}$ and 
    $10^4$~GeV$^{-4}$, respectively, while the three-body contacts $c_D$ and $c_E$ are dimensionless.
    }
  \begin{tabular}{|c|c|c|} 
    \hline
    $\mathbf{x}$ & \xspace$\Delta$NNLO$_{\rm GO}(394)$\xspace ($\mathbf{x}_0$) & Optimal ($\mathbf{x}_{\textrm{opt}}$)\\
    \hline
    $\tilde{C}_{^1{\rm{S}_0}}$ & $-0.3388$  &  $-0.3727$ \\
    $\tilde{C}_{^3{\rm{S}_1}}$ & $-0.2598$  &  $-0.2694$ \\
    \hline
    $C_{^1{\rm{S}_0}}$ & $+2.5054$  &  $+2.7555$ \\
    $C_{^3{\rm{P}_0}}$ & $+0.7005$  &  $+0.6304$ \\
    $C_{^1{\rm{P}_1}}$ & $-0.3880$  &  $-0.3492$ \\
    $C_{^3{\rm{P}_1}}$ & $-0.9649$  &  $-1.0613$ \\
    $C_{^3{\rm{S}_1}}$ & $+1.0022$  &  $+0.9614$ \\
    $C_{^3{\rm{S}_1}-^3{\rm{D}_1}}$ & $+0.4525$  &  $+0.4766$ \\
    $C_{^3{\rm{P}_2}}$ & $-0.8831$  &  $-0.8098$ \\
    \hline
    $c_D$ & $+0.0814$  &  $-2.0000$ \\
    $c_E$ & $-0.0024$  &  $-0.6198$ \\
    \hline
  \end{tabular}
  \label{table:lecs}
\end{table}

\section{Results} 
\label{sec:mass-model-results}

\subsection{Renormalization and Naturalness }

In an effective field theory, coupling constants are expected to be of natural size, i.e. of order one when dimensions have been properly accounted for~\cite{lepage1997}. Following Refs.~\cite{manohar1984,georgi1993,Furnstahl1997} we extract powers of pion decay constant $f_\pi$ and the breakdown scale  $\Lambda_b$ to arrive at dimensionless low-energy constants. 
This means multiplying the low-energy constants by $f_\pi^2$ at leading order and by $f_\pi^2 \Lambda_b^2$
at next-to-leading order. We also divide the dimensionless low-energy constants by the usual factor $4\pi$ from angular integrals. 
 
\begin{figure}[htb]
  \centering
  \includegraphics[width=\columnwidth]{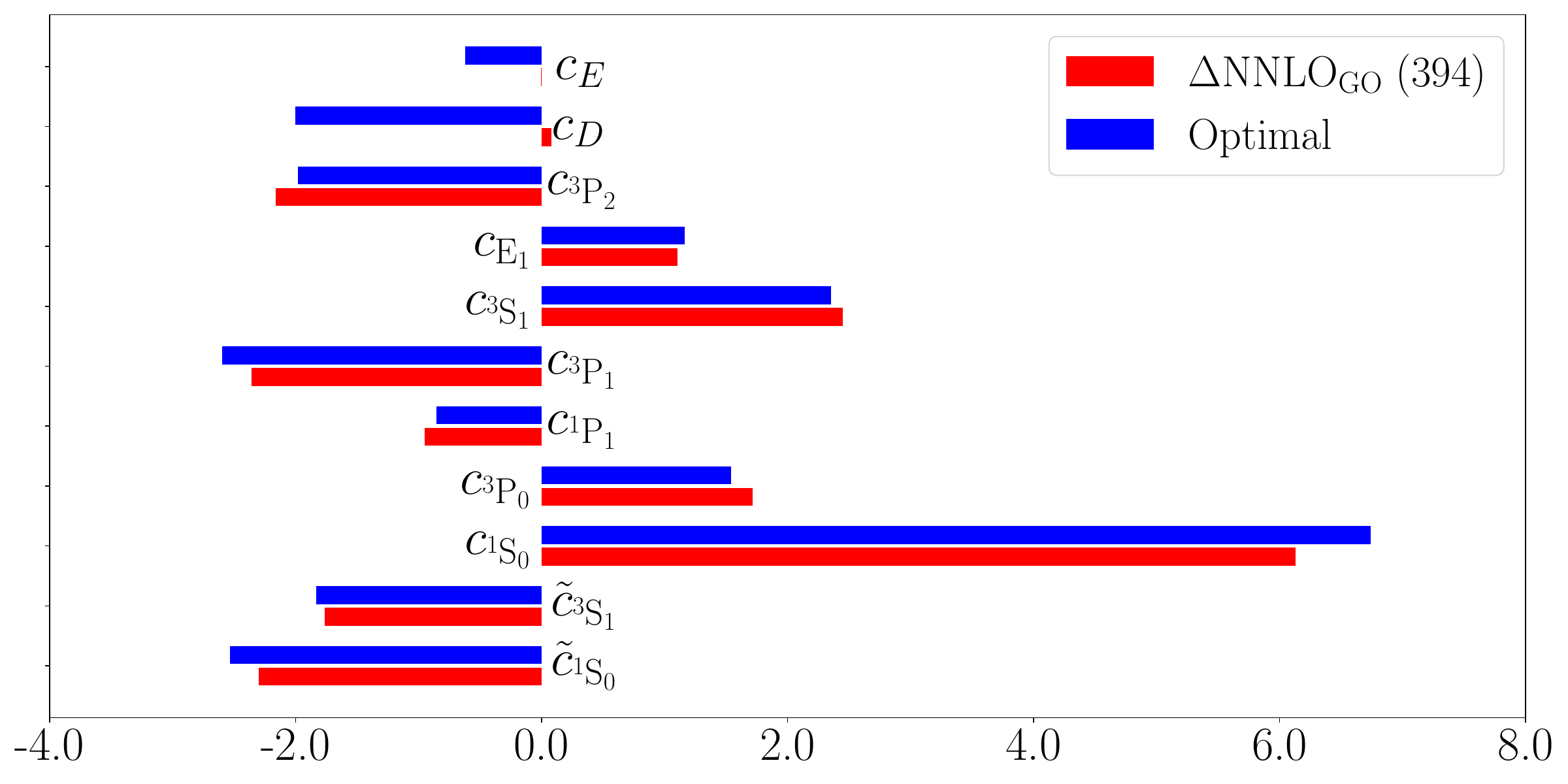}
  \caption{Dimensionless low-energy constants for the original \xspace$\Delta$NNLO$_{\rm GO}(394)$\xspace interaction(red) and the optimal interaction that renormalizes the Hartree-Fock energy (blue).
  }
  \label{fig:optimal-lecs}
\end{figure}

The results are shown in Fig.~\ref{fig:optimal-lecs} as red and blue bars for the dimensionless versions of the original and renormalized low-energy constants, $\mathbf{x}_0$ and $\mathbf{x}_{\rm opt}$, respectively. 
We see that most low-energy constants are of natural size. 
Exceptions are the three-body contacts of the \xspace$\Delta$NNLO$_{\rm GO}(394)$\xspace interaction, which are unnaturally small in $\mathbf{x}_0$
and the next-to-leading order $s$-wave contact $C_{^1S_0}$, which is a bit larger (by about a factor of two or so) than one would expect, both for the original and renormalized interaction. We note that the renormalization yields optimal three-body contacts that are of natural size.

\begin{figure*}
      \includegraphics[width=0.32\textwidth]{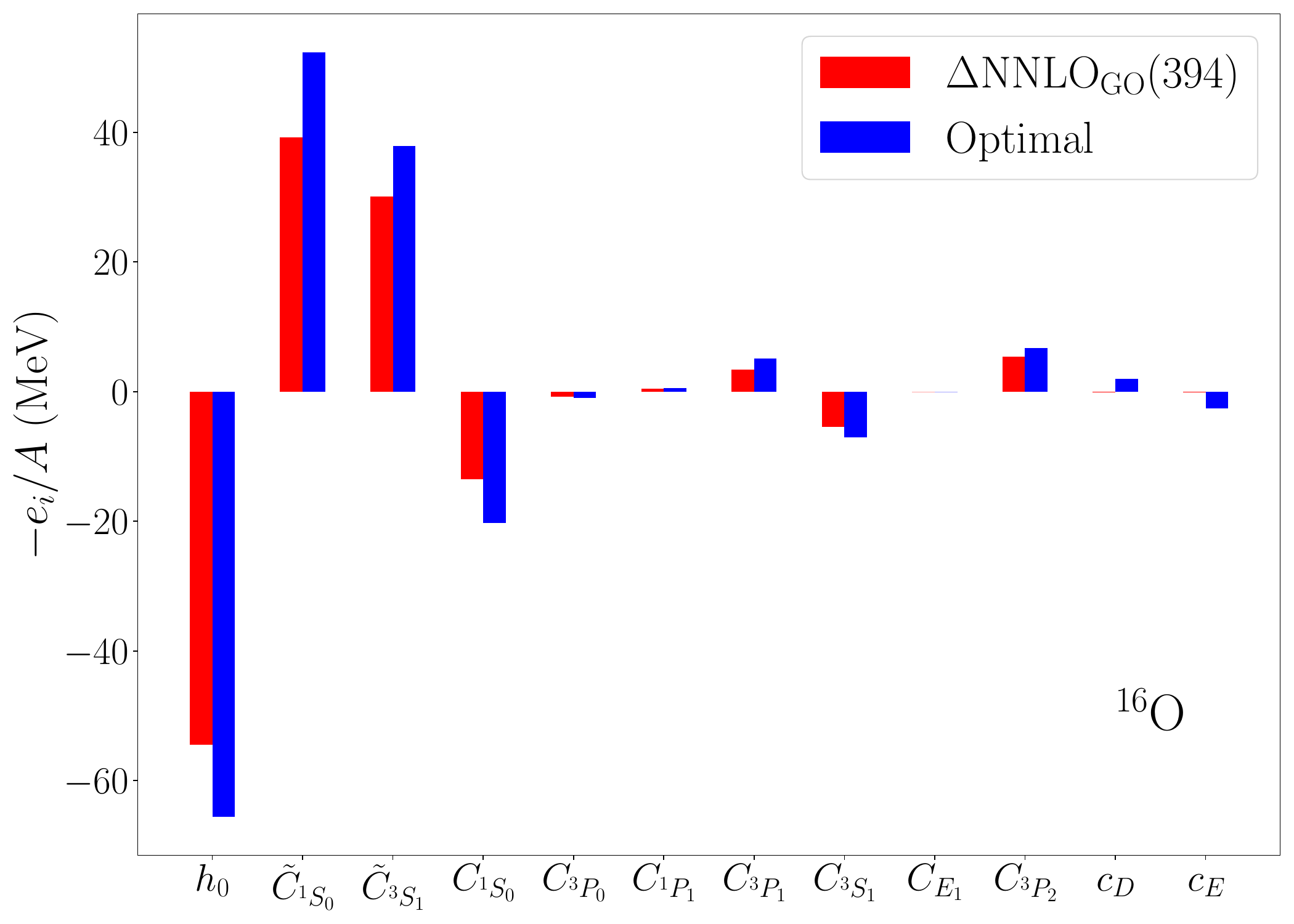}
      \includegraphics[width=0.32\textwidth]{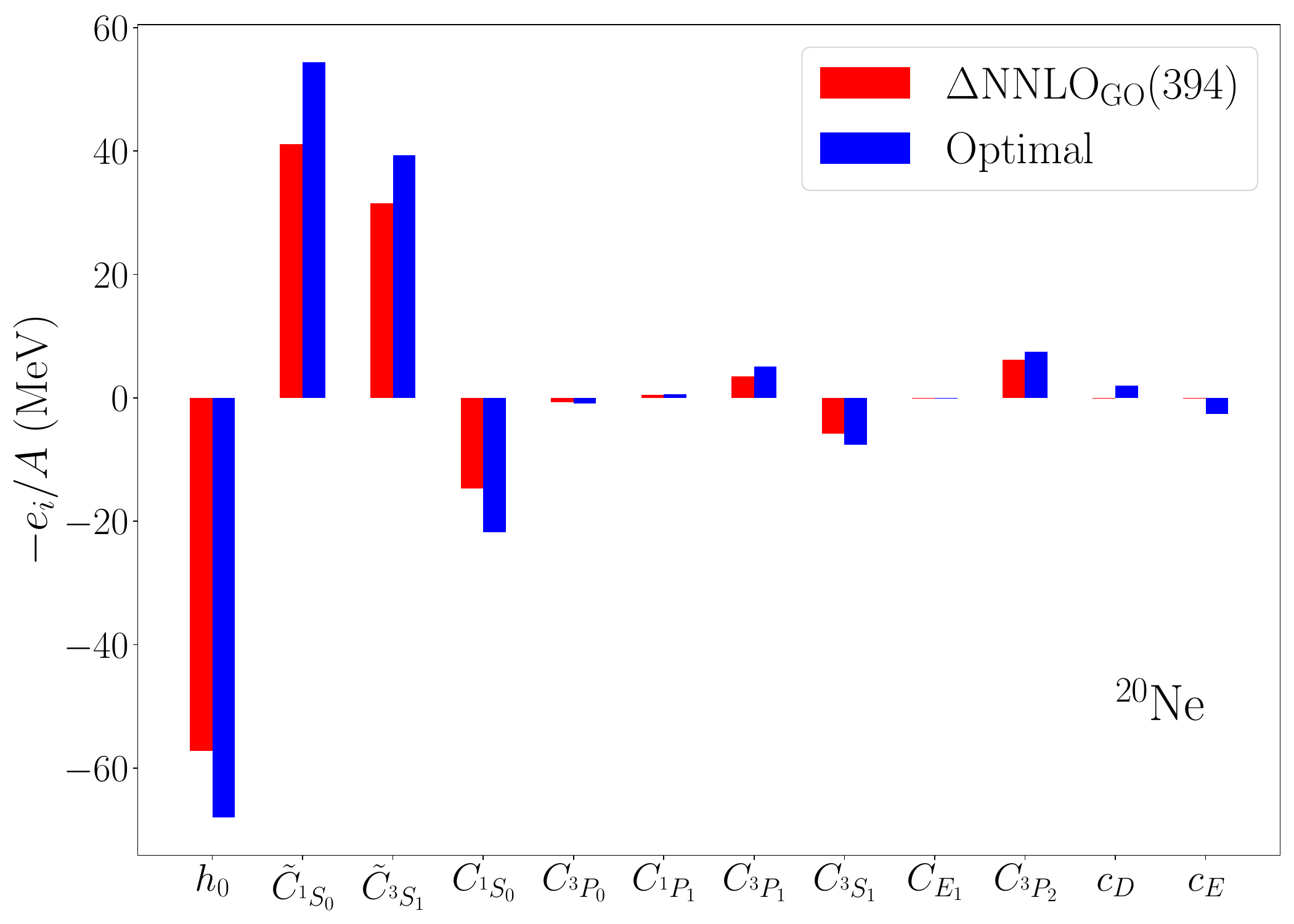}
      \includegraphics[width=0.32\textwidth]{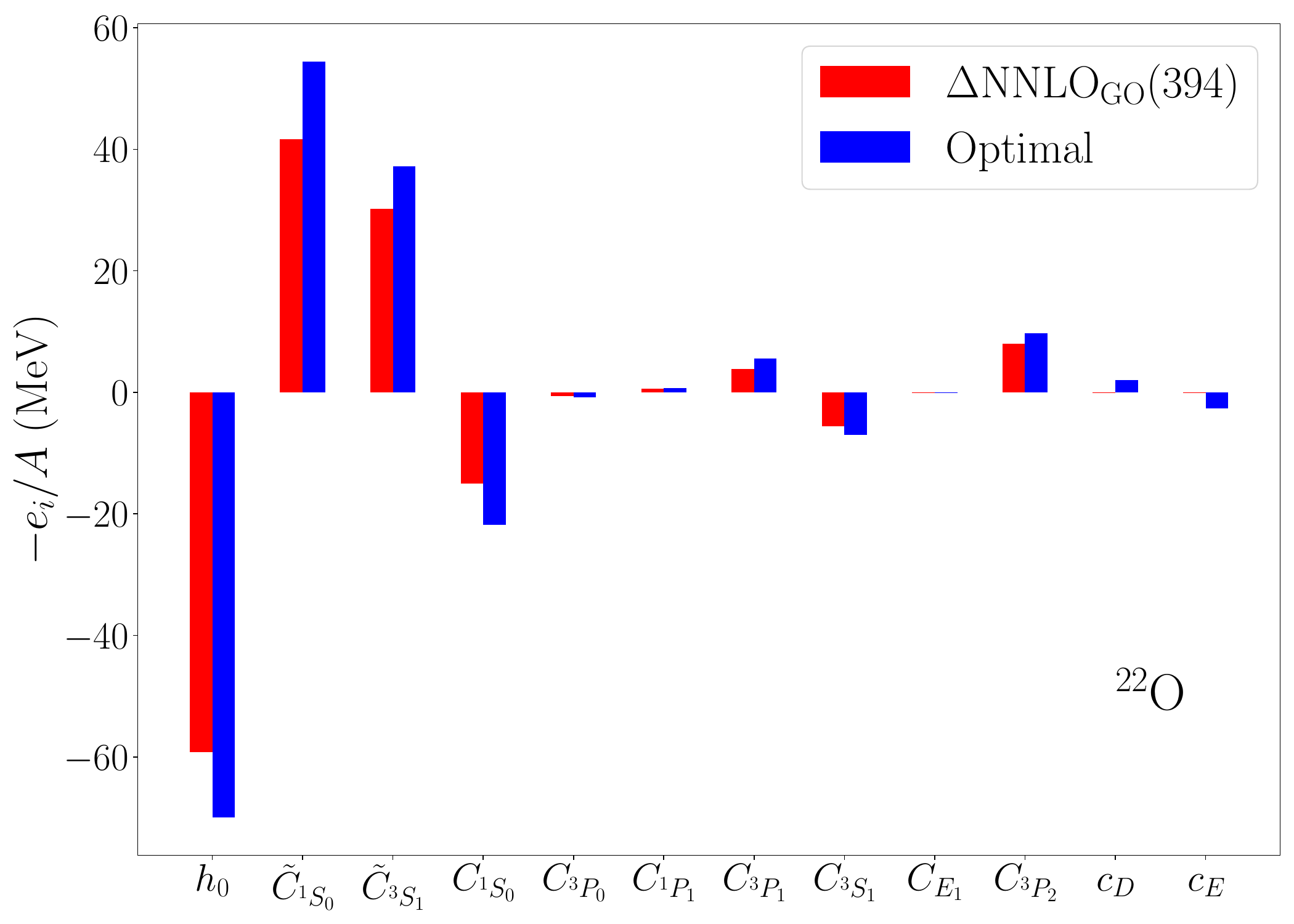}
      \includegraphics[width=0.32\textwidth]{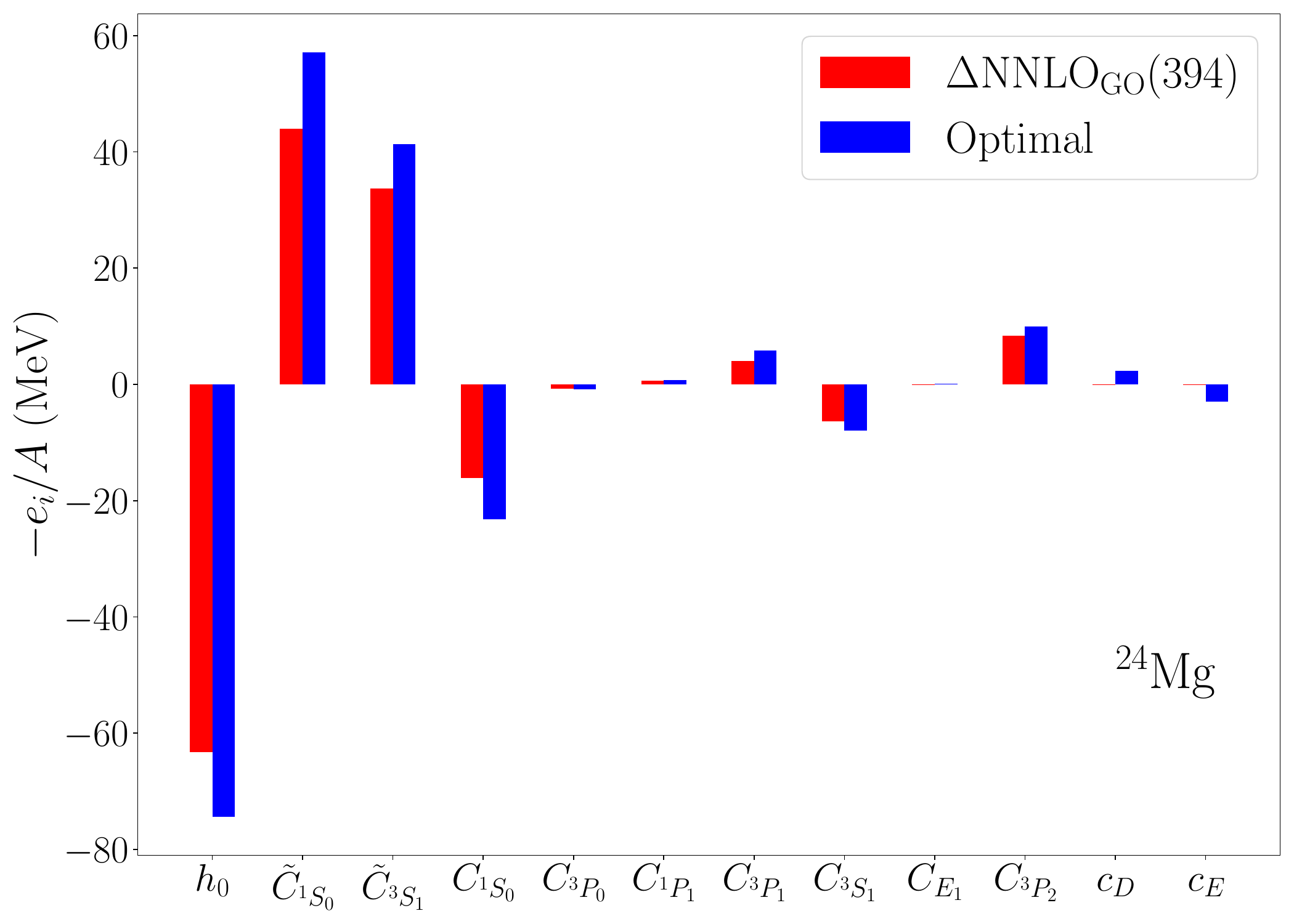} 
      \includegraphics[width=0.32\textwidth]{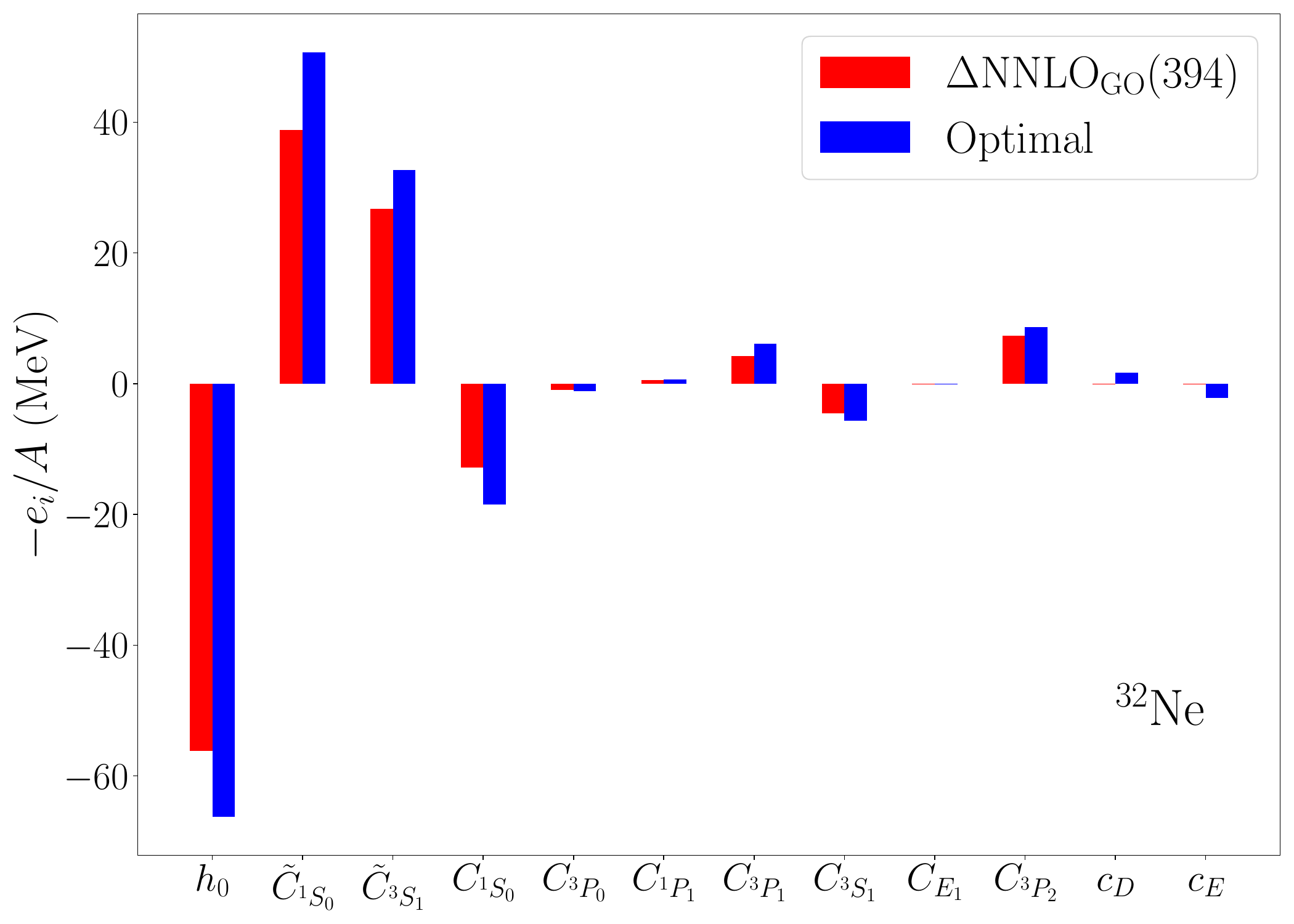}
      \includegraphics[width=0.32\textwidth]{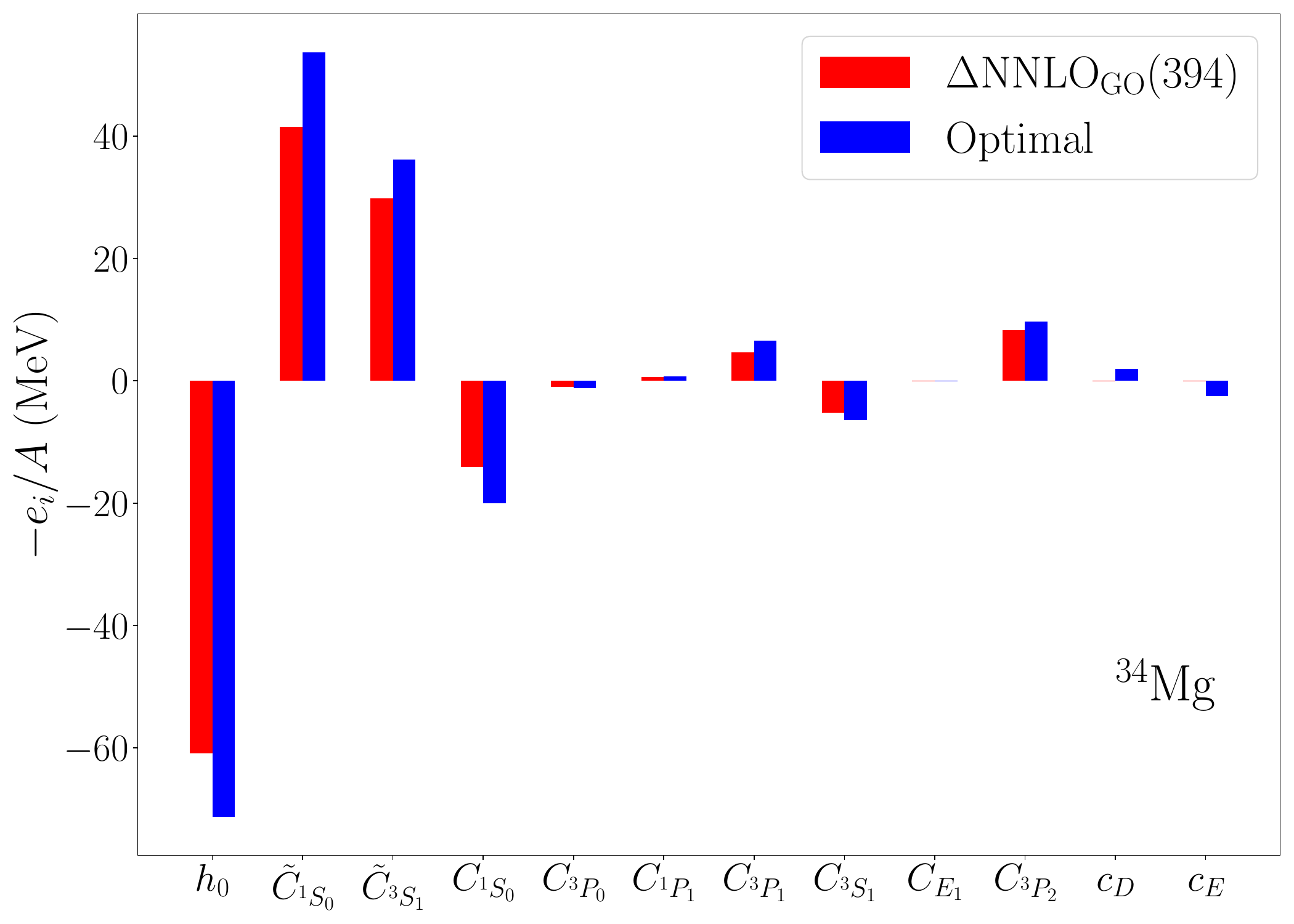} 
      \includegraphics[width=0.32\textwidth]{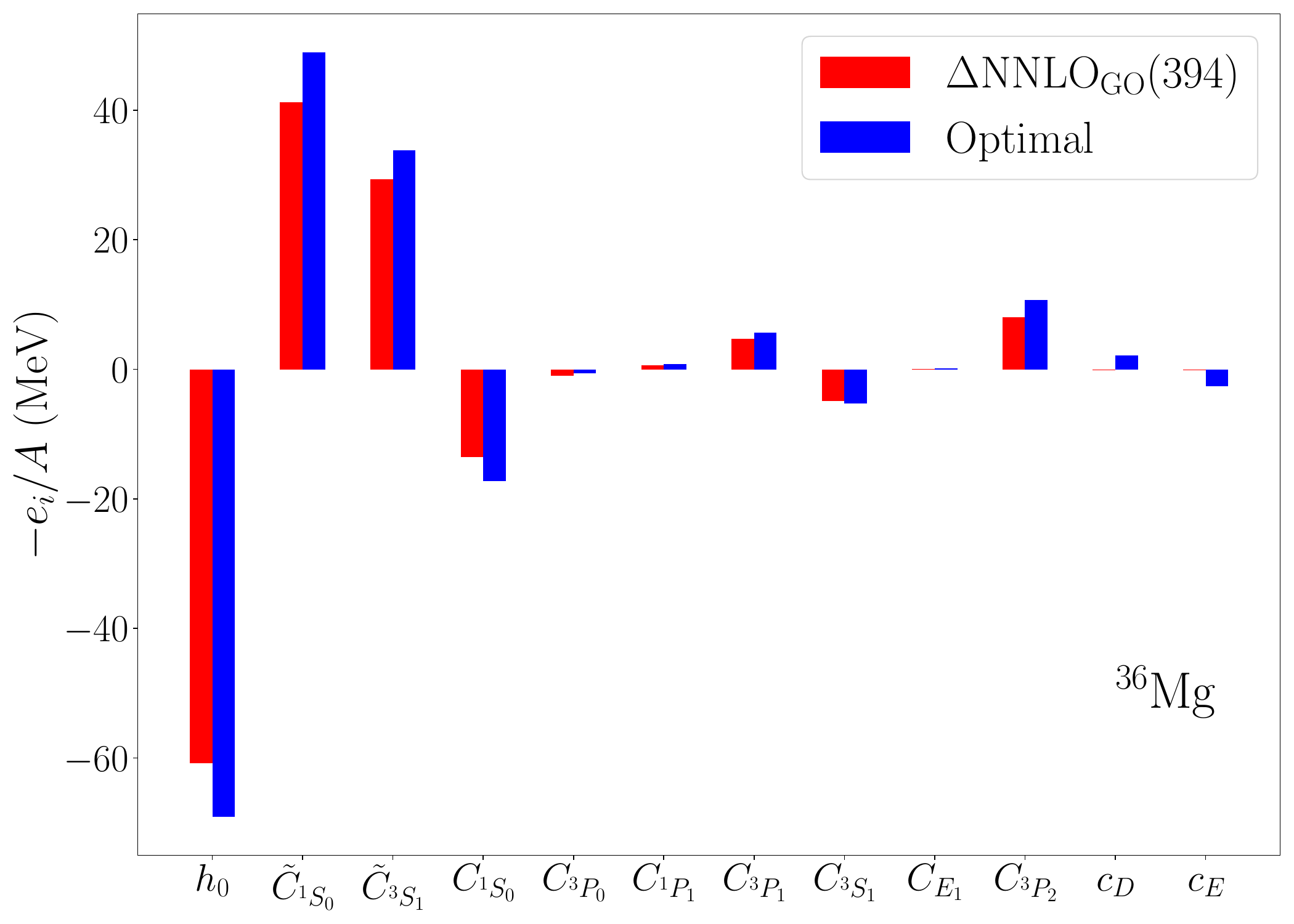}
      \includegraphics[width=0.32\textwidth]{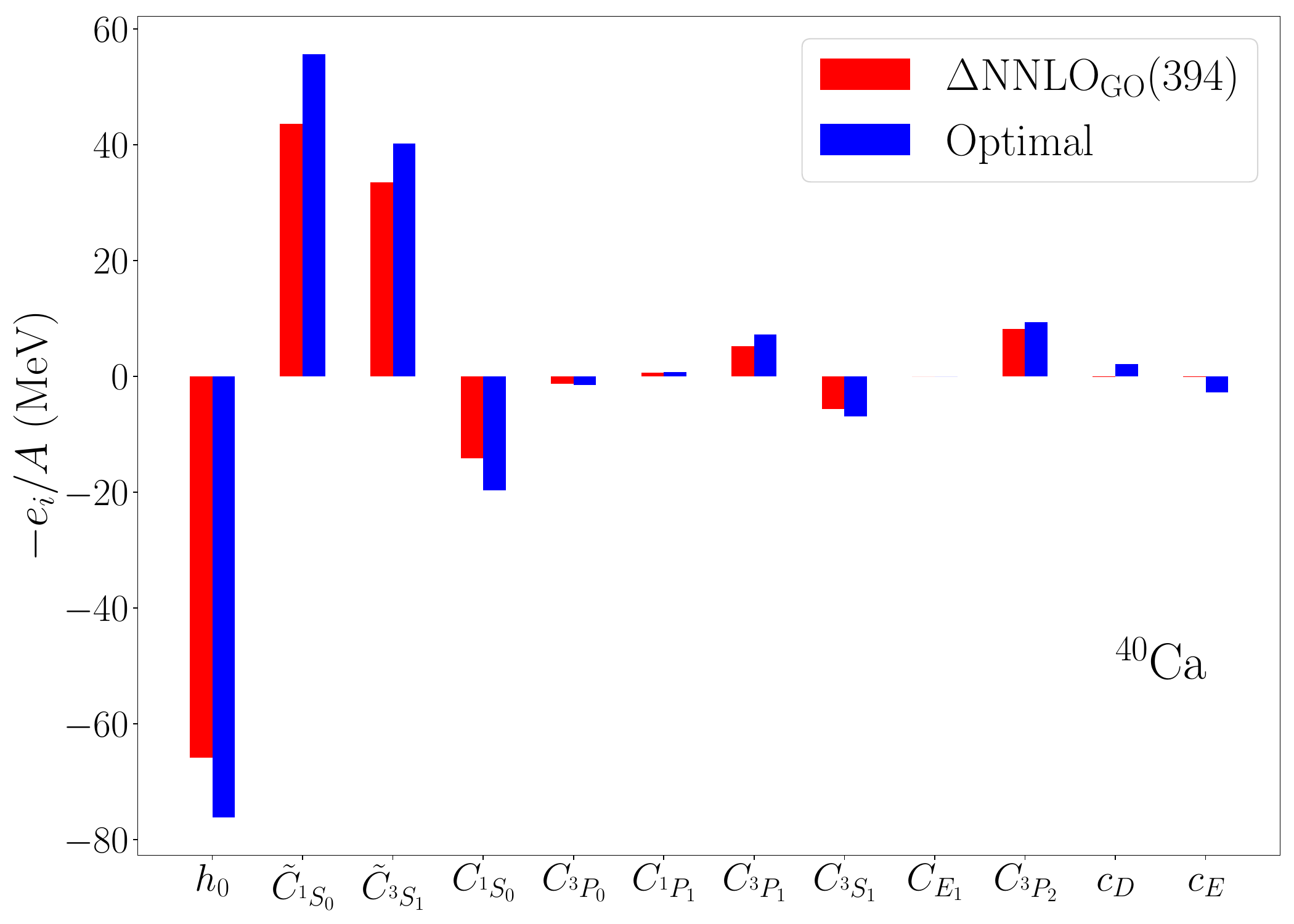} 
      \includegraphics[width=0.32\textwidth]{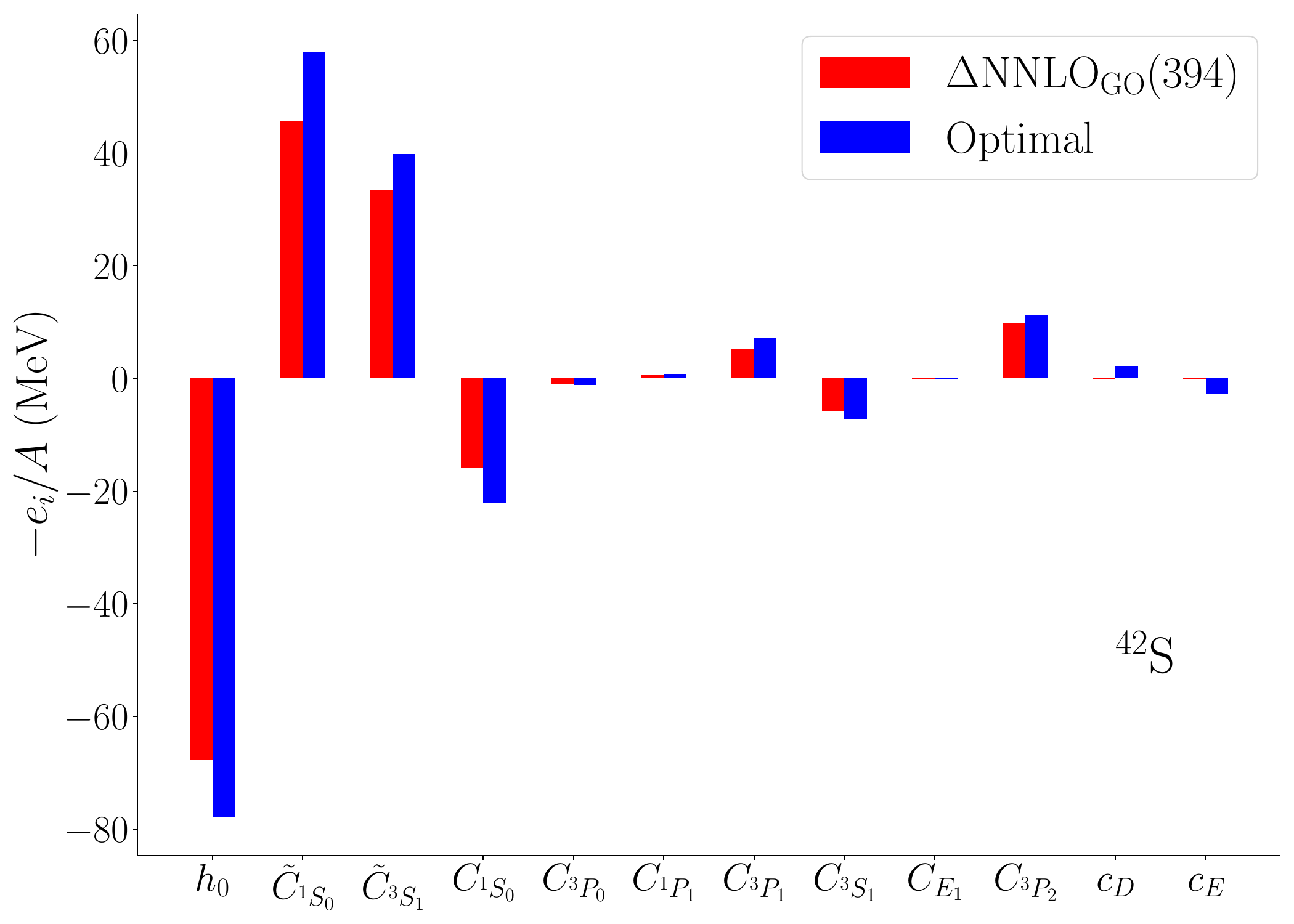} 
  \caption{Contributions to the Hartree-Fock energy per nucleon from the individual Hamiltonian terms $h_i$  for the nuclei $^{16,22}$O, $^{20,32}$Ne, $^{24,34,36}$Mg $^{40}$Ca and $^{42}$S used to calibrate the mass model. The red and blue bars shows results for the original and renormalized \xspace$\Delta$NNLO$_{\rm GO}(394)$\xspace interaction with low-energy constants $\mathbf{x}_0$ and $\mathbf{x}_{\rm opt}$, respectively.}
   \label{fig:individual-contributions1}
\end{figure*}

\begin{figure*} 
      \includegraphics[width=0.32\textwidth]{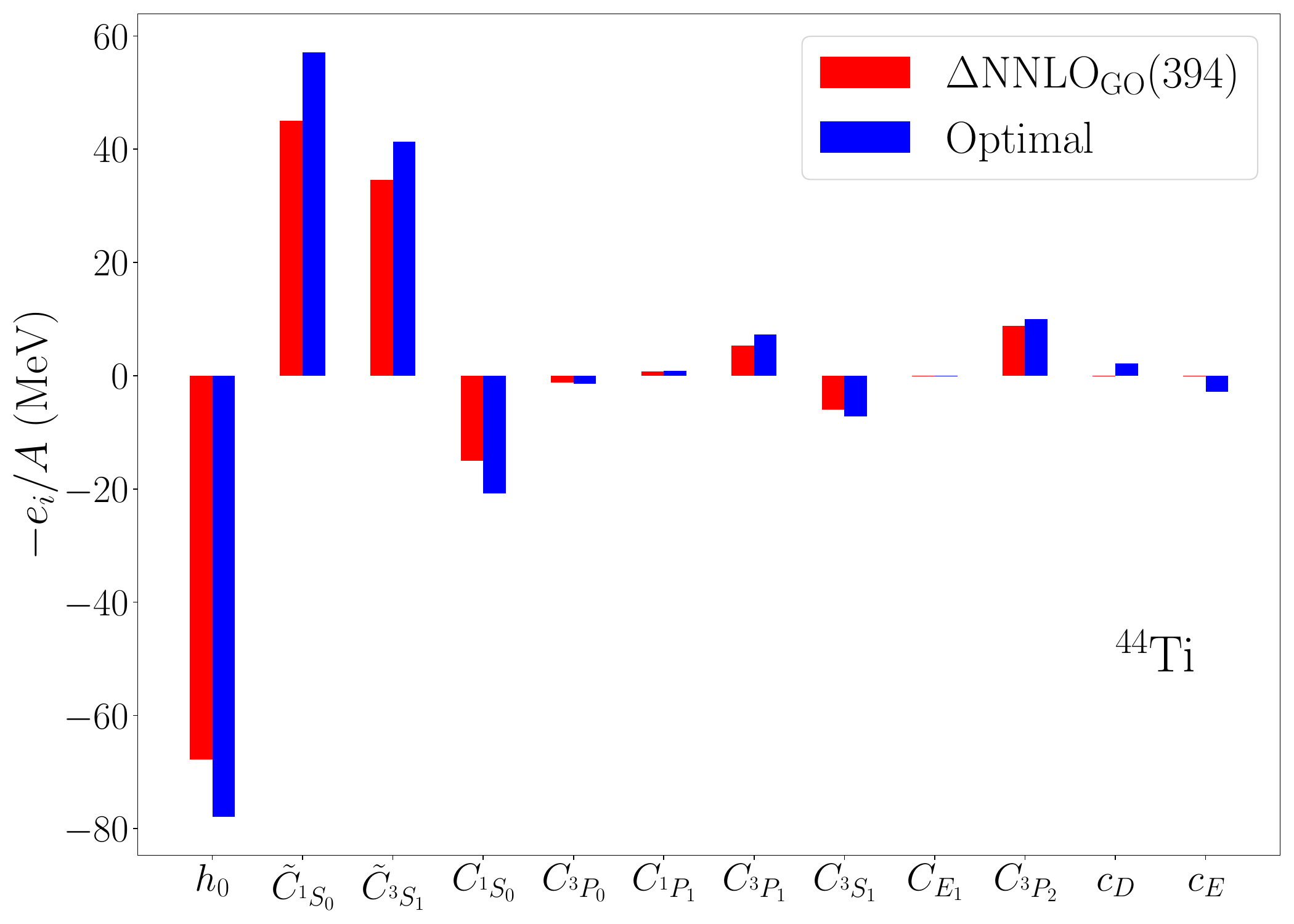} 
      \includegraphics[width=0.32\textwidth]{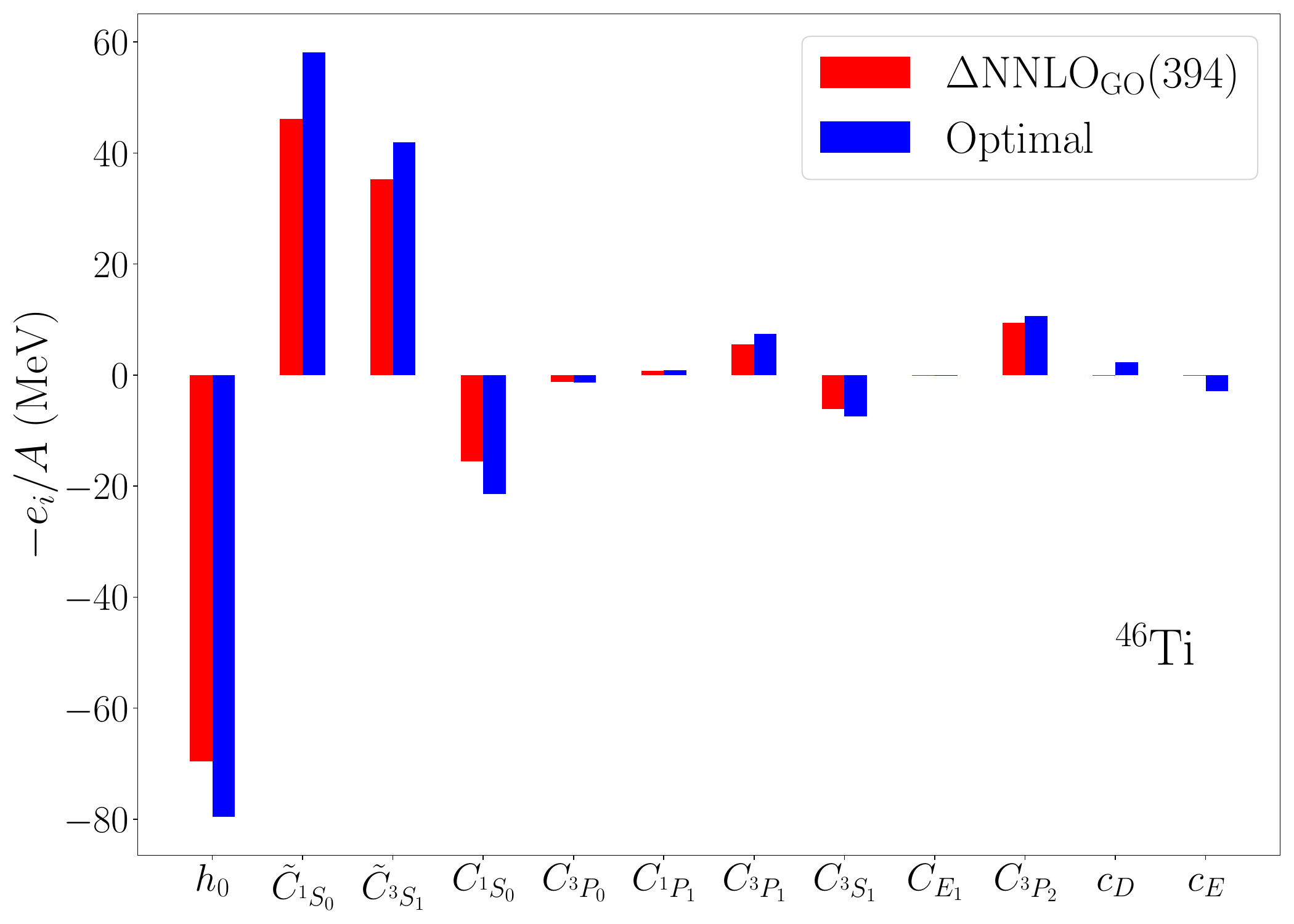} 
      \includegraphics[width=0.32\textwidth]{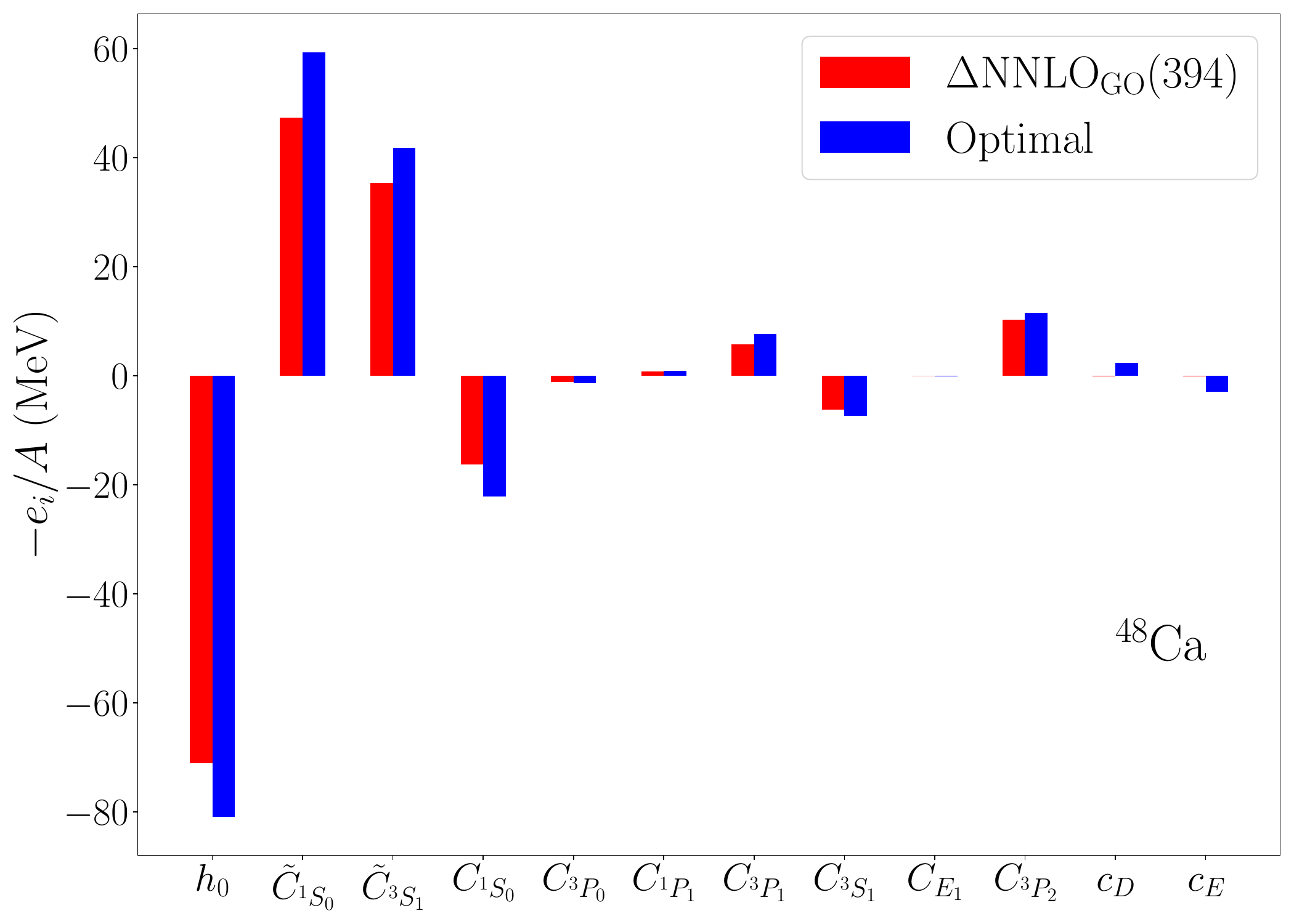} 
      \includegraphics[width=0.32\textwidth]{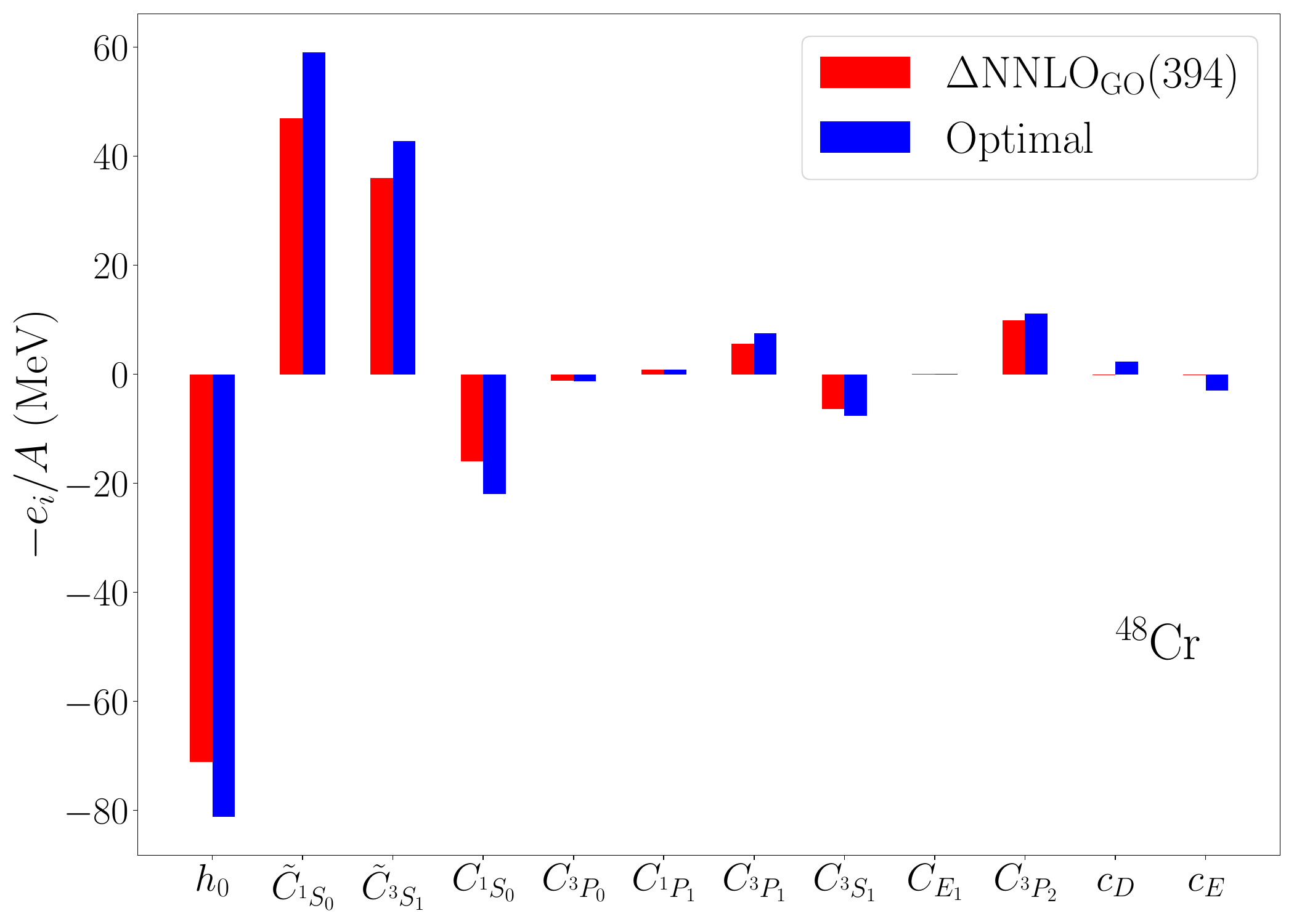} 
      \includegraphics[width=0.32\textwidth]{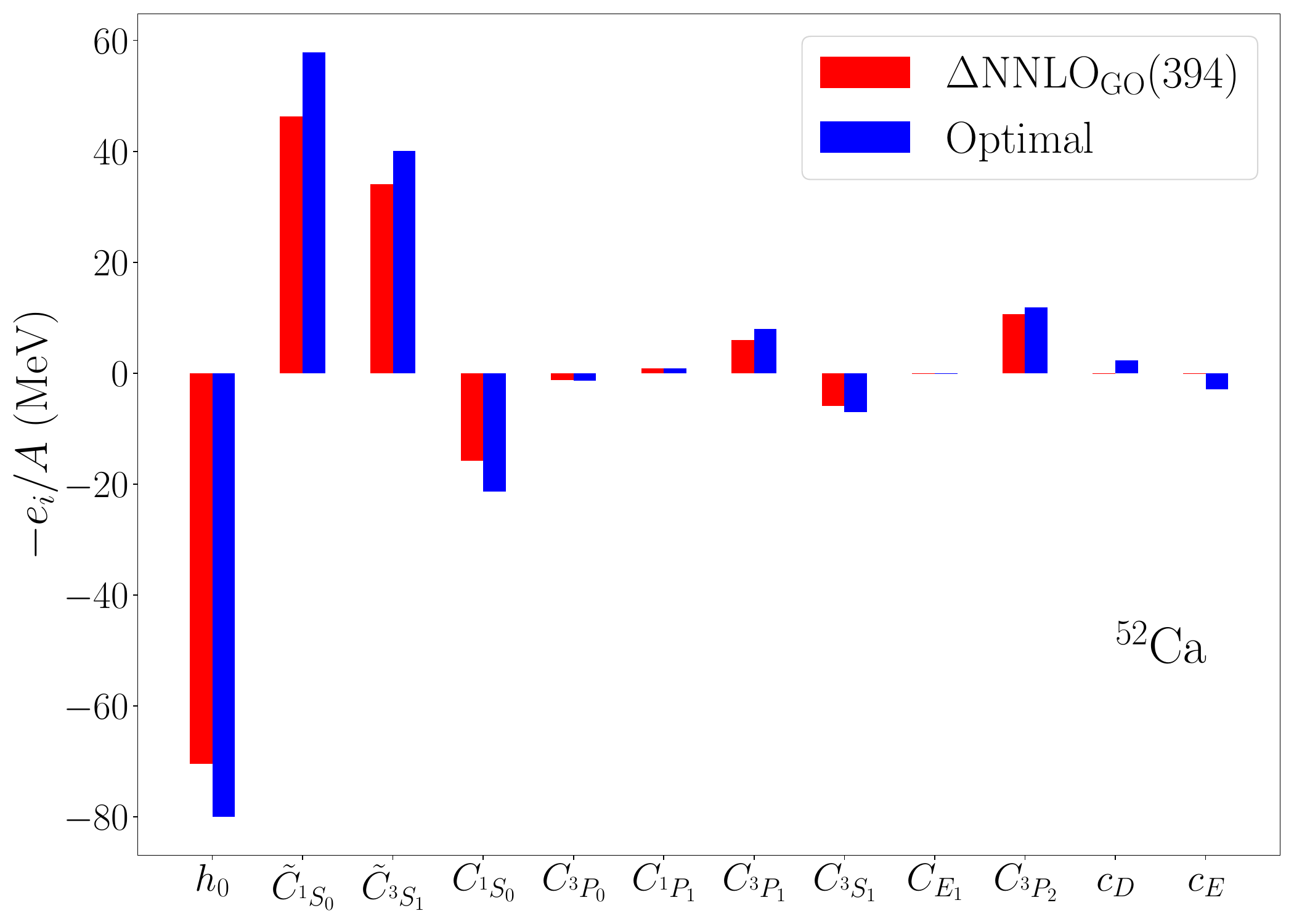} 
      \includegraphics[width=0.32\textwidth]{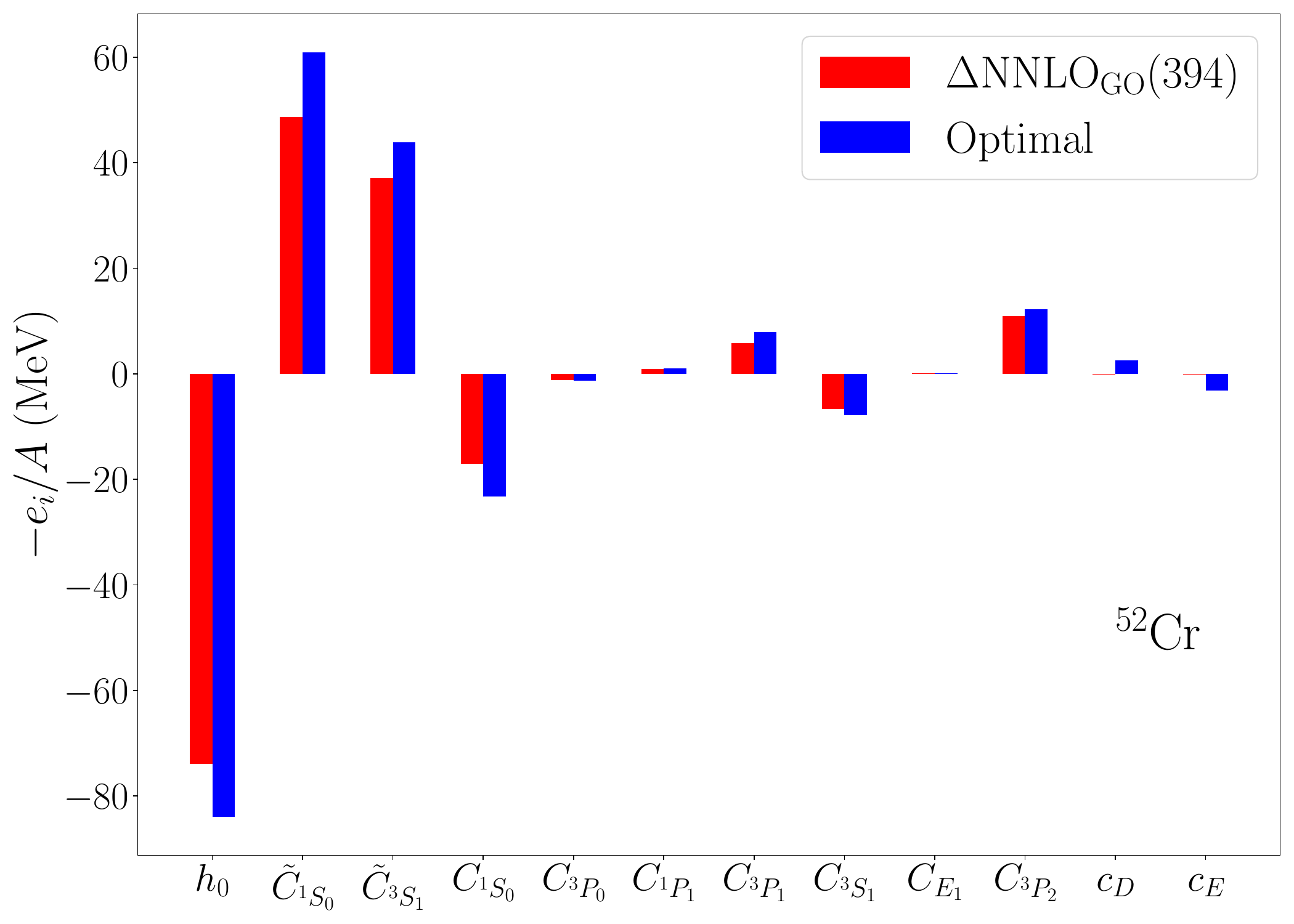} 
      \includegraphics[width=0.32\textwidth]{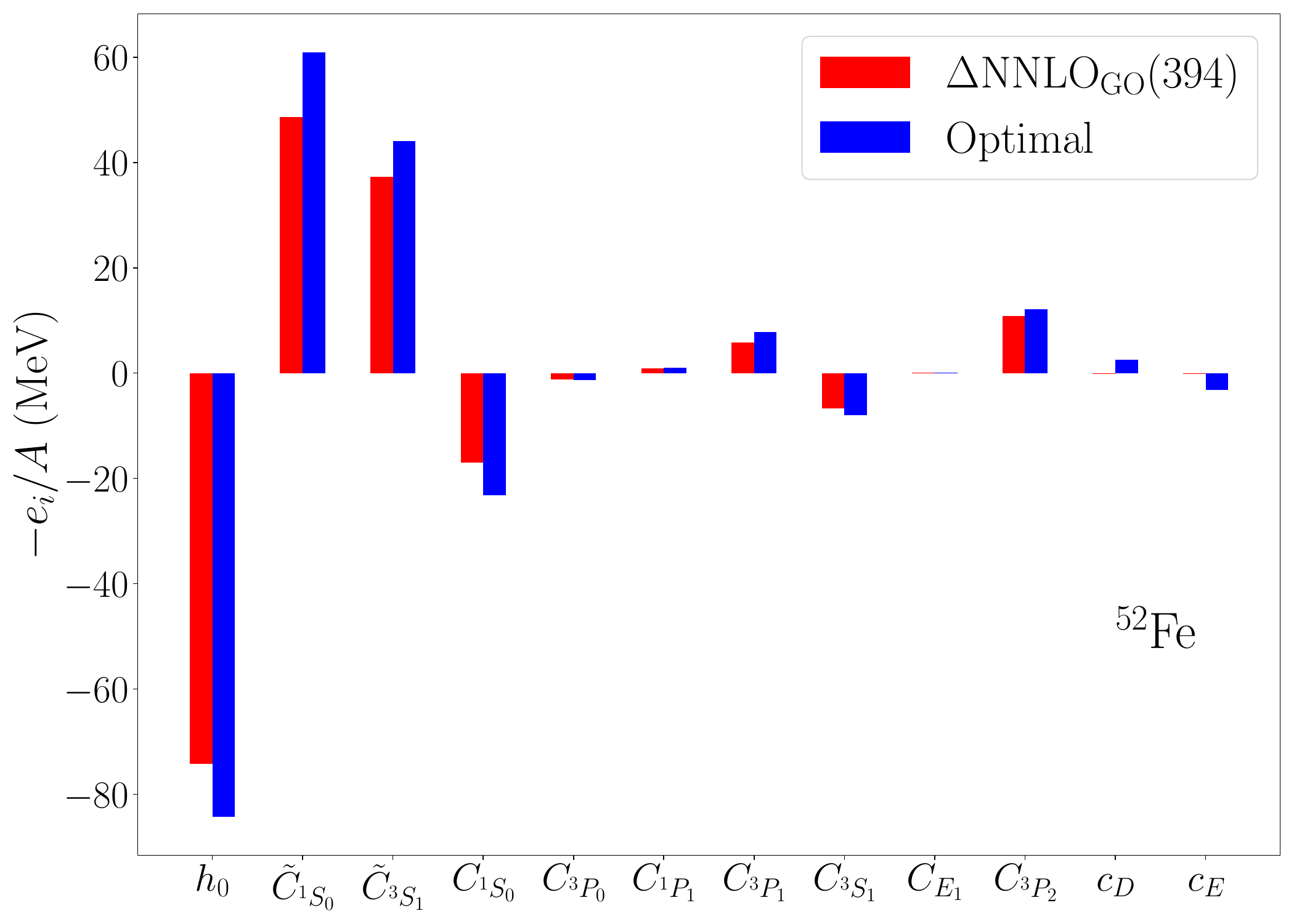}
      \includegraphics[width=0.32\textwidth]{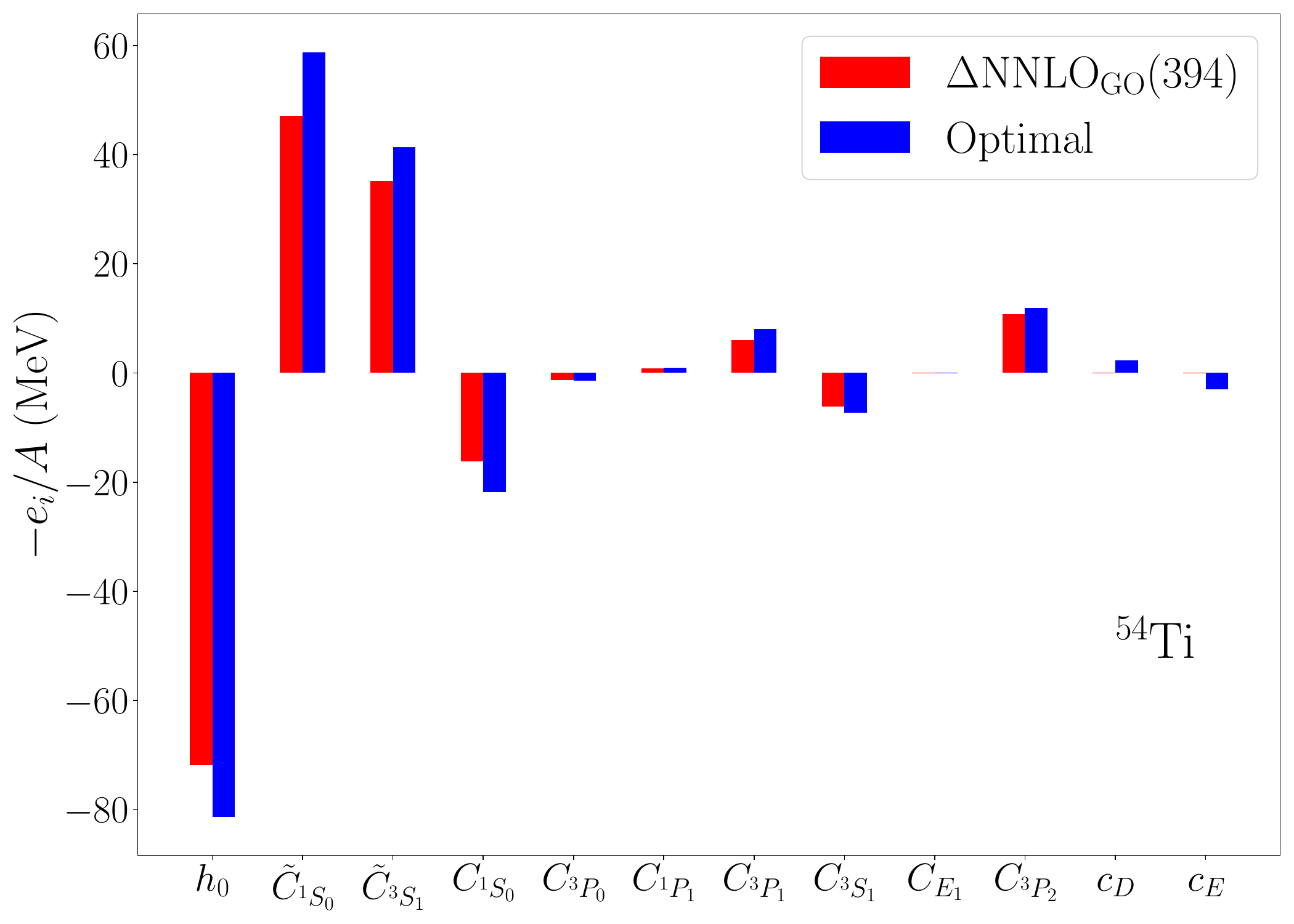}
      \includegraphics[width=0.32\textwidth]{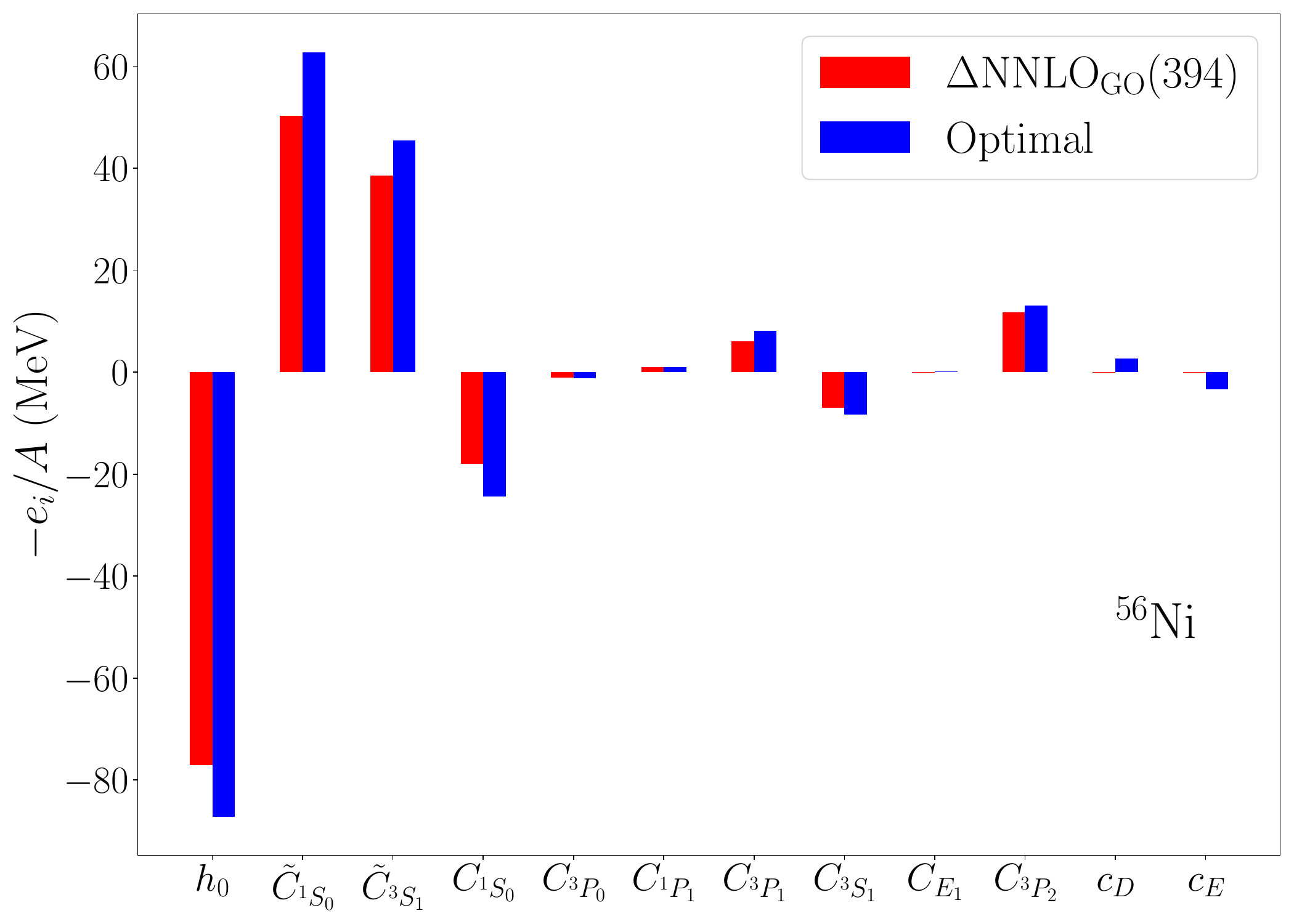} 
  \caption{Same as Fig.~\ref{fig:individual-contributions1} but for the nuclei $^{44,46,54}$Ti, $^{48,52}$Ca, $^{48,52}$Cr, $^{52}$Fe and $^{56}$Ni.}
   \label{fig:individual-contributions2}
\end{figure*}

We also analyzed how the individual terms $h_i$ with $i=0,\ldots, 11$ of the Hamiltonian~(\ref{eq:h-matrices}) contributed to the Hartree-Fock ground-state energy. 
For a given nucleus the energy contribution of $h_i$ is 
\begin{equation}
  \begin{aligned}
    \label{eq:individual-contributions}
    e_i
    &\equiv x_{i} \times
    \frac
    {\bra{\rm{L}}{h^{\textrm{RB}}_i}\ket{\rm{R}}}
    {\bra{\rm{L}}{N^{\textrm{RB}}}\ket{\rm{R}}}
  \end{aligned}\,.
\end{equation}
Here, $x_{0}=1$ and $\ket{\rm{L}}$ and $\ket{\rm{R}}$ are the left  and right eigenvectors of nonsymmetric eigenvalue problem~\eqref{eq:emulated-hf-energy}. Of course, we have
\begin{align}
  \sum_{i=0}^{11} e_i
  =
  E^{\rm{RB}}_0\,.
\end{align}
For the nuclei used in the calibration, the 12 contributions $e_i$ are shown in Figs.~\ref{fig:individual-contributions1} and \ref{fig:individual-contributions2}. 
The constant part $h_0$ consists of the kinetic energy and the pion-exchange contributions not renormalized here. It yields a repulsion.
Consistent with the power counting in chiral effective field theory\xspace, the dominant contribution to binding energy comes from the leading $S$-wave contacts
($\tilde{C}_{^1\rm{S}_0}$
and
$\tilde{C}_{^3\rm{S}_1}$). This is in line with the results of the global sensitivity analysis in Ref.~\cite{ekstrom2019}.
We also see that $c_D$ and $c_E$ make substantial contributions, compared to the \xspace$\Delta$NNLO$_{\rm GO}(394)$\xspace case, now that they are natural in size.  
We note that positive (negative) values in Table~\ref{table:lecs} yield repulsive (attractive) contributions to the energy, with the exception of the $c_E$ term, where the convention is reversed.

\subsection{Predictions of the Mass Model}

\begin{figure}[htb]
    \centering
    \includegraphics[width=\columnwidth]{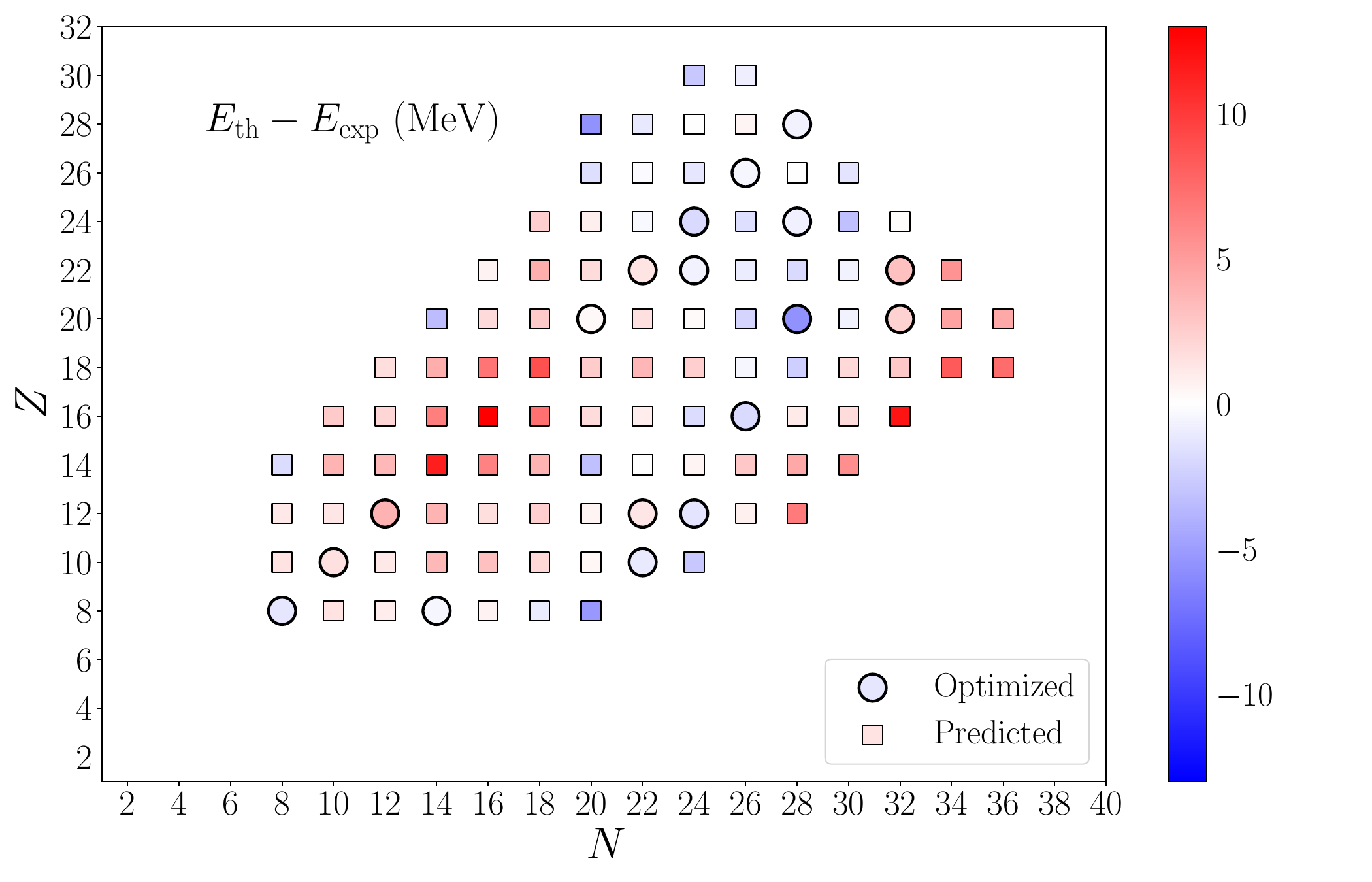}
    \caption{
      Differences between the ground-state energies from the mass model and experimental values for the 107 nuclei computed in this paper shown in the nuclear chart.  
    Positive residuals (red) correspond to under-binding and negative (blue) to over-binding.
    }
    \label{fig:residuals-nuc-chart-all}
\end{figure}

We use the calibrated mass model to compute the ground-state energies for 107 even-even nuclei from oxygen to zink with mass numbers $A\le 56$. Figure~\ref{fig:residuals-nuc-chart-all} shows the residuals, i.e. the differences between theoretical and experimental ground-state energies. Circles mark the nuclei used in the calibration and squares mark the predictions. 
We found a RMS deviation of $3.5$~MeV for the full set of 107 nuclei. The largest residuals are $11.4$,  $15.8$, and $12.0$~MeV for \nuclide{Si}{28}, \nuclide{S}{32}, and \nuclide{S}{48}, respectively.

For comparison, a fit of the semi-empirical mass formula by Bethe and Weizs\"acker to the 107 nuclei 
yields a RMS deviation of $2.07$~MeV. In contrast, \textcite{stroberg2021} performed \emph{ab initio} computations of the ground-state energies for about 700 nuclei with $2\le A\le 56$ and found a RMS deviation of 3.3~MeV. Those computations were based on a Hamiltonian from chiral effective field theory that yields accurate binding energies. Thus, a mass model based on chiral interactions should probably be more accurate than the one obtained in this work.

\begin{figure}[htb]
  \centering
  \includegraphics[width=\columnwidth]{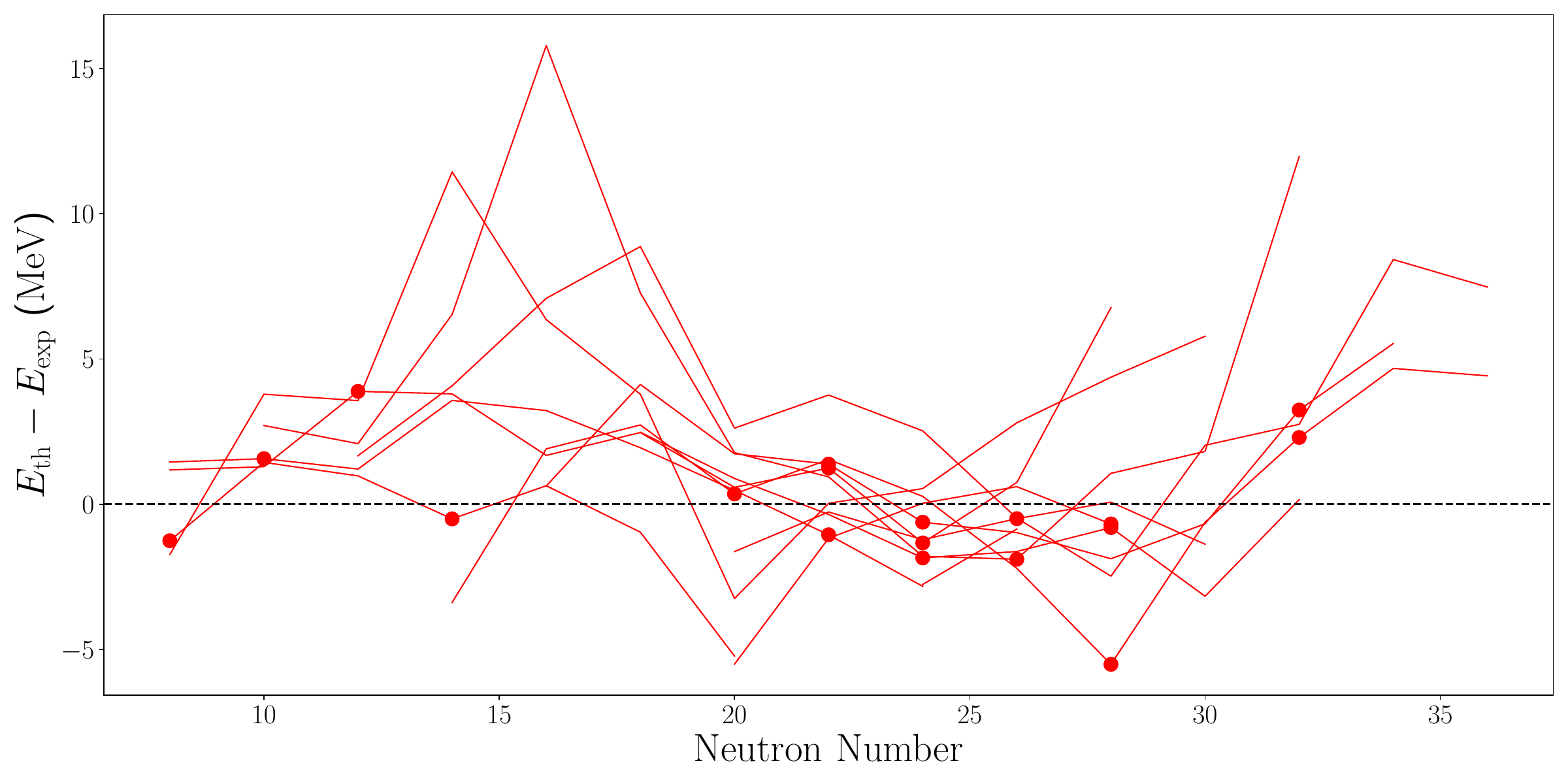}
  \caption{
    Differences of the ground-state energy between mass-model results and experimental values for the 107 nuclei of this paper, shown as a function of neutron number.
    Positive residuals correspond to under binding.
    Solid lines connect isotopes, and solid dots mark the ones included in the calibration stage. 
    }
    \label{fig:residuals-isotope-chains}
  \end{figure}

The residuals are also shown in Fig.~\ref{fig:residuals-isotope-chains} for isotope chains as a function of neutron number.
We see the three outlying nuclei with residuals larger than $10$~MeV here as well.
We offer some comments on these larger residuals in the next section.

\begin{figure}[htb]
    \setlength{\tabcolsep}{0pt}
    \begin{tabular}{cc}
      \includegraphics[width=0.25\textwidth]{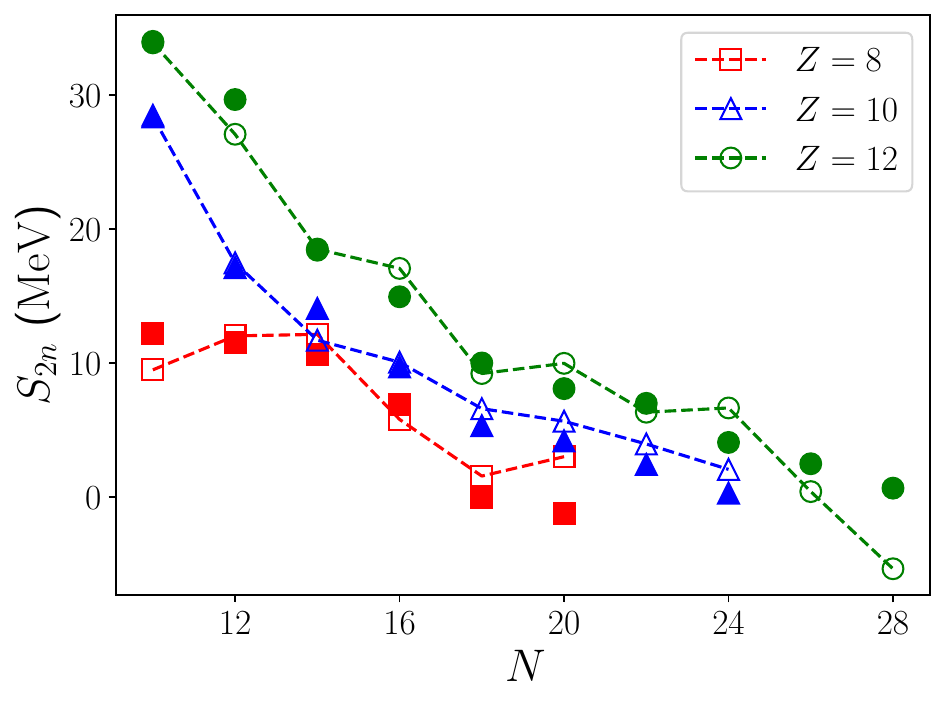} &
      \includegraphics[width=0.25\textwidth]{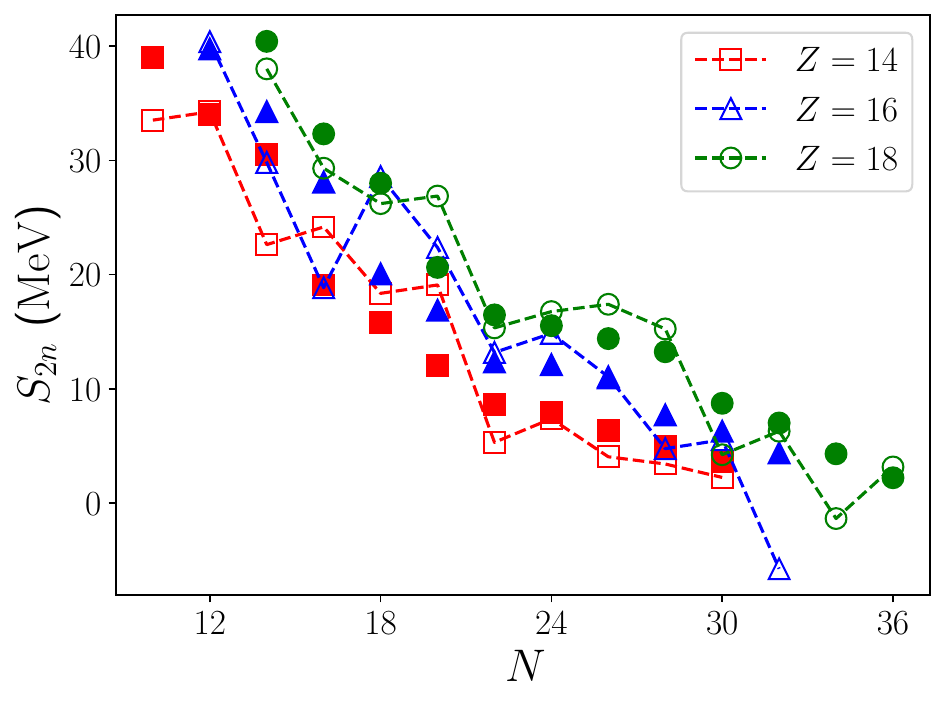} \\
      \includegraphics[width=0.25\textwidth]{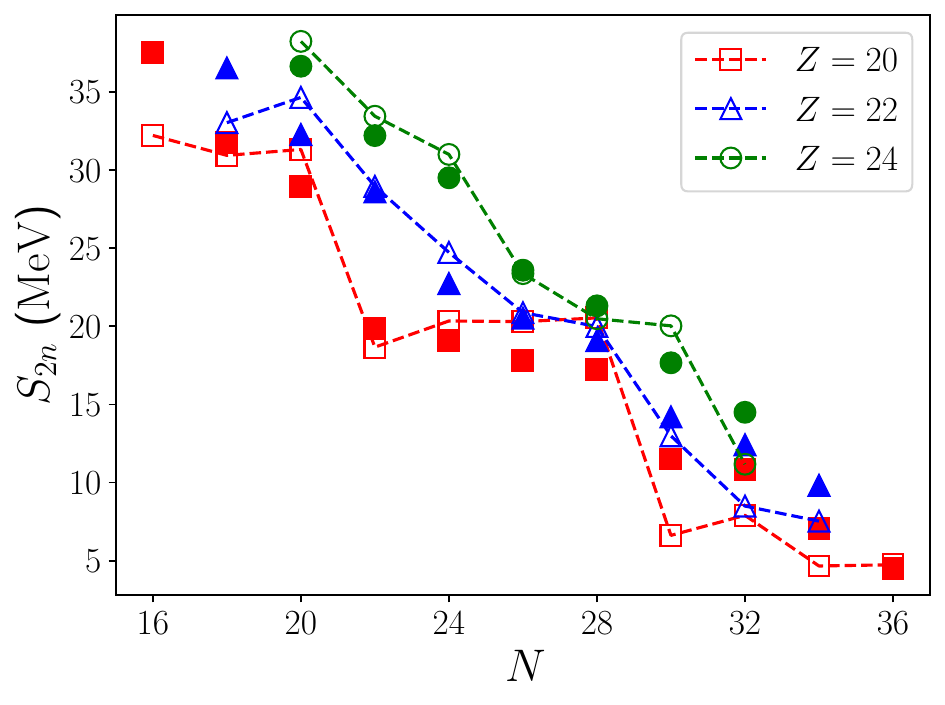} &
      \includegraphics[width=0.25\textwidth]{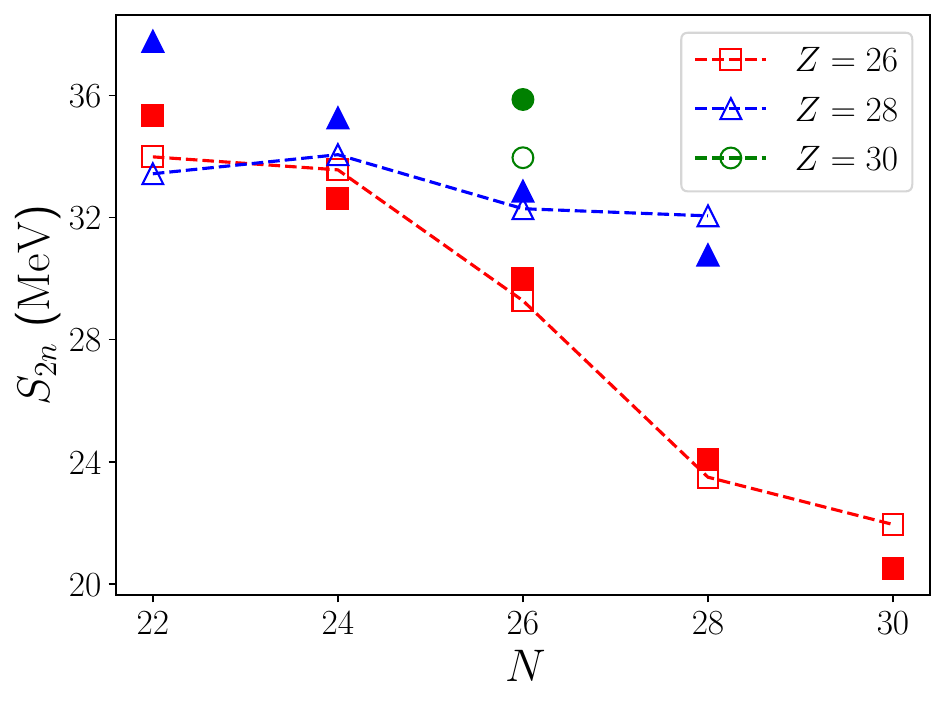} 
  \end{tabular}
  \caption{
  Two-neutron separation energies from the mass model (hollow markers connected by dashed lines along isotopes) as a function of neutron number compared to experiment (solid markers). Different elements shown with differently shaped markers. 
  }
   \label{fig:two-neutron-separation}
\end{figure}

We also computed two-neutron separation energies $S_{2n}$ and show the results in Fig.~\ref{fig:two-neutron-separation}. The RMS deviation is $3.17$~MeV. This is a bit smaller than the RMS error on binding energies, and suggests that the latter errors are correlated.  
However, the correlation is not nearly as strong as in the \emph{ab initio} computations~\cite{stroberg2021}, where RMS errors for $S_{2n}$ are 0.7 to 1.4~MeV (and significantly smaller than the RMS error of 3.3~MeV on binding energies). Thus, the \emph{ab initio} promise (or expectation), i.e. capturing local trends more accurately~\cite{kortelainen2021}, is not fully realized in the Hartree-Fock mass model. 

The results obtained for the binding and two-neutron separation energies suggest that
a more accurate modeling is desirable. We suspect that long-range correlations are lacking in our approach. Those would come from angular momentum projection, from including effects of superfluidity and pairing in a quasi-particle approach followed by number projections, and possibly by generator coordinate methods for nuclei that are neither semimagic nor rigidly deformed. The latter statement is consistent with the finding that the binding energies of \nuclide{Si}{28}, \nuclide{S}{32}, and \nuclide{S}{48} are least accurate.

\section{Summary and Discussion} 
\label{sec:discussion}

We constructed a nuclear mass model  based on Hartree-Fock computations of Hamiltonian from chiral effective field theory\xspace. 
To achieve this, we renormalized the short-range \xspace {low-energy constants}\xspace appearing in a delta-full chiral interaction including two- and three-body contributions up to NNLO. 
The renormalization was done such that Hartree-Fock computations yield accurate ground-state energies for 18 nuclei used in the calibration.
For this calibration set, the resulting RMS deviation was $2.17$~MeV. We checked that the renormalized \xspace {low-energy constants}\xspace are natural in size, the only exception being the subleading $S$-wave contact which is probably too large by a factor of about two, both in the original and renormalized potential. 
Using the calibrated mass model on a set of 107 nuclei with $N,Z\ge 8$ and $A\le 56$ yields an RMS deviation of $3.5$~MeV. 

There are several avenues to improve the accuracy of a mean-field-based mass model. First, one could compute mean-field energies for oblate and prolate configurations and take the minimum as the ground-state energy. Second, one could include superfluid correlations by using Hartree-Fock-Bogoliubov methods. Third, one could project onto good quantum numbers (angular momentum, and proton and neutron numbers). Fourth, one could even mix product states corresponding to e.g. different deformations to more accurately capture non-rigid deformation.  All of these effects are expected to be relevant because they are of the scale of 1~MeV~\cite{coelloperez2024,papenbrock2024}. (Including effects of triaxiality, in contrast, only yields tens of keV~\cite{moller2006} and is probably not relevant at the precision we are concerned about.) 

When looking back at Figs.~\ref{fig:individual-contributions1} and \ref{fig:individual-contributions2}, one wonders how relevant pion physics is for nuclear binding. After all, pion-exchange contributions are mostly included in $h_0$ and they do not overcome the repulsive kinetic energy. A mass model based on pion-less effective field theory, however, would also include four-nucleon contact-forces~\cite{bazak2019} at next-to-leading order. This is computationally less attractive. 

It would also be interesting to study the order-by-order convergence of renormalized Hamiltonians from chiral effective field theory, and to also include contact terms that enter at next-to-next-to-next-to leading order. The tools developed in this study would allow this.

\acknowledgments
We thank Zhonghao Sun and Lars Zurek for useful discussions. This work was supported by the U.S. Department of Energy, Office of
Science, Office of Nuclear Physics, under Award Nos.~DE-FG02-96ER40963, DE-SC0024465, and under SciDAC-5 NUCLEI, and by the U.S.
Department of Energy, Office of Science, Office
of Advanced Scientific Computing Research SciDAC program under Contract
Nos.\ DE-AC02-06CH11357 and DE-AC02-05CH11231.
Computer time was provided by the Innovative and Novel Computational Impact on Theory and Experiment (INCITE) program. This research used resources of the Oak Ridge Leadership
Computing Facility located at Oak Ridge National Laboratory, which is
supported by the Office of Science of the Department of Energy under
contract No. DE-AC05-00OR22725. 
CM additionally acknowledges that a portion of this manuscript was written 
under the auspices of the U.S. Department of Energy by Lawrence Livermore National Laboratory under Contract DE-AC52-07NA27344.
\bibliography{master}

\begin{thebibliography}{78}%
\makeatletter
\providecommand \@ifxundefined [1]{%
 \@ifx{#1\undefined}
}%
\providecommand \@ifnum [1]{%
 \ifnum #1\expandafter \@firstoftwo
 \else \expandafter \@secondoftwo
 \fi
}%
\providecommand \@ifx [1]{%
 \ifx #1\expandafter \@firstoftwo
 \else \expandafter \@secondoftwo
 \fi
}%
\providecommand \natexlab [1]{#1}%
\providecommand \enquote  [1]{``#1''}%
\providecommand \bibnamefont  [1]{#1}%
\providecommand \bibfnamefont [1]{#1}%
\providecommand \citenamefont [1]{#1}%
\providecommand \href@noop [0]{\@secondoftwo}%
\providecommand \href [0]{\begingroup \@sanitize@url \@href}%
\providecommand \@href[1]{\@@startlink{#1}\@@href}%
\providecommand \@@href[1]{\endgroup#1\@@endlink}%
\providecommand \@sanitize@url [0]{\catcode `\\12\catcode `\$12\catcode `\&12\catcode `\#12\catcode `\^12\catcode `\_12\catcode `\%12\relax}%
\providecommand \@@startlink[1]{}%
\providecommand \@@endlink[0]{}%
\providecommand \url  [0]{\begingroup\@sanitize@url \@url }%
\providecommand \@url [1]{\endgroup\@href {#1}{\urlprefix }}%
\providecommand \urlprefix  [0]{URL }%
\providecommand \Eprint [0]{\href }%
\providecommand \doibase [0]{https://doi.org/}%
\providecommand \selectlanguage [0]{\@gobble}%
\providecommand \bibinfo  [0]{\@secondoftwo}%
\providecommand \bibfield  [0]{\@secondoftwo}%
\providecommand \translation [1]{[#1]}%
\providecommand \BibitemOpen [0]{}%
\providecommand \bibitemStop [0]{}%
\providecommand \bibitemNoStop [0]{.\EOS\space}%
\providecommand \EOS [0]{\spacefactor3000\relax}%
\providecommand \BibitemShut  [1]{\csname bibitem#1\endcsname}%
\let\auto@bib@innerbib\@empty
\bibitem [{\citenamefont {Thoennessen}(2016)}]{thoennessen2016}%
  \BibitemOpen
  \bibfield  {author} {\bibinfo {author} {\bibfnamefont {M.}~\bibnamefont {Thoennessen}},\ }\href {https://doi.org/10.1007/978-3-319-31763-2} {\emph {\bibinfo {title} {The Discovery of Isotopes}}}\ (\bibinfo  {publisher} {Springer},\ \bibinfo {address} {Cham, Switzerland},\ \bibinfo {year} {2016})\BibitemShut {NoStop}%
\bibitem [{\citenamefont {Erler}\ \emph {et~al.}(2012)\citenamefont {Erler}, \citenamefont {Birge}, \citenamefont {Kortelainen}, \citenamefont {Nazarewicz}, \citenamefont {Olsen}, \citenamefont {Perhac},\ and\ \citenamefont {Stoitsov}}]{erler2012}%
  \BibitemOpen
  \bibfield  {author} {\bibinfo {author} {\bibfnamefont {J.}~\bibnamefont {Erler}}, \bibinfo {author} {\bibfnamefont {N.}~\bibnamefont {Birge}}, \bibinfo {author} {\bibfnamefont {M.}~\bibnamefont {Kortelainen}}, \bibinfo {author} {\bibfnamefont {W.}~\bibnamefont {Nazarewicz}}, \bibinfo {author} {\bibfnamefont {E.}~\bibnamefont {Olsen}}, \bibinfo {author} {\bibfnamefont {A.~M.}\ \bibnamefont {Perhac}},\ and\ \bibinfo {author} {\bibfnamefont {M.}~\bibnamefont {Stoitsov}},\ }\bibfield  {title} {\bibinfo {title} {The limits of the nuclear landscape},\ }\href {https://doi.org/10.1038/nature11188} {\bibfield  {journal} {\bibinfo  {journal} {Nature}\ }\textbf {\bibinfo {volume} {486}},\ \bibinfo {pages} {509 } (\bibinfo {year} {2012})}\BibitemShut {NoStop}%
\bibitem [{\citenamefont {{Blaum}}\ \emph {et~al.}(2013)\citenamefont {{Blaum}}, \citenamefont {{Dilling}},\ and\ \citenamefont {{N{\"o}rtersh{\"a}user}}}]{blaum2013}%
  \BibitemOpen
  \bibfield  {author} {\bibinfo {author} {\bibfnamefont {K.}~\bibnamefont {{Blaum}}}, \bibinfo {author} {\bibfnamefont {J.}~\bibnamefont {{Dilling}}},\ and\ \bibinfo {author} {\bibfnamefont {W.}~\bibnamefont {{N{\"o}rtersh{\"a}user}}},\ }\bibfield  {title} {\bibinfo {title} {{Precision atomic physics techniques for nuclear physics with radioactive beams}},\ }\href {https://doi.org/10.1088/0031-8949/2013/T152/014017} {\bibfield  {journal} {\bibinfo  {journal} {Phys. Scr.}\ }\textbf {\bibinfo {volume} {T152}},\ \bibinfo {eid} {014017} (\bibinfo {year} {2013})}\BibitemShut {NoStop}%
\bibitem [{\citenamefont {Horowitz}\ \emph {et~al.}(2019)\citenamefont {Horowitz} \emph {et~al.}}]{Horowitz:2018ndv}%
  \BibitemOpen
  \bibfield  {author} {\bibinfo {author} {\bibfnamefont {C.~J.}\ \bibnamefont {Horowitz}} \emph {et~al.},\ }\bibfield  {title} {\bibinfo {title} {{$r$-Process Nucleosynthesis: Connecting Rare-Isotope Beam Facilities with the Cosmos}},\ }\href {https://doi.org/10.1088/1361-6471/ab0849} {\bibfield  {journal} {\bibinfo  {journal} {J. Phys. G}\ }\textbf {\bibinfo {volume} {46}},\ \bibinfo {pages} {083001} (\bibinfo {year} {2019})},\ \Eprint {https://arxiv.org/abs/1805.04637} {arXiv:1805.04637 [astro-ph.SR]} \BibitemShut {NoStop}%
\bibitem [{\citenamefont {Sprouse}\ \emph {et~al.}(2020)\citenamefont {Sprouse}, \citenamefont {Navarro~Perez}, \citenamefont {Surman}, \citenamefont {Mumpower}, \citenamefont {McLaughlin},\ and\ \citenamefont {Schunck}}]{sprouse2020}%
  \BibitemOpen
  \bibfield  {author} {\bibinfo {author} {\bibfnamefont {T.~M.}\ \bibnamefont {Sprouse}}, \bibinfo {author} {\bibfnamefont {R.}~\bibnamefont {Navarro~Perez}}, \bibinfo {author} {\bibfnamefont {R.}~\bibnamefont {Surman}}, \bibinfo {author} {\bibfnamefont {M.~R.}\ \bibnamefont {Mumpower}}, \bibinfo {author} {\bibfnamefont {G.~C.}\ \bibnamefont {McLaughlin}},\ and\ \bibinfo {author} {\bibfnamefont {N.}~\bibnamefont {Schunck}},\ }\bibfield  {title} {\bibinfo {title} {Propagation of statistical uncertainties of skyrme mass models to simulations of $r$-process nucleosynthesis},\ }\href {https://doi.org/10.1103/PhysRevC.101.055803} {\bibfield  {journal} {\bibinfo  {journal} {Phys. Rev. C}\ }\textbf {\bibinfo {volume} {101}},\ \bibinfo {pages} {055803} (\bibinfo {year} {2020})}\BibitemShut {NoStop}%
\bibitem [{\citenamefont {Holmbeck}\ \emph {et~al.}(2023)\citenamefont {Holmbeck}, \citenamefont {Sprouse},\ and\ \citenamefont {Mumpower}}]{Holmbeck:2023bjs}%
  \BibitemOpen
  \bibfield  {author} {\bibinfo {author} {\bibfnamefont {E.~M.}\ \bibnamefont {Holmbeck}}, \bibinfo {author} {\bibfnamefont {T.~M.}\ \bibnamefont {Sprouse}},\ and\ \bibinfo {author} {\bibfnamefont {M.~R.}\ \bibnamefont {Mumpower}},\ }\bibfield  {title} {\bibinfo {title} {{Nucleosynthesis and observation of the heaviest elements}},\ }\href {https://doi.org/10.1140/epja/s10050-023-00927-7} {\bibfield  {journal} {\bibinfo  {journal} {Eur. Phys. J. A}\ }\textbf {\bibinfo {volume} {59}},\ \bibinfo {pages} {28} (\bibinfo {year} {2023})},\ \Eprint {https://arxiv.org/abs/2304.01850} {arXiv:2304.01850 [nucl-th]} \BibitemShut {NoStop}%
\bibitem [{\citenamefont {Weizs{\"{a}}cker}(1935)}]{Weizsacker1935}%
  \BibitemOpen
  \bibfield  {author} {\bibinfo {author} {\bibfnamefont {C.~F.}\ \bibnamefont {Weizs{\"{a}}cker}},\ }\bibfield  {title} {\bibinfo {title} {{Zur Theorie der Kernmassen}},\ }\href {https://doi.org/10.1007/BF01337700} {\bibfield  {journal} {\bibinfo  {journal} {Zeitschrift f{\"{u}}r Physik}\ }\textbf {\bibinfo {volume} {96}},\ \bibinfo {pages} {431} (\bibinfo {year} {1935})}\BibitemShut {NoStop}%
\bibitem [{\citenamefont {Bethe}(1937)}]{Bethe1937}%
  \BibitemOpen
  \bibfield  {author} {\bibinfo {author} {\bibfnamefont {H.~A.}\ \bibnamefont {Bethe}},\ }\bibfield  {title} {\bibinfo {title} {{Nuclear Physics B. Nuclear Dynamics, Theoretical}},\ }\href {https://doi.org/10.1103/RevModPhys.9.69} {\bibfield  {journal} {\bibinfo  {journal} {Reviews of Modern Physics}\ }\textbf {\bibinfo {volume} {9}},\ \bibinfo {pages} {69} (\bibinfo {year} {1937})}\BibitemShut {NoStop}%
\bibitem [{\citenamefont {Kejzlar}\ \emph {et~al.}(2020)\citenamefont {Kejzlar}, \citenamefont {Neufcourt}, \citenamefont {Nazarewicz},\ and\ \citenamefont {Reinhard}}]{Kejzlar:2020vla}%
  \BibitemOpen
  \bibfield  {author} {\bibinfo {author} {\bibfnamefont {V.}~\bibnamefont {Kejzlar}}, \bibinfo {author} {\bibfnamefont {L.}~\bibnamefont {Neufcourt}}, \bibinfo {author} {\bibfnamefont {W.}~\bibnamefont {Nazarewicz}},\ and\ \bibinfo {author} {\bibfnamefont {P.-G.}\ \bibnamefont {Reinhard}},\ }\bibfield  {title} {\bibinfo {title} {{Statistical aspects of nuclear mass models}},\ }\href {https://doi.org/10.1088/1361-6471/ab907c} {\bibfield  {journal} {\bibinfo  {journal} {J. Phys. G}\ }\textbf {\bibinfo {volume} {47}},\ \bibinfo {pages} {094001} (\bibinfo {year} {2020})},\ \Eprint {https://arxiv.org/abs/2002.04151} {arXiv:2002.04151 [nucl-th]} \BibitemShut {NoStop}%
\bibitem [{\citenamefont {M\"oller}\ \emph {et~al.}(2016)\citenamefont {M\"oller}, \citenamefont {Sierk}, \citenamefont {Ichikawa},\ and\ \citenamefont {Sagawa}}]{moller2016}%
  \BibitemOpen
  \bibfield  {author} {\bibinfo {author} {\bibfnamefont {P.}~\bibnamefont {M\"oller}}, \bibinfo {author} {\bibfnamefont {A.~J.}\ \bibnamefont {Sierk}}, \bibinfo {author} {\bibfnamefont {T.}~\bibnamefont {Ichikawa}},\ and\ \bibinfo {author} {\bibfnamefont {H.}~\bibnamefont {Sagawa}},\ }\bibfield  {title} {\bibinfo {title} {Nuclear ground-state masses and deformations: Frdm(2012)},\ }\href {https://doi.org/10.1016/j.adt.2015.10.002} {\bibfield  {journal} {\bibinfo  {journal} {Atomic Data and Nuclear Data Tables}\ }\textbf {\bibinfo {volume} {109-110}},\ \bibinfo {pages} {1} (\bibinfo {year} {2016})}\BibitemShut {NoStop}%
\bibitem [{\citenamefont {Kortelainen}\ \emph {et~al.}(2010)\citenamefont {Kortelainen}, \citenamefont {Lesinski}, \citenamefont {Mor\'e}, \citenamefont {Nazarewicz}, \citenamefont {Sarich}, \citenamefont {Schunck}, \citenamefont {Stoitsov},\ and\ \citenamefont {Wild}}]{Kortelainen2010}%
  \BibitemOpen
  \bibfield  {author} {\bibinfo {author} {\bibfnamefont {M.}~\bibnamefont {Kortelainen}}, \bibinfo {author} {\bibfnamefont {T.}~\bibnamefont {Lesinski}}, \bibinfo {author} {\bibfnamefont {J.}~\bibnamefont {Mor\'e}}, \bibinfo {author} {\bibfnamefont {W.}~\bibnamefont {Nazarewicz}}, \bibinfo {author} {\bibfnamefont {J.}~\bibnamefont {Sarich}}, \bibinfo {author} {\bibfnamefont {N.}~\bibnamefont {Schunck}}, \bibinfo {author} {\bibfnamefont {M.~V.}\ \bibnamefont {Stoitsov}},\ and\ \bibinfo {author} {\bibfnamefont {S.}~\bibnamefont {Wild}},\ }\bibfield  {title} {\bibinfo {title} {Nuclear energy density optimization},\ }\href {https://doi.org/10.1103/PhysRevC.82.024313} {\bibfield  {journal} {\bibinfo  {journal} {Phys. Rev. C}\ }\textbf {\bibinfo {volume} {82}},\ \bibinfo {pages} {024313} (\bibinfo {year} {2010})}\BibitemShut {NoStop}%
\bibitem [{\citenamefont {Goriely}\ \emph {et~al.}(2013)\citenamefont {Goriely}, \citenamefont {Chamel},\ and\ \citenamefont {Pearson}}]{goriely2013}%
  \BibitemOpen
  \bibfield  {author} {\bibinfo {author} {\bibfnamefont {S.}~\bibnamefont {Goriely}}, \bibinfo {author} {\bibfnamefont {N.}~\bibnamefont {Chamel}},\ and\ \bibinfo {author} {\bibfnamefont {J.~M.}\ \bibnamefont {Pearson}},\ }\bibfield  {title} {\bibinfo {title} {Further explorations of skyrme-hartree-fock-bogoliubov mass formulas. xiii. the 2012 atomic mass evaluation and the symmetry coefficient},\ }\href {https://doi.org/10.1103/PhysRevC.88.024308} {\bibfield  {journal} {\bibinfo  {journal} {Phys. Rev. C}\ }\textbf {\bibinfo {volume} {88}},\ \bibinfo {pages} {024308} (\bibinfo {year} {2013})}\BibitemShut {NoStop}%
\bibitem [{\citenamefont {Lunney}\ \emph {et~al.}(2003)\citenamefont {Lunney}, \citenamefont {Pearson},\ and\ \citenamefont {Thibault}}]{Lunney2003}%
  \BibitemOpen
  \bibfield  {author} {\bibinfo {author} {\bibfnamefont {D.}~\bibnamefont {Lunney}}, \bibinfo {author} {\bibfnamefont {J.~M.}\ \bibnamefont {Pearson}},\ and\ \bibinfo {author} {\bibfnamefont {C.}~\bibnamefont {Thibault}},\ }\bibfield  {title} {\bibinfo {title} {{Recent trends in the determination of nuclear masses}},\ }\href {https://doi.org/10.1103/RevModPhys.75.1021} {\bibfield  {journal} {\bibinfo  {journal} {Reviews of Modern Physics}\ }\textbf {\bibinfo {volume} {75}},\ \bibinfo {pages} {1021} (\bibinfo {year} {2003})}\BibitemShut {NoStop}%
\bibitem [{\citenamefont {Pearson}\ and\ \citenamefont {Goriely}(2006)}]{Pearson2006}%
  \BibitemOpen
  \bibfield  {author} {\bibinfo {author} {\bibfnamefont {J.~M.}\ \bibnamefont {Pearson}}\ and\ \bibinfo {author} {\bibfnamefont {S.}~\bibnamefont {Goriely}},\ }\bibfield  {title} {\bibinfo {title} {{Nuclear mass formulas for astrophysics}},\ }\href {https://doi.org/10.1016/j.nuclphysa.2004.06.005} {\bibfield  {journal} {\bibinfo  {journal} {Nuclear Physics A}\ }\textbf {\bibinfo {volume} {777}},\ \bibinfo {pages} {623} (\bibinfo {year} {2006})}\BibitemShut {NoStop}%
\bibitem [{\citenamefont {Stone}\ and\ \citenamefont {Reinhard}(2007)}]{Stone2007}%
  \BibitemOpen
  \bibfield  {author} {\bibinfo {author} {\bibfnamefont {J.~R.}\ \bibnamefont {Stone}}\ and\ \bibinfo {author} {\bibfnamefont {P.~G.}\ \bibnamefont {Reinhard}},\ }\bibfield  {title} {\bibinfo {title} {{The Skyrme interaction in finite nuclei and nuclear matter}},\ }\href {https://doi.org/10.1016/j.ppnp.2006.07.001} {\bibfield  {journal} {\bibinfo  {journal} {Progress in Particle and Nuclear Physics}\ }\textbf {\bibinfo {volume} {58}},\ \bibinfo {pages} {587} (\bibinfo {year} {2007})},\ \Eprint {https://arxiv.org/abs/0607002} {arXiv:0607002 [nucl-th]} \BibitemShut {NoStop}%
\bibitem [{\citenamefont {Grasso}(2019)}]{Grasso2019}%
  \BibitemOpen
  \bibfield  {author} {\bibinfo {author} {\bibfnamefont {M.}~\bibnamefont {Grasso}},\ }\bibfield  {title} {\bibinfo {title} {{Effective density functionals beyond mean field}},\ }\href {https://doi.org/10.1016/j.ppnp.2019.02.002} {\bibfield  {journal} {\bibinfo  {journal} {Progress in Particle and Nuclear Physics}\ }\textbf {\bibinfo {volume} {106}},\ \bibinfo {pages} {256} (\bibinfo {year} {2019})},\ \Eprint {https://arxiv.org/abs/1811.01039} {arXiv:1811.01039} \BibitemShut {NoStop}%
\bibitem [{\citenamefont {Robledo}\ \emph {et~al.}(2018)\citenamefont {Robledo}, \citenamefont {Rodríguez},\ and\ \citenamefont {Rodríguez-Guzmán}}]{Robledo2019}%
  \BibitemOpen
  \bibfield  {author} {\bibinfo {author} {\bibfnamefont {L.~M.}\ \bibnamefont {Robledo}}, \bibinfo {author} {\bibfnamefont {T.~R.}\ \bibnamefont {Rodríguez}},\ and\ \bibinfo {author} {\bibfnamefont {R.~R.}\ \bibnamefont {Rodríguez-Guzmán}},\ }\bibfield  {title} {\bibinfo {title} {{Mean field and beyond description of nuclear structure with the Gogny force: a review}},\ }\href {https://doi.org/10.1088/1361-6471/aadebd} {\bibfield  {journal} {\bibinfo  {journal} {J. Phys. G: Nucl. Part. Phys.}\ }\textbf {\bibinfo {volume} {46}},\ \bibinfo {pages} {013001} (\bibinfo {year} {2018})}\BibitemShut {NoStop}%
\bibitem [{\citenamefont {Duflo}\ and\ \citenamefont {Zuker}(1995)}]{Duflo1995}%
  \BibitemOpen
  \bibfield  {author} {\bibinfo {author} {\bibfnamefont {J.}~\bibnamefont {Duflo}}\ and\ \bibinfo {author} {\bibfnamefont {A.~P.}\ \bibnamefont {Zuker}},\ }\bibfield  {title} {\bibinfo {title} {{Microscopic Mass Formulas}},\ }\href {https://doi.org/10.1103/PhysRevC.52.R23} {\bibfield  {journal} {\bibinfo  {journal} {Physical Review C}\ }\textbf {\bibinfo {volume} {52}},\ \bibinfo {pages} {23} (\bibinfo {year} {1995})},\ \Eprint {https://arxiv.org/abs/9505011} {arXiv:9505011 [nucl-th]} \BibitemShut {NoStop}%
\bibitem [{\citenamefont {Carlsson}\ \emph {et~al.}(2008)\citenamefont {Carlsson}, \citenamefont {Dobaczewski},\ and\ \citenamefont {Kortelainen}}]{carlsson2008}%
  \BibitemOpen
  \bibfield  {author} {\bibinfo {author} {\bibfnamefont {B.~G.}\ \bibnamefont {Carlsson}}, \bibinfo {author} {\bibfnamefont {J.}~\bibnamefont {Dobaczewski}},\ and\ \bibinfo {author} {\bibfnamefont {M.}~\bibnamefont {Kortelainen}},\ }\bibfield  {title} {\bibinfo {title} {Local nuclear energy density functional at next-to-next-to-next-to-leading order},\ }\href {https://doi.org/10.1103/PhysRevC.78.044326} {\bibfield  {journal} {\bibinfo  {journal} {Phys. Rev. C}\ }\textbf {\bibinfo {volume} {78}},\ \bibinfo {pages} {044326} (\bibinfo {year} {2008})}\BibitemShut {NoStop}%
\bibitem [{\citenamefont {Carlsson}\ \emph {et~al.}(2010)\citenamefont {Carlsson}, \citenamefont {Dobaczewski},\ and\ \citenamefont {Kortelainen}}]{carlsson2010}%
  \BibitemOpen
  \bibfield  {author} {\bibinfo {author} {\bibfnamefont {B.~G.}\ \bibnamefont {Carlsson}}, \bibinfo {author} {\bibfnamefont {J.}~\bibnamefont {Dobaczewski}},\ and\ \bibinfo {author} {\bibfnamefont {M.}~\bibnamefont {Kortelainen}},\ }\bibfield  {title} {\bibinfo {title} {Erratum: Local nuclear energy density functional at next-to-next-to-next-to-leading order [phys. rev. c 78, 044326 (2008)]},\ }\href {https://doi.org/10.1103/PhysRevC.81.029904} {\bibfield  {journal} {\bibinfo  {journal} {Phys. Rev. C}\ }\textbf {\bibinfo {volume} {81}},\ \bibinfo {pages} {029904} (\bibinfo {year} {2010})}\BibitemShut {NoStop}%
\bibitem [{\citenamefont {Raimondi}\ \emph {et~al.}(2014)\citenamefont {Raimondi}, \citenamefont {Bennaceur},\ and\ \citenamefont {Dobaczewski}}]{Raimondi2014}%
  \BibitemOpen
  \bibfield  {author} {\bibinfo {author} {\bibfnamefont {F.}~\bibnamefont {Raimondi}}, \bibinfo {author} {\bibfnamefont {K.}~\bibnamefont {Bennaceur}},\ and\ \bibinfo {author} {\bibfnamefont {J.}~\bibnamefont {Dobaczewski}},\ }\bibfield  {title} {\bibinfo {title} {Nonlocal energy density functionals for low-energy nuclear structure},\ }\href {https://doi.org/10.1088/0954-3899/41/5/055112} {\bibfield  {journal} {\bibinfo  {journal} {J. Phys. G: Nucl. Part. Phys.}\ }\textbf {\bibinfo {volume} {41}},\ \bibinfo {pages} {055112} (\bibinfo {year} {2014})}\BibitemShut {NoStop}%
\bibitem [{\citenamefont {Sobiczewski}\ and\ \citenamefont {Litvinov}(2014)}]{Sobiczewski2014}%
  \BibitemOpen
  \bibfield  {author} {\bibinfo {author} {\bibfnamefont {A.}~\bibnamefont {Sobiczewski}}\ and\ \bibinfo {author} {\bibfnamefont {Y.~A.}\ \bibnamefont {Litvinov}},\ }\bibfield  {title} {\bibinfo {title} {{Predictive power of nuclear-mass models}},\ }\href {https://doi.org/10.1103/PhysRevC.90.017302} {\bibfield  {journal} {\bibinfo  {journal} {Physical Review C}\ }\textbf {\bibinfo {volume} {90}},\ \bibinfo {pages} {2} (\bibinfo {year} {2014})}\BibitemShut {NoStop}%
\bibitem [{\citenamefont {Kejzlar}\ \emph {et~al.}(2023)\citenamefont {Kejzlar}, \citenamefont {Neufcourt},\ and\ \citenamefont {Nazarewicz}}]{Kejzlar:2023tlm}%
  \BibitemOpen
  \bibfield  {author} {\bibinfo {author} {\bibfnamefont {V.}~\bibnamefont {Kejzlar}}, \bibinfo {author} {\bibfnamefont {L.}~\bibnamefont {Neufcourt}},\ and\ \bibinfo {author} {\bibfnamefont {W.}~\bibnamefont {Nazarewicz}},\ }\bibfield  {title} {\bibinfo {title} {{Local Bayesian Dirichlet mixing of imperfect models}},\ }\href {https://doi.org/10.1038/s41598-023-46568-0} {\bibfield  {journal} {\bibinfo  {journal} {Sci. Rep.}\ }\textbf {\bibinfo {volume} {13}},\ \bibinfo {pages} {19600} (\bibinfo {year} {2023})},\ \Eprint {https://arxiv.org/abs/2311.01596} {arXiv:2311.01596 [stat.ME]} \BibitemShut {NoStop}%
\bibitem [{\citenamefont {Mumpower}\ \emph {et~al.}(2015)\citenamefont {Mumpower}, \citenamefont {Surman}, \citenamefont {Fang}, \citenamefont {Beard}, \citenamefont {M{\"{o}}ller}, \citenamefont {Kawano},\ and\ \citenamefont {Aprahamian}}]{Mumpower2015}%
  \BibitemOpen
  \bibfield  {author} {\bibinfo {author} {\bibfnamefont {M.~R.}\ \bibnamefont {Mumpower}}, \bibinfo {author} {\bibfnamefont {R.}~\bibnamefont {Surman}}, \bibinfo {author} {\bibfnamefont {D.~L.}\ \bibnamefont {Fang}}, \bibinfo {author} {\bibfnamefont {M.}~\bibnamefont {Beard}}, \bibinfo {author} {\bibfnamefont {P.}~\bibnamefont {M{\"{o}}ller}}, \bibinfo {author} {\bibfnamefont {T.}~\bibnamefont {Kawano}},\ and\ \bibinfo {author} {\bibfnamefont {A.}~\bibnamefont {Aprahamian}},\ }\bibfield  {title} {\bibinfo {title} {{Impact of individual nuclear masses on r -process abundances}},\ }\href {https://doi.org/10.1103/PhysRevC.92.035807} {\bibfield  {journal} {\bibinfo  {journal} {Physical Review C}\ }\textbf {\bibinfo {volume} {92}},\ \bibinfo {pages} {1} (\bibinfo {year} {2015})},\ \Eprint {https://arxiv.org/abs/1505.07789} {arXiv:1505.07789} \BibitemShut {NoStop}%
\bibitem [{\citenamefont {Mumpower}\ \emph {et~al.}(2016)\citenamefont {Mumpower}, \citenamefont {Surman}, \citenamefont {McLaughlin},\ and\ \citenamefont {Aprahamian}}]{Mumpower:2015ova}%
  \BibitemOpen
  \bibfield  {author} {\bibinfo {author} {\bibfnamefont {M.~R.}\ \bibnamefont {Mumpower}}, \bibinfo {author} {\bibfnamefont {R.}~\bibnamefont {Surman}}, \bibinfo {author} {\bibfnamefont {G.~C.}\ \bibnamefont {McLaughlin}},\ and\ \bibinfo {author} {\bibfnamefont {A.}~\bibnamefont {Aprahamian}},\ }\bibfield  {title} {\bibinfo {title} {{The impact of individual nuclear properties on $r$-process nucleosynthesis}},\ }\href {https://doi.org/10.1016/j.ppnp.2015.09.001} {\bibfield  {journal} {\bibinfo  {journal} {Prog. Part. Nucl. Phys.}\ }\textbf {\bibinfo {volume} {86}},\ \bibinfo {pages} {86} (\bibinfo {year} {2016})},\ \bibinfo {note} {[Erratum: Prog.Part.Nucl.Phys. 87, 116--116 (2016)]},\ \Eprint {https://arxiv.org/abs/1508.07352} {arXiv:1508.07352 [nucl-th]} \BibitemShut {NoStop}%
\bibitem [{\citenamefont {Saito}\ \emph {et~al.}(2024)\citenamefont {Saito}, \citenamefont {Dillmann}, \citenamefont {Kruecken}, \citenamefont {Mumpower},\ and\ \citenamefont {Surman}}]{Saito:2023seh}%
  \BibitemOpen
  \bibfield  {author} {\bibinfo {author} {\bibfnamefont {Y.}~\bibnamefont {Saito}}, \bibinfo {author} {\bibfnamefont {I.}~\bibnamefont {Dillmann}}, \bibinfo {author} {\bibfnamefont {R.}~\bibnamefont {Kruecken}}, \bibinfo {author} {\bibfnamefont {M.~R.}\ \bibnamefont {Mumpower}},\ and\ \bibinfo {author} {\bibfnamefont {R.}~\bibnamefont {Surman}},\ }\bibfield  {title} {\bibinfo {title} {{Uncertainty quantification of mass models using ensemble Bayesian model averaging}},\ }\href {https://doi.org/10.1103/PhysRevC.109.054301} {\bibfield  {journal} {\bibinfo  {journal} {Phys. Rev. C}\ }\textbf {\bibinfo {volume} {109}},\ \bibinfo {pages} {054301} (\bibinfo {year} {2024})},\ \Eprint {https://arxiv.org/abs/2305.01782} {arXiv:2305.01782 [nucl-th]} \BibitemShut {NoStop}%
\bibitem [{\citenamefont {Ekstr\"om}\ \emph {et~al.}(2023)\citenamefont {Ekstr\"om}, \citenamefont {Forss\'en}, \citenamefont {Hagen}, \citenamefont {Jansen}, \citenamefont {Jiang},\ and\ \citenamefont {Papenbrock}}]{Ekstrom:2022yea}%
  \BibitemOpen
  \bibfield  {author} {\bibinfo {author} {\bibfnamefont {A.}~\bibnamefont {Ekstr\"om}}, \bibinfo {author} {\bibfnamefont {C.}~\bibnamefont {Forss\'en}}, \bibinfo {author} {\bibfnamefont {G.}~\bibnamefont {Hagen}}, \bibinfo {author} {\bibfnamefont {G.~R.}\ \bibnamefont {Jansen}}, \bibinfo {author} {\bibfnamefont {W.}~\bibnamefont {Jiang}},\ and\ \bibinfo {author} {\bibfnamefont {T.}~\bibnamefont {Papenbrock}},\ }\bibfield  {title} {\bibinfo {title} {{What is ab initio in nuclear theory?}},\ }\href {https://doi.org/10.3389/fphy.2023.1129094} {\bibfield  {journal} {\bibinfo  {journal} {Front. Phys.}\ }\textbf {\bibinfo {volume} {11}},\ \bibinfo {pages} {1129094} (\bibinfo {year} {2023})}\BibitemShut {NoStop}%
\bibitem [{\citenamefont {Stroberg}\ \emph {et~al.}(2021)\citenamefont {Stroberg}, \citenamefont {Holt}, \citenamefont {Schwenk},\ and\ \citenamefont {Simonis}}]{stroberg2021}%
  \BibitemOpen
  \bibfield  {author} {\bibinfo {author} {\bibfnamefont {S.~R.}\ \bibnamefont {Stroberg}}, \bibinfo {author} {\bibfnamefont {J.~D.}\ \bibnamefont {Holt}}, \bibinfo {author} {\bibfnamefont {A.}~\bibnamefont {Schwenk}},\ and\ \bibinfo {author} {\bibfnamefont {J.}~\bibnamefont {Simonis}},\ }\bibfield  {title} {\bibinfo {title} {Ab initio limits of atomic nuclei},\ }\href {https://doi.org/10.1103/PhysRevLett.126.022501} {\bibfield  {journal} {\bibinfo  {journal} {Phys. Rev. Lett.}\ }\textbf {\bibinfo {volume} {126}},\ \bibinfo {pages} {022501} (\bibinfo {year} {2021})}\BibitemShut {NoStop}%
\bibitem [{\citenamefont {{Scalesi}}\ \emph {et~al.}(2024)\citenamefont {{Scalesi}}, \citenamefont {{Duguet}}, \citenamefont {{Demol}}, \citenamefont {{Frosini}}, \citenamefont {{Som{\`a}}},\ and\ \citenamefont {{Tichai}}}]{scalesi2024}%
  \BibitemOpen
  \bibfield  {author} {\bibinfo {author} {\bibfnamefont {A.}~\bibnamefont {{Scalesi}}}, \bibinfo {author} {\bibfnamefont {T.}~\bibnamefont {{Duguet}}}, \bibinfo {author} {\bibfnamefont {P.}~\bibnamefont {{Demol}}}, \bibinfo {author} {\bibfnamefont {M.}~\bibnamefont {{Frosini}}}, \bibinfo {author} {\bibfnamefont {V.}~\bibnamefont {{Som{\`a}}}},\ and\ \bibinfo {author} {\bibfnamefont {A.}~\bibnamefont {{Tichai}}},\ }\bibfield  {title} {\bibinfo {title} {{Impact of correlations on nuclear binding energies: Ab initio calculations of singly and doubly open-shell nuclei}},\ }\href {https://doi.org/10.1140/epja/s10050-024-01424-1} {\bibfield  {journal} {\bibinfo  {journal} {European Physical Journal A}\ }\textbf {\bibinfo {volume} {60}},\ \bibinfo {pages} {209} (\bibinfo {year} {2024})}\BibitemShut {NoStop}%
\bibitem [{\citenamefont {{Koszor{\'u}s}}\ \emph {et~al.}(2021)\citenamefont {{Koszor{\'u}s}}, \citenamefont {{Yang}}, \citenamefont {{Jiang}}, \citenamefont {{Novario}}, \citenamefont {{Bai}}, \citenamefont {{Billowes}}, \citenamefont {{Binnersley}}, \citenamefont {{Bissell}}, \citenamefont {{Cocolios}}, \citenamefont {{Cooper}}, \citenamefont {{de Groote}}, \citenamefont {{Ekstr{\"o}m}}, \citenamefont {{Flanagan}}, \citenamefont {{Forss{\'e}n}}, \citenamefont {{Franchoo}}, \citenamefont {{Ruiz}}, \citenamefont {{Gustafsson}}, \citenamefont {{Hagen}}, \citenamefont {{Jansen}}, \citenamefont {{Kanellakopoulos}}, \citenamefont {{Kortelainen}}, \citenamefont {{Nazarewicz}}, \citenamefont {{Neyens}}, \citenamefont {{Papenbrock}}, \citenamefont {{Reinhard}}, \citenamefont {{Ricketts}}, \citenamefont {{Sahoo}}, \citenamefont {{Vernon}},\ and\ \citenamefont {{Wilkins}}}]{koszorus2021}%
  \BibitemOpen
  \bibfield  {author} {\bibinfo {author} {\bibfnamefont {{\'A}.}~\bibnamefont {{Koszor{\'u}s}}}, \bibinfo {author} {\bibfnamefont {X.~F.}\ \bibnamefont {{Yang}}}, \bibinfo {author} {\bibfnamefont {W.~G.}\ \bibnamefont {{Jiang}}}, \bibinfo {author} {\bibfnamefont {S.~J.}\ \bibnamefont {{Novario}}}, \bibinfo {author} {\bibfnamefont {S.~W.}\ \bibnamefont {{Bai}}}, \bibinfo {author} {\bibfnamefont {J.}~\bibnamefont {{Billowes}}}, \bibinfo {author} {\bibfnamefont {C.~L.}\ \bibnamefont {{Binnersley}}}, \bibinfo {author} {\bibfnamefont {M.~L.}\ \bibnamefont {{Bissell}}}, \bibinfo {author} {\bibfnamefont {T.~E.}\ \bibnamefont {{Cocolios}}}, \bibinfo {author} {\bibfnamefont {B.~S.}\ \bibnamefont {{Cooper}}}, \bibinfo {author} {\bibfnamefont {R.~P.}\ \bibnamefont {{de Groote}}}, \bibinfo {author} {\bibfnamefont {A.}~\bibnamefont {{Ekstr{\"o}m}}}, \bibinfo {author} {\bibfnamefont {K.~T.}\ \bibnamefont {{Flanagan}}}, \bibinfo {author} {\bibfnamefont {C.}~\bibnamefont {{Forss{\'e}n}}}, \bibinfo {author} {\bibfnamefont
  {S.}~\bibnamefont {{Franchoo}}}, \bibinfo {author} {\bibfnamefont {R.~F.~G.}\ \bibnamefont {{Ruiz}}}, \bibinfo {author} {\bibfnamefont {F.~P.}\ \bibnamefont {{Gustafsson}}}, \bibinfo {author} {\bibfnamefont {G.}~\bibnamefont {{Hagen}}}, \bibinfo {author} {\bibfnamefont {G.~R.}\ \bibnamefont {{Jansen}}}, \bibinfo {author} {\bibfnamefont {A.}~\bibnamefont {{Kanellakopoulos}}}, \bibinfo {author} {\bibfnamefont {M.}~\bibnamefont {{Kortelainen}}}, \bibinfo {author} {\bibfnamefont {W.}~\bibnamefont {{Nazarewicz}}}, \bibinfo {author} {\bibfnamefont {G.}~\bibnamefont {{Neyens}}}, \bibinfo {author} {\bibfnamefont {T.}~\bibnamefont {{Papenbrock}}}, \bibinfo {author} {\bibfnamefont {P.~G.}\ \bibnamefont {{Reinhard}}}, \bibinfo {author} {\bibfnamefont {C.~M.}\ \bibnamefont {{Ricketts}}}, \bibinfo {author} {\bibfnamefont {B.~K.}\ \bibnamefont {{Sahoo}}}, \bibinfo {author} {\bibfnamefont {A.~R.}\ \bibnamefont {{Vernon}}},\ and\ \bibinfo {author} {\bibfnamefont {S.~G.}\ \bibnamefont {{Wilkins}}},\ }\bibfield  {title}
  {\bibinfo {title} {{Charge radii of exotic potassium isotopes challenge nuclear theory and the magic character of N = 32}},\ }\href {https://doi.org/10.1038/s41567-020-01136-5} {\bibfield  {journal} {\bibinfo  {journal} {Nature Physics}\ }\textbf {\bibinfo {volume} {17}},\ \bibinfo {pages} {439} (\bibinfo {year} {2021})}\BibitemShut {NoStop}%
\bibitem [{\citenamefont {Kortelainen}\ \emph {et~al.}(2021)\citenamefont {Kortelainen}, \citenamefont {Sun}, \citenamefont {Hagen}, \citenamefont {Nazarewicz}, \citenamefont {Papenbrock},\ and\ \citenamefont {Reinhard}}]{kortelainen2021}%
  \BibitemOpen
  \bibfield  {author} {\bibinfo {author} {\bibfnamefont {M.}~\bibnamefont {Kortelainen}}, \bibinfo {author} {\bibfnamefont {Z.}~\bibnamefont {Sun}}, \bibinfo {author} {\bibfnamefont {G.}~\bibnamefont {Hagen}}, \bibinfo {author} {\bibfnamefont {W.}~\bibnamefont {Nazarewicz}}, \bibinfo {author} {\bibfnamefont {T.}~\bibnamefont {Papenbrock}},\ and\ \bibinfo {author} {\bibfnamefont {P.-G.}\ \bibnamefont {Reinhard}},\ }\href@noop {} {\bibinfo {title} {Universal trend of charge radii of even-even ca-zn nuclei}} (\bibinfo {year} {2021}),\ \Eprint {https://arxiv.org/abs/2111.12464} {arXiv:2111.12464 [nucl-th]} \BibitemShut {NoStop}%
\bibitem [{\citenamefont {Stoitsov}\ \emph {et~al.}(2010)\citenamefont {Stoitsov}, \citenamefont {Kortelainen}, \citenamefont {Bogner}, \citenamefont {Duguet}, \citenamefont {Furnstahl}, \citenamefont {Gebremariam},\ and\ \citenamefont {Schunck}}]{Stoitsov2010}%
  \BibitemOpen
  \bibfield  {author} {\bibinfo {author} {\bibfnamefont {M.}~\bibnamefont {Stoitsov}}, \bibinfo {author} {\bibfnamefont {M.}~\bibnamefont {Kortelainen}}, \bibinfo {author} {\bibfnamefont {S.~K.}\ \bibnamefont {Bogner}}, \bibinfo {author} {\bibfnamefont {T.}~\bibnamefont {Duguet}}, \bibinfo {author} {\bibfnamefont {R.~J.}\ \bibnamefont {Furnstahl}}, \bibinfo {author} {\bibfnamefont {B.}~\bibnamefont {Gebremariam}},\ and\ \bibinfo {author} {\bibfnamefont {N.}~\bibnamefont {Schunck}},\ }\bibfield  {title} {\bibinfo {title} {{Microscopically based energy density functionals for nuclei using the density matrix expansion: Implementation and pre-optimization}},\ }\href {https://doi.org/10.1103/PhysRevC.82.054307} {\bibfield  {journal} {\bibinfo  {journal} {Physical Review C}\ }\textbf {\bibinfo {volume} {82}},\ \bibinfo {pages} {1} (\bibinfo {year} {2010})},\ \Eprint {https://arxiv.org/abs/1009.3452} {arXiv:1009.3452} \BibitemShut {NoStop}%
\bibitem [{\citenamefont {Gebremariam}\ \emph {et~al.}(2011)\citenamefont {Gebremariam}, \citenamefont {Bogner},\ and\ \citenamefont {Duguet}}]{gebremariam2011}%
  \BibitemOpen
  \bibfield  {author} {\bibinfo {author} {\bibfnamefont {B.}~\bibnamefont {Gebremariam}}, \bibinfo {author} {\bibfnamefont {S.~K.}\ \bibnamefont {Bogner}},\ and\ \bibinfo {author} {\bibfnamefont {T.}~\bibnamefont {Duguet}},\ }\bibfield  {title} {\bibinfo {title} {Microscopically-constrained fock energy density functionals from chiral effective field theory. i. two-nucleon interactions},\ }\href {https://doi.org/10.1016/j.nuclphysa.2010.12.009} {\bibfield  {journal} {\bibinfo  {journal} {Nuclear Physics A}\ }\textbf {\bibinfo {volume} {851}},\ \bibinfo {pages} {17} (\bibinfo {year} {2011})}\BibitemShut {NoStop}%
\bibitem [{\citenamefont {{Navarro P{\'{e}}rez}}\ \emph {et~al.}(2018)\citenamefont {{Navarro P{\'{e}}rez}}, \citenamefont {Schunck}, \citenamefont {Dyhdalo}, \citenamefont {Furnstahl},\ and\ \citenamefont {Bogner}}]{NavarroPerez2018}%
  \BibitemOpen
  \bibfield  {author} {\bibinfo {author} {\bibfnamefont {R.}~\bibnamefont {{Navarro P{\'{e}}rez}}}, \bibinfo {author} {\bibfnamefont {N.}~\bibnamefont {Schunck}}, \bibinfo {author} {\bibfnamefont {A.}~\bibnamefont {Dyhdalo}}, \bibinfo {author} {\bibfnamefont {R.~J.}\ \bibnamefont {Furnstahl}},\ and\ \bibinfo {author} {\bibfnamefont {S.~K.}\ \bibnamefont {Bogner}},\ }\bibfield  {title} {\bibinfo {title} {{Microscopically based energy density functionals for nuclei using the density matrix expansion. II. Full optimization and validation}},\ }\href {https://doi.org/10.1103/PhysRevC.97.054304} {\bibfield  {journal} {\bibinfo  {journal} {Physical Review C}\ }\textbf {\bibinfo {volume} {97}},\ \bibinfo {pages} {054304} (\bibinfo {year} {2018})},\ \Eprint {https://arxiv.org/abs/1801.08615} {arXiv:1801.08615} \BibitemShut {NoStop}%
\bibitem [{\citenamefont {Salvioni}\ \emph {et~al.}(2020)\citenamefont {Salvioni}, \citenamefont {Dobaczewski}, \citenamefont {Barbieri}, \citenamefont {Carlsson}, \citenamefont {Idini},\ and\ \citenamefont {Pastore}}]{Salvioni2020}%
  \BibitemOpen
  \bibfield  {author} {\bibinfo {author} {\bibfnamefont {G.}~\bibnamefont {Salvioni}}, \bibinfo {author} {\bibfnamefont {J.}~\bibnamefont {Dobaczewski}}, \bibinfo {author} {\bibfnamefont {C.}~\bibnamefont {Barbieri}}, \bibinfo {author} {\bibfnamefont {G.}~\bibnamefont {Carlsson}}, \bibinfo {author} {\bibfnamefont {A.}~\bibnamefont {Idini}},\ and\ \bibinfo {author} {\bibfnamefont {A.}~\bibnamefont {Pastore}},\ }\bibfield  {title} {\bibinfo {title} {Model nuclear energy density functionals derived from ab initio calculations},\ }\href {https://doi.org/10.1088/1361-6471/ab8d8e} {\bibfield  {journal} {\bibinfo  {journal} {J. Phys. G: Nucl. Part. Phys.}\ }\textbf {\bibinfo {volume} {47}},\ \bibinfo {pages} {085107} (\bibinfo {year} {2020})}\BibitemShut {NoStop}%
\bibitem [{\citenamefont {Zurek}\ \emph {et~al.}(2021)\citenamefont {Zurek}, \citenamefont {{Coello P{\'{e}}rez}}, \citenamefont {Bogner}, \citenamefont {Furnstahl},\ and\ \citenamefont {Schwenk}}]{Zurek2021}%
  \BibitemOpen
  \bibfield  {author} {\bibinfo {author} {\bibfnamefont {L.}~\bibnamefont {Zurek}}, \bibinfo {author} {\bibfnamefont {E.~A.}\ \bibnamefont {{Coello P{\'{e}}rez}}}, \bibinfo {author} {\bibfnamefont {S.~K.}\ \bibnamefont {Bogner}}, \bibinfo {author} {\bibfnamefont {R.~J.}\ \bibnamefont {Furnstahl}},\ and\ \bibinfo {author} {\bibfnamefont {A.}~\bibnamefont {Schwenk}},\ }\bibfield  {title} {\bibinfo {title} {{Comparing different density-matrix expansions for long-range pion exchange}},\ }\href {https://doi.org/10.1103/PhysRevC.103.014325} {\bibfield  {journal} {\bibinfo  {journal} {Physical Review C}\ }\textbf {\bibinfo {volume} {103}},\ \bibinfo {pages} {014325} (\bibinfo {year} {2021})},\ \Eprint {https://arxiv.org/abs/2010.12518} {arXiv:2010.12518} \BibitemShut {NoStop}%
\bibitem [{\citenamefont {Duguet}\ \emph {et~al.}(2023)\citenamefont {Duguet}, \citenamefont {Ebran}, \citenamefont {Frosini}, \citenamefont {Hergert},\ and\ \citenamefont {Som{\`{a}}}}]{Duguet2023a}%
  \BibitemOpen
  \bibfield  {author} {\bibinfo {author} {\bibfnamefont {T.}~\bibnamefont {Duguet}}, \bibinfo {author} {\bibfnamefont {J.~P.}\ \bibnamefont {Ebran}}, \bibinfo {author} {\bibfnamefont {M.}~\bibnamefont {Frosini}}, \bibinfo {author} {\bibfnamefont {H.}~\bibnamefont {Hergert}},\ and\ \bibinfo {author} {\bibfnamefont {V.}~\bibnamefont {Som{\`{a}}}},\ }\bibfield  {title} {\bibinfo {title} {{Rooting the EDF method into the ab initio framework: PGCM-PT formalism based on MR-IMSRG pre-processed Hamiltonians}},\ }\href {https://doi.org/10.1140/epja/s10050-023-00914-y} {\bibfield  {journal} {\bibinfo  {journal} {European Physical Journal A}\ }\textbf {\bibinfo {volume} {59}},\ \bibinfo {pages} {1} (\bibinfo {year} {2023})},\ \Eprint {https://arxiv.org/abs/2209.03424} {arXiv:2209.03424} \BibitemShut {NoStop}%
\bibitem [{\citenamefont {Zurek}\ \emph {et~al.}(2024)\citenamefont {Zurek}, \citenamefont {Bogner}, \citenamefont {Furnstahl}, \citenamefont {{Navarro P{\'{e}}rez}}, \citenamefont {Schunck},\ and\ \citenamefont {Schwenk}}]{Zurek2024}%
  \BibitemOpen
  \bibfield  {author} {\bibinfo {author} {\bibfnamefont {L.}~\bibnamefont {Zurek}}, \bibinfo {author} {\bibfnamefont {S.~K.}\ \bibnamefont {Bogner}}, \bibinfo {author} {\bibfnamefont {R.~J.}\ \bibnamefont {Furnstahl}}, \bibinfo {author} {\bibfnamefont {R.}~\bibnamefont {{Navarro P{\'{e}}rez}}}, \bibinfo {author} {\bibfnamefont {N.}~\bibnamefont {Schunck}},\ and\ \bibinfo {author} {\bibfnamefont {A.}~\bibnamefont {Schwenk}},\ }\bibfield  {title} {\bibinfo {title} {{Optimized nuclear energy density functionals including long-range pion contributions}},\ }\href {https://doi.org/10.1103/PhysRevC.109.014319} {\bibfield  {journal} {\bibinfo  {journal} {Physical Review C}\ }\textbf {\bibinfo {volume} {109}},\ \bibinfo {pages} {1} (\bibinfo {year} {2024})},\ \Eprint {https://arxiv.org/abs/2307.13568} {arXiv:2307.13568} \BibitemShut {NoStop}%
\bibitem [{\citenamefont {Zurek}(2025)}]{Zurek:2025rvt}%
  \BibitemOpen
  \bibfield  {author} {\bibinfo {author} {\bibfnamefont {L.}~\bibnamefont {Zurek}},\ }\bibfield  {title} {\bibinfo {title} {{Towards Nuclear Energy Density Functionals from First Principles}},\ }\href {https://doi.org/10.5506/APhysPolBSupp.18.2-A5} {\bibfield  {journal} {\bibinfo  {journal} {Acta Phys. Polon. Supp.}\ }\textbf {\bibinfo {volume} {18}},\ \bibinfo {pages} {2} (\bibinfo {year} {2025})},\ \Eprint {https://arxiv.org/abs/2504.10164} {arXiv:2504.10164 [nucl-th]} \BibitemShut {NoStop}%
\bibitem [{\citenamefont {Negele}\ and\ \citenamefont {Vautherin}(1972)}]{negele1972}%
  \BibitemOpen
  \bibfield  {author} {\bibinfo {author} {\bibfnamefont {J.~W.}\ \bibnamefont {Negele}}\ and\ \bibinfo {author} {\bibfnamefont {D.}~\bibnamefont {Vautherin}},\ }\bibfield  {title} {\bibinfo {title} {Density-matrix expansion for an effective nuclear hamiltonian},\ }\href {https://doi.org/10.1103/PhysRevC.5.1472} {\bibfield  {journal} {\bibinfo  {journal} {Phys. Rev. C}\ }\textbf {\bibinfo {volume} {5}},\ \bibinfo {pages} {1472} (\bibinfo {year} {1972})}\BibitemShut {NoStop}%
\bibitem [{\citenamefont {Negele}\ and\ \citenamefont {Vautherin}(1975)}]{negele1975}%
  \BibitemOpen
  \bibfield  {author} {\bibinfo {author} {\bibfnamefont {J.~W.}\ \bibnamefont {Negele}}\ and\ \bibinfo {author} {\bibfnamefont {D.}~\bibnamefont {Vautherin}},\ }\bibfield  {title} {\bibinfo {title} {Density-matrix expansion for an effective nuclear hamiltonian. ii},\ }\href {https://doi.org/10.1103/PhysRevC.11.1031} {\bibfield  {journal} {\bibinfo  {journal} {Phys. Rev. C}\ }\textbf {\bibinfo {volume} {11}},\ \bibinfo {pages} {1031} (\bibinfo {year} {1975})}\BibitemShut {NoStop}%
\bibitem [{\citenamefont {Bogner}\ \emph {et~al.}(2009)\citenamefont {Bogner}, \citenamefont {Furnstahl},\ and\ \citenamefont {Platter}}]{Bogner:2008kj}%
  \BibitemOpen
  \bibfield  {author} {\bibinfo {author} {\bibfnamefont {S.~K.}\ \bibnamefont {Bogner}}, \bibinfo {author} {\bibfnamefont {R.~J.}\ \bibnamefont {Furnstahl}},\ and\ \bibinfo {author} {\bibfnamefont {L.}~\bibnamefont {Platter}},\ }\bibfield  {title} {\bibinfo {title} {{Density Matrix Expansion for Low-Momentum Interactions}},\ }\href {https://doi.org/10.1140/epja/i2008-10695-1} {\bibfield  {journal} {\bibinfo  {journal} {Eur. Phys. J. A}\ }\textbf {\bibinfo {volume} {39}},\ \bibinfo {pages} {219} (\bibinfo {year} {2009})},\ \Eprint {https://arxiv.org/abs/0811.4198} {arXiv:0811.4198 [nucl-th]} \BibitemShut {NoStop}%
\bibitem [{\citenamefont {Skyrme}(1956)}]{skyrme1956}%
  \BibitemOpen
  \bibfield  {author} {\bibinfo {author} {\bibfnamefont {T.~H.~R.}\ \bibnamefont {Skyrme}},\ }\bibfield  {title} {\bibinfo {title} {Cvii. the nuclear surface},\ }\href {https://doi.org/10.1080/14786435608238186} {\bibfield  {journal} {\bibinfo  {journal} {The Philosophical Magazine: A Journal of Theoretical Experimental and Applied Physics}\ }\textbf {\bibinfo {volume} {1}},\ \bibinfo {pages} {1043} (\bibinfo {year} {1956})}\BibitemShut {NoStop}%
\bibitem [{\citenamefont {Skyrme}(1958)}]{skyrme1958}%
  \BibitemOpen
  \bibfield  {author} {\bibinfo {author} {\bibfnamefont {T.~H.~R.}\ \bibnamefont {Skyrme}},\ }\bibfield  {title} {\bibinfo {title} {The effective nuclear potential},\ }\href {https://doi.org/10.1016/0029-5582(58)90345-6} {\bibfield  {journal} {\bibinfo  {journal} {Nuclear Physics}\ }\textbf {\bibinfo {volume} {9}},\ \bibinfo {pages} {615} (\bibinfo {year} {1958})}\BibitemShut {NoStop}%
\bibitem [{\citenamefont {Sun}\ \emph {et~al.}(2022)\citenamefont {Sun}, \citenamefont {Bell}, \citenamefont {Hagen},\ and\ \citenamefont {Papenbrock}}]{Sun2022b}%
  \BibitemOpen
  \bibfield  {author} {\bibinfo {author} {\bibfnamefont {Z.~H.}\ \bibnamefont {Sun}}, \bibinfo {author} {\bibfnamefont {C.~A.}\ \bibnamefont {Bell}}, \bibinfo {author} {\bibfnamefont {G.}~\bibnamefont {Hagen}},\ and\ \bibinfo {author} {\bibfnamefont {T.}~\bibnamefont {Papenbrock}},\ }\bibfield  {title} {\bibinfo {title} {{How to Renormalize Coupled Cluster Theory}},\ }\href {https://doi.org/10.1103/PhysRevC.106.L061302} {\bibfield  {journal} {\bibinfo  {journal} {Physical Review C}\ }\textbf {\bibinfo {volume} {106}},\ \bibinfo {pages} {L061302} (\bibinfo {year} {2022})},\ \Eprint {https://arxiv.org/abs/2205.12990} {arXiv:2205.12990} \BibitemShut {NoStop}%
\bibitem [{\citenamefont {Hagen}\ \emph {et~al.}(2014)\citenamefont {Hagen}, \citenamefont {Papenbrock}, \citenamefont {Hjorth-Jensen},\ and\ \citenamefont {Dean}}]{hagen2014}%
  \BibitemOpen
  \bibfield  {author} {\bibinfo {author} {\bibfnamefont {G.}~\bibnamefont {Hagen}}, \bibinfo {author} {\bibfnamefont {T.}~\bibnamefont {Papenbrock}}, \bibinfo {author} {\bibfnamefont {M.}~\bibnamefont {Hjorth-Jensen}},\ and\ \bibinfo {author} {\bibfnamefont {D.~J.}\ \bibnamefont {Dean}},\ }\bibfield  {title} {\bibinfo {title} {Coupled-cluster computations of atomic nuclei},\ }\href {https://doi.org/10.1088/0034-4885/77/9/096302} {\bibfield  {journal} {\bibinfo  {journal} {Rep. Prog. Phys.}\ }\textbf {\bibinfo {volume} {77}},\ \bibinfo {pages} {096302} (\bibinfo {year} {2014})}\BibitemShut {NoStop}%
\bibitem [{\citenamefont {Sun}\ \emph {et~al.}(2025)\citenamefont {Sun}, \citenamefont {Ekstr\"om}, \citenamefont {Forss\'en}, \citenamefont {Hagen}, \citenamefont {Jansen},\ and\ \citenamefont {Papenbrock}}]{sun2024}%
  \BibitemOpen
  \bibfield  {author} {\bibinfo {author} {\bibfnamefont {Z.~H.}\ \bibnamefont {Sun}}, \bibinfo {author} {\bibfnamefont {A.}~\bibnamefont {Ekstr\"om}}, \bibinfo {author} {\bibfnamefont {C.}~\bibnamefont {Forss\'en}}, \bibinfo {author} {\bibfnamefont {G.}~\bibnamefont {Hagen}}, \bibinfo {author} {\bibfnamefont {G.~R.}\ \bibnamefont {Jansen}},\ and\ \bibinfo {author} {\bibfnamefont {T.}~\bibnamefont {Papenbrock}},\ }\bibfield  {title} {\bibinfo {title} {Multiscale physics of atomic nuclei from first principles},\ }\href {https://doi.org/10.1103/PhysRevX.15.011028} {\bibfield  {journal} {\bibinfo  {journal} {Phys. Rev. X}\ }\textbf {\bibinfo {volume} {15}},\ \bibinfo {pages} {011028} (\bibinfo {year} {2025})}\BibitemShut {NoStop}%
\bibitem [{\citenamefont {van Kolck}(1994)}]{vankolck1994}%
  \BibitemOpen
  \bibfield  {author} {\bibinfo {author} {\bibfnamefont {U.}~\bibnamefont {van Kolck}},\ }\bibfield  {title} {\bibinfo {title} {{Few-nucleon forces from chiral Lagrangians}},\ }\href {https://doi.org/10.1103/PhysRevC.49.2932} {\bibfield  {journal} {\bibinfo  {journal} {Phys. Rev. C}\ }\textbf {\bibinfo {volume} {49}},\ \bibinfo {pages} {2932} (\bibinfo {year} {1994})}\BibitemShut {NoStop}%
\bibitem [{\citenamefont {Epelbaum}\ \emph {et~al.}(2009)\citenamefont {Epelbaum}, \citenamefont {Hammer},\ and\ \citenamefont {Mei\ss{}ner}}]{epelbaum2009}%
  \BibitemOpen
  \bibfield  {author} {\bibinfo {author} {\bibfnamefont {E.}~\bibnamefont {Epelbaum}}, \bibinfo {author} {\bibfnamefont {H.-W.}\ \bibnamefont {Hammer}},\ and\ \bibinfo {author} {\bibfnamefont {U.-G.}\ \bibnamefont {Mei\ss{}ner}},\ }\bibfield  {title} {\bibinfo {title} {Modern theory of nuclear forces},\ }\href {https://doi.org/10.1103/RevModPhys.81.1773} {\bibfield  {journal} {\bibinfo  {journal} {Rev. Mod. Phys.}\ }\textbf {\bibinfo {volume} {81}},\ \bibinfo {pages} {1773} (\bibinfo {year} {2009})}\BibitemShut {NoStop}%
\bibitem [{\citenamefont {Hammer}\ \emph {et~al.}(2020)\citenamefont {Hammer}, \citenamefont {K\"onig},\ and\ \citenamefont {van Kolck}}]{Hammer2020}%
  \BibitemOpen
  \bibfield  {author} {\bibinfo {author} {\bibfnamefont {H.-W.}\ \bibnamefont {Hammer}}, \bibinfo {author} {\bibfnamefont {S.}~\bibnamefont {K\"onig}},\ and\ \bibinfo {author} {\bibfnamefont {U.}~\bibnamefont {van Kolck}},\ }\bibfield  {title} {\bibinfo {title} {Nuclear effective field theory: Status and perspectives},\ }\href {https://doi.org/10.1103/RevModPhys.92.025004} {\bibfield  {journal} {\bibinfo  {journal} {Rev. Mod. Phys.}\ }\textbf {\bibinfo {volume} {92}},\ \bibinfo {pages} {025004} (\bibinfo {year} {2020})}\BibitemShut {NoStop}%
\bibitem [{\citenamefont {Machleidt}\ and\ \citenamefont {Entem}(2011)}]{machleidt2011}%
  \BibitemOpen
  \bibfield  {author} {\bibinfo {author} {\bibfnamefont {R.}~\bibnamefont {Machleidt}}\ and\ \bibinfo {author} {\bibfnamefont {D.}~\bibnamefont {Entem}},\ }\bibfield  {title} {\bibinfo {title} {Chiral effective field theory and nuclear forces},\ }\href {https://doi.org/10.1016/j.physrep.2011.02.001} {\bibfield  {journal} {\bibinfo  {journal} {Phys. Rep.}\ }\textbf {\bibinfo {volume} {503}},\ \bibinfo {pages} {1 } (\bibinfo {year} {2011})}\BibitemShut {NoStop}%
\bibitem [{\citenamefont {Epelbaum}\ \emph {et~al.}(2002)\citenamefont {Epelbaum}, \citenamefont {Nogga}, \citenamefont {Gl\"ockle}, \citenamefont {Kamada}, \citenamefont {Mei\ss{}ner},\ and\ \citenamefont {Wita\l{}a}}]{epelbaum2002}%
  \BibitemOpen
  \bibfield  {author} {\bibinfo {author} {\bibfnamefont {E.}~\bibnamefont {Epelbaum}}, \bibinfo {author} {\bibfnamefont {A.}~\bibnamefont {Nogga}}, \bibinfo {author} {\bibfnamefont {W.}~\bibnamefont {Gl\"ockle}}, \bibinfo {author} {\bibfnamefont {H.}~\bibnamefont {Kamada}}, \bibinfo {author} {\bibfnamefont {U.-G.}\ \bibnamefont {Mei\ss{}ner}},\ and\ \bibinfo {author} {\bibfnamefont {H.}~\bibnamefont {Wita\l{}a}},\ }\bibfield  {title} {\bibinfo {title} {Three-nucleon forces from chiral effective field theory},\ }\href {https://doi.org/10.1103/PhysRevC.66.064001} {\bibfield  {journal} {\bibinfo  {journal} {Phys. Rev. C}\ }\textbf {\bibinfo {volume} {66}},\ \bibinfo {pages} {064001} (\bibinfo {year} {2002})}\BibitemShut {NoStop}%
\bibitem [{\citenamefont {Ekstr\"om}\ \emph {et~al.}(2018)\citenamefont {Ekstr\"om}, \citenamefont {Hagen}, \citenamefont {Morris}, \citenamefont {Papenbrock},\ and\ \citenamefont {Schwartz}}]{ekstrom2018}%
  \BibitemOpen
  \bibfield  {author} {\bibinfo {author} {\bibfnamefont {A.}~\bibnamefont {Ekstr\"om}}, \bibinfo {author} {\bibfnamefont {G.}~\bibnamefont {Hagen}}, \bibinfo {author} {\bibfnamefont {T.~D.}\ \bibnamefont {Morris}}, \bibinfo {author} {\bibfnamefont {T.}~\bibnamefont {Papenbrock}},\ and\ \bibinfo {author} {\bibfnamefont {P.~D.}\ \bibnamefont {Schwartz}},\ }\bibfield  {title} {\bibinfo {title} {$\mathrm{\ensuremath{\Delta}}$ isobars and nuclear saturation},\ }\href {https://doi.org/10.1103/PhysRevC.97.024332} {\bibfield  {journal} {\bibinfo  {journal} {Phys. Rev. C}\ }\textbf {\bibinfo {volume} {97}},\ \bibinfo {pages} {024332} (\bibinfo {year} {2018})}\BibitemShut {NoStop}%
\bibitem [{\citenamefont {Jiang}\ \emph {et~al.}(2020)\citenamefont {Jiang}, \citenamefont {Ekstr\"om}, \citenamefont {Forss\'en}, \citenamefont {Hagen}, \citenamefont {Jansen},\ and\ \citenamefont {Papenbrock}}]{jiang2020}%
  \BibitemOpen
  \bibfield  {author} {\bibinfo {author} {\bibfnamefont {W.~G.}\ \bibnamefont {Jiang}}, \bibinfo {author} {\bibfnamefont {A.}~\bibnamefont {Ekstr\"om}}, \bibinfo {author} {\bibfnamefont {C.}~\bibnamefont {Forss\'en}}, \bibinfo {author} {\bibfnamefont {G.}~\bibnamefont {Hagen}}, \bibinfo {author} {\bibfnamefont {G.~R.}\ \bibnamefont {Jansen}},\ and\ \bibinfo {author} {\bibfnamefont {T.}~\bibnamefont {Papenbrock}},\ }\bibfield  {title} {\bibinfo {title} {Accurate bulk properties of nuclei from $a=2$ to $\ensuremath{\infty}$ from potentials with $\mathrm{\ensuremath{\Delta}}$ isobars},\ }\href {https://doi.org/10.1103/PhysRevC.102.054301} {\bibfield  {journal} {\bibinfo  {journal} {Phys. Rev. C}\ }\textbf {\bibinfo {volume} {102}},\ \bibinfo {pages} {054301} (\bibinfo {year} {2020})}\BibitemShut {NoStop}%
\bibitem [{\citenamefont {Hu}\ \emph {et~al.}(2024{\natexlab{a}})\citenamefont {Hu}, \citenamefont {Sun}, \citenamefont {Hagen},\ and\ \citenamefont {Papenbrock}}]{hu2024}%
  \BibitemOpen
  \bibfield  {author} {\bibinfo {author} {\bibfnamefont {B.~S.}\ \bibnamefont {Hu}}, \bibinfo {author} {\bibfnamefont {Z.~H.}\ \bibnamefont {Sun}}, \bibinfo {author} {\bibfnamefont {G.}~\bibnamefont {Hagen}},\ and\ \bibinfo {author} {\bibfnamefont {T.}~\bibnamefont {Papenbrock}},\ }\bibfield  {title} {\bibinfo {title} {Ab initio computations of strongly deformed nuclei near $^{80}\mathrm{Zr}$},\ }\href {https://doi.org/10.1103/PhysRevC.110.L011302} {\bibfield  {journal} {\bibinfo  {journal} {Phys. Rev. C}\ }\textbf {\bibinfo {volume} {110}},\ \bibinfo {pages} {L011302} (\bibinfo {year} {2024}{\natexlab{a}})}\BibitemShut {NoStop}%
\bibitem [{\citenamefont {Hu}\ \emph {et~al.}(2024{\natexlab{b}})\citenamefont {Hu}, \citenamefont {Sun}, \citenamefont {Hagen}, \citenamefont {Jansen},\ and\ \citenamefont {Papenbrock}}]{hu2024b}%
  \BibitemOpen
  \bibfield  {author} {\bibinfo {author} {\bibfnamefont {B.~S.}\ \bibnamefont {Hu}}, \bibinfo {author} {\bibfnamefont {Z.~H.}\ \bibnamefont {Sun}}, \bibinfo {author} {\bibfnamefont {G.}~\bibnamefont {Hagen}}, \bibinfo {author} {\bibfnamefont {G.~R.}\ \bibnamefont {Jansen}},\ and\ \bibinfo {author} {\bibfnamefont {T.}~\bibnamefont {Papenbrock}},\ }\bibfield  {title} {\bibinfo {title} {Ab initio computations from 78ni towards 70ca along neutron number $n=50$},\ }\href {https://doi.org/10.1016/j.physletb.2024.139010} {\bibfield  {journal} {\bibinfo  {journal} {Physics Letters B}\ }\textbf {\bibinfo {volume} {858}},\ \bibinfo {pages} {139010} (\bibinfo {year} {2024}{\natexlab{b}})}\BibitemShut {NoStop}%
\bibitem [{\citenamefont {Hagen}\ \emph {et~al.}(2007)\citenamefont {Hagen}, \citenamefont {Papenbrock}, \citenamefont {Dean}, \citenamefont {Schwenk}, \citenamefont {Nogga}, \citenamefont {W\l{}och},\ and\ \citenamefont {Piecuch}}]{hagen2007a}%
  \BibitemOpen
  \bibfield  {author} {\bibinfo {author} {\bibfnamefont {G.}~\bibnamefont {Hagen}}, \bibinfo {author} {\bibfnamefont {T.}~\bibnamefont {Papenbrock}}, \bibinfo {author} {\bibfnamefont {D.~J.}\ \bibnamefont {Dean}}, \bibinfo {author} {\bibfnamefont {A.}~\bibnamefont {Schwenk}}, \bibinfo {author} {\bibfnamefont {A.}~\bibnamefont {Nogga}}, \bibinfo {author} {\bibfnamefont {M.}~\bibnamefont {W\l{}och}},\ and\ \bibinfo {author} {\bibfnamefont {P.}~\bibnamefont {Piecuch}},\ }\bibfield  {title} {\bibinfo {title} {{Coupled-cluster theory for three-body Hamiltonians}},\ }\href {https://doi.org/10.1103/PhysRevC.76.034302} {\bibfield  {journal} {\bibinfo  {journal} {Phys. Rev. C}\ }\textbf {\bibinfo {volume} {76}},\ \bibinfo {pages} {034302} (\bibinfo {year} {2007})}\BibitemShut {NoStop}%
\bibitem [{\citenamefont {Roth}\ \emph {et~al.}(2012)\citenamefont {Roth}, \citenamefont {Binder}, \citenamefont {Vobig}, \citenamefont {Calci}, \citenamefont {Langhammer},\ and\ \citenamefont {Navr\'atil}}]{roth2012}%
  \BibitemOpen
  \bibfield  {author} {\bibinfo {author} {\bibfnamefont {R.}~\bibnamefont {Roth}}, \bibinfo {author} {\bibfnamefont {S.}~\bibnamefont {Binder}}, \bibinfo {author} {\bibfnamefont {K.}~\bibnamefont {Vobig}}, \bibinfo {author} {\bibfnamefont {A.}~\bibnamefont {Calci}}, \bibinfo {author} {\bibfnamefont {J.}~\bibnamefont {Langhammer}},\ and\ \bibinfo {author} {\bibfnamefont {P.}~\bibnamefont {Navr\'atil}},\ }\bibfield  {title} {\bibinfo {title} {{Medium-Mass Nuclei with Normal-Ordered Chiral $NN\mathbf{+}3N$ Interactions}},\ }\href {https://doi.org/10.1103/PhysRevLett.109.052501} {\bibfield  {journal} {\bibinfo  {journal} {Phys. Rev. Lett.}\ }\textbf {\bibinfo {volume} {109}},\ \bibinfo {pages} {052501} (\bibinfo {year} {2012})}\BibitemShut {NoStop}%
\bibitem [{\citenamefont {Frosini}\ \emph {et~al.}(2021)\citenamefont {Frosini}, \citenamefont {Duguet}, \citenamefont {Bally}, \citenamefont {Beaujeault-Taudi\`ere}, \citenamefont {Ebran},\ and\ \citenamefont {Som\`a}}]{Frosini:2021tuj}%
  \BibitemOpen
  \bibfield  {author} {\bibinfo {author} {\bibfnamefont {M.}~\bibnamefont {Frosini}}, \bibinfo {author} {\bibfnamefont {T.}~\bibnamefont {Duguet}}, \bibinfo {author} {\bibfnamefont {B.}~\bibnamefont {Bally}}, \bibinfo {author} {\bibfnamefont {Y.}~\bibnamefont {Beaujeault-Taudi\`ere}}, \bibinfo {author} {\bibfnamefont {J.~P.}\ \bibnamefont {Ebran}},\ and\ \bibinfo {author} {\bibfnamefont {V.}~\bibnamefont {Som\`a}},\ }\bibfield  {title} {\bibinfo {title} {{In-medium $k$-body reduction of $n$-body operators: A flexible symmetry-conserving approach based on the sole one-body density matrix}},\ }\href {https://doi.org/10.1140/epja/s10050-021-00458-z} {\bibfield  {journal} {\bibinfo  {journal} {Eur. Phys. J. A}\ }\textbf {\bibinfo {volume} {57}},\ \bibinfo {pages} {151} (\bibinfo {year} {2021})}\BibitemShut {NoStop}%
\bibitem [{\citenamefont {Benner}\ \emph {et~al.}(2017)\citenamefont {Benner}, \citenamefont {Ohlberger}, \citenamefont {Patera}, \citenamefont {Rozza},\ and\ \citenamefont {Urban}}]{Benner2017}%
  \BibitemOpen
  \bibinfo {editor} {\bibfnamefont {P.}~\bibnamefont {Benner}}, \bibinfo {editor} {\bibfnamefont {M.}~\bibnamefont {Ohlberger}}, \bibinfo {editor} {\bibfnamefont {A.}~\bibnamefont {Patera}}, \bibinfo {editor} {\bibfnamefont {G.}~\bibnamefont {Rozza}},\ and\ \bibinfo {editor} {\bibfnamefont {K.}~\bibnamefont {Urban}},\ eds.,\ \href {https://doi.org/10.1007/978-3-319-58786-8} {\emph {\bibinfo {title} {Model Reduction of Parametrized Systems}}},\ \bibinfo {edition} {1st}\ ed.,\ \bibinfo {series} {MS\&A}, Vol.~\bibinfo {volume} {17}\ (\bibinfo  {publisher} {Springer International Publishing},\ \bibinfo {address} {Cham},\ \bibinfo {year} {2017})\ pp.\ \bibinfo {pages} {XII, 504}\BibitemShut {NoStop}%
\bibitem [{\citenamefont {Frame}\ \emph {et~al.}(2018)\citenamefont {Frame}, \citenamefont {He}, \citenamefont {Ipsen}, \citenamefont {Lee}, \citenamefont {Lee},\ and\ \citenamefont {Rrapaj}}]{frame2018}%
  \BibitemOpen
  \bibfield  {author} {\bibinfo {author} {\bibfnamefont {D.}~\bibnamefont {Frame}}, \bibinfo {author} {\bibfnamefont {R.}~\bibnamefont {He}}, \bibinfo {author} {\bibfnamefont {I.}~\bibnamefont {Ipsen}}, \bibinfo {author} {\bibfnamefont {D.}~\bibnamefont {Lee}}, \bibinfo {author} {\bibfnamefont {D.}~\bibnamefont {Lee}},\ and\ \bibinfo {author} {\bibfnamefont {E.}~\bibnamefont {Rrapaj}},\ }\bibfield  {title} {\bibinfo {title} {Eigenvector continuation with subspace learning},\ }\href {https://doi.org/10.1103/PhysRevLett.121.032501} {\bibfield  {journal} {\bibinfo  {journal} {Phys. Rev. Lett.}\ }\textbf {\bibinfo {volume} {121}},\ \bibinfo {pages} {032501} (\bibinfo {year} {2018})}\BibitemShut {NoStop}%
\bibitem [{\citenamefont {Ekstr\"om}\ and\ \citenamefont {Hagen}(2019)}]{ekstrom2019}%
  \BibitemOpen
  \bibfield  {author} {\bibinfo {author} {\bibfnamefont {A.}~\bibnamefont {Ekstr\"om}}\ and\ \bibinfo {author} {\bibfnamefont {G.}~\bibnamefont {Hagen}},\ }\bibfield  {title} {\bibinfo {title} {Global sensitivity analysis of bulk properties of an atomic nucleus},\ }\href {https://doi.org/10.1103/PhysRevLett.123.252501} {\bibfield  {journal} {\bibinfo  {journal} {Phys. Rev. Lett.}\ }\textbf {\bibinfo {volume} {123}},\ \bibinfo {pages} {252501} (\bibinfo {year} {2019})}\BibitemShut {NoStop}%
\bibitem [{\citenamefont {K\"onig}\ \emph {et~al.}(2020)\citenamefont {K\"onig}, \citenamefont {Ekstr\"om}, \citenamefont {Hebeler}, \citenamefont {Lee},\ and\ \citenamefont {Schwenk}}]{Konig:2019adq}%
  \BibitemOpen
  \bibfield  {author} {\bibinfo {author} {\bibfnamefont {S.}~\bibnamefont {K\"onig}}, \bibinfo {author} {\bibfnamefont {A.}~\bibnamefont {Ekstr\"om}}, \bibinfo {author} {\bibfnamefont {K.}~\bibnamefont {Hebeler}}, \bibinfo {author} {\bibfnamefont {D.}~\bibnamefont {Lee}},\ and\ \bibinfo {author} {\bibfnamefont {A.}~\bibnamefont {Schwenk}},\ }\bibfield  {title} {\bibinfo {title} {{Eigenvector Continuation as an Efficient and Accurate Emulator for Uncertainty Quantification}},\ }\href {https://doi.org/10.1016/j.physletb.2020.135814} {\bibfield  {journal} {\bibinfo  {journal} {Phys. Lett. B}\ }\textbf {\bibinfo {volume} {810}},\ \bibinfo {pages} {135814} (\bibinfo {year} {2020})},\ \Eprint {https://arxiv.org/abs/1909.08446} {arXiv:1909.08446 [nucl-th]} \BibitemShut {NoStop}%
\bibitem [{\citenamefont {Drischler}\ \emph {et~al.}(2021)\citenamefont {Drischler}, \citenamefont {Quinonez}, \citenamefont {Giuliani}, \citenamefont {Lovell},\ and\ \citenamefont {Nunes}}]{Drischler2021a}%
  \BibitemOpen
  \bibfield  {author} {\bibinfo {author} {\bibfnamefont {C.}~\bibnamefont {Drischler}}, \bibinfo {author} {\bibfnamefont {M.}~\bibnamefont {Quinonez}}, \bibinfo {author} {\bibfnamefont {P.~G.}\ \bibnamefont {Giuliani}}, \bibinfo {author} {\bibfnamefont {A.~E.}\ \bibnamefont {Lovell}},\ and\ \bibinfo {author} {\bibfnamefont {F.~M.}\ \bibnamefont {Nunes}},\ }\bibfield  {title} {\bibinfo {title} {{Toward emulating nuclear reactions using eigenvector continuation}},\ }\href {https://doi.org/10.1016/j.physletb.2021.136777} {\bibfield  {journal} {\bibinfo  {journal} {Physics Letters, Section B: Nuclear, Elementary Particle and High-Energy Physics}\ }\textbf {\bibinfo {volume} {823}},\ \bibinfo {pages} {136777} (\bibinfo {year} {2021})},\ \Eprint {https://arxiv.org/abs/2108.08269} {arXiv:2108.08269} \BibitemShut {NoStop}%
\bibitem [{\citenamefont {Drischler}\ \emph {et~al.}(2022)\citenamefont {Drischler}, \citenamefont {Melendez}, \citenamefont {Furnstahl}, \citenamefont {Garcia},\ and\ \citenamefont {Zhang}}]{Drischler2022}%
  \BibitemOpen
  \bibfield  {author} {\bibinfo {author} {\bibfnamefont {C.}~\bibnamefont {Drischler}}, \bibinfo {author} {\bibfnamefont {J.~A.}\ \bibnamefont {Melendez}}, \bibinfo {author} {\bibfnamefont {R.~J.}\ \bibnamefont {Furnstahl}}, \bibinfo {author} {\bibfnamefont {A.~J.}\ \bibnamefont {Garcia}},\ and\ \bibinfo {author} {\bibfnamefont {X.}~\bibnamefont {Zhang}},\ }\bibfield  {title} {\bibinfo {title} {{BUQEYE guide to projection-based emulators in nuclear physics}},\ }\href {https://doi.org/10.3389/fphy.2022.1092931} {\bibfield  {journal} {\bibinfo  {journal} {Frontiers in Physics}\ }\textbf {\bibinfo {volume} {10}},\ \bibinfo {pages} {1} (\bibinfo {year} {2022})},\ \Eprint {https://arxiv.org/abs/2212.04912} {arXiv:2212.04912} \BibitemShut {NoStop}%
\bibitem [{\citenamefont {Yapa}\ and\ \citenamefont {K\"onig}(2022)}]{Yapa2022}%
  \BibitemOpen
  \bibfield  {author} {\bibinfo {author} {\bibfnamefont {N.}~\bibnamefont {Yapa}}\ and\ \bibinfo {author} {\bibfnamefont {S.}~\bibnamefont {K\"onig}},\ }\bibfield  {title} {\bibinfo {title} {Volume extrapolation via eigenvector continuation},\ }\href {https://doi.org/10.1103/PhysRevC.106.014309} {\bibfield  {journal} {\bibinfo  {journal} {Phys. Rev. C}\ }\textbf {\bibinfo {volume} {106}},\ \bibinfo {pages} {014309} (\bibinfo {year} {2022})}\BibitemShut {NoStop}%
\bibitem [{\citenamefont {Duguet}\ \emph {et~al.}(2024)\citenamefont {Duguet}, \citenamefont {Ekstr\"om}, \citenamefont {Furnstahl}, \citenamefont {K\"onig},\ and\ \citenamefont {Lee}}]{Duguet2023b}%
  \BibitemOpen
  \bibfield  {author} {\bibinfo {author} {\bibfnamefont {T.}~\bibnamefont {Duguet}}, \bibinfo {author} {\bibfnamefont {A.}~\bibnamefont {Ekstr\"om}}, \bibinfo {author} {\bibfnamefont {R.~J.}\ \bibnamefont {Furnstahl}}, \bibinfo {author} {\bibfnamefont {S.}~\bibnamefont {K\"onig}},\ and\ \bibinfo {author} {\bibfnamefont {D.}~\bibnamefont {Lee}},\ }\bibfield  {title} {\bibinfo {title} {Colloquium: Eigenvector continuation and projection-based emulators},\ }\href {https://doi.org/10.1103/RevModPhys.96.031002} {\bibfield  {journal} {\bibinfo  {journal} {Rev. Mod. Phys.}\ }\textbf {\bibinfo {volume} {96}},\ \bibinfo {pages} {031002} (\bibinfo {year} {2024})}\BibitemShut {NoStop}%
\bibitem [{\citenamefont {Hebeler}\ \emph {et~al.}(2011)\citenamefont {Hebeler}, \citenamefont {Bogner}, \citenamefont {Furnstahl}, \citenamefont {Nogga},\ and\ \citenamefont {Schwenk}}]{hebeler2011}%
  \BibitemOpen
  \bibfield  {author} {\bibinfo {author} {\bibfnamefont {K.}~\bibnamefont {Hebeler}}, \bibinfo {author} {\bibfnamefont {S.~K.}\ \bibnamefont {Bogner}}, \bibinfo {author} {\bibfnamefont {R.~J.}\ \bibnamefont {Furnstahl}}, \bibinfo {author} {\bibfnamefont {A.}~\bibnamefont {Nogga}},\ and\ \bibinfo {author} {\bibfnamefont {A.}~\bibnamefont {Schwenk}},\ }\bibfield  {title} {\bibinfo {title} {Improved nuclear matter calculations from chiral low-momentum interactions},\ }\href {https://doi.org/10.1103/PhysRevC.83.031301} {\bibfield  {journal} {\bibinfo  {journal} {Phys. Rev. C}\ }\textbf {\bibinfo {volume} {83}},\ \bibinfo {pages} {031301} (\bibinfo {year} {2011})}\BibitemShut {NoStop}%
\bibitem [{\citenamefont {Virtanen}\ \emph {et~al.}(2020)\citenamefont {Virtanen}, \citenamefont {Gommers}, \citenamefont {Oliphant}, \citenamefont {Haberland}, \citenamefont {Reddy}, \citenamefont {Cournapeau}, \citenamefont {Burovski}, \citenamefont {Peterson}, \citenamefont {Weckesser}, \citenamefont {Bright} \emph {et~al.}}]{virtanen2020scipy}%
  \BibitemOpen
  \bibfield  {author} {\bibinfo {author} {\bibfnamefont {P.}~\bibnamefont {Virtanen}}, \bibinfo {author} {\bibfnamefont {R.}~\bibnamefont {Gommers}}, \bibinfo {author} {\bibfnamefont {T.~E.}\ \bibnamefont {Oliphant}}, \bibinfo {author} {\bibfnamefont {M.}~\bibnamefont {Haberland}}, \bibinfo {author} {\bibfnamefont {T.}~\bibnamefont {Reddy}}, \bibinfo {author} {\bibfnamefont {D.}~\bibnamefont {Cournapeau}}, \bibinfo {author} {\bibfnamefont {E.}~\bibnamefont {Burovski}}, \bibinfo {author} {\bibfnamefont {P.}~\bibnamefont {Peterson}}, \bibinfo {author} {\bibfnamefont {W.}~\bibnamefont {Weckesser}}, \bibinfo {author} {\bibfnamefont {J.}~\bibnamefont {Bright}}, \emph {et~al.},\ }\bibfield  {title} {\bibinfo {title} {Scipy 1.0: fundamental algorithms for scientific computing in python},\ }\href {https://doi.org/10.1038/s41592-019-0686-2} {\bibfield  {journal} {\bibinfo  {journal} {Nature Methods}\ }\textbf {\bibinfo {volume} {17}},\ \bibinfo {pages} {261} (\bibinfo {year} {2020})}\BibitemShut {NoStop}%
\bibitem [{\citenamefont {Mor{\'e}}\ \emph {et~al.}(1980)\citenamefont {Mor{\'e}}, \citenamefont {Garbow},\ and\ \citenamefont {Hillstrom}}]{more1980user}%
  \BibitemOpen
  \bibfield  {author} {\bibinfo {author} {\bibfnamefont {J.~J.}\ \bibnamefont {Mor{\'e}}}, \bibinfo {author} {\bibfnamefont {B.~S.}\ \bibnamefont {Garbow}},\ and\ \bibinfo {author} {\bibfnamefont {K.~E.}\ \bibnamefont {Hillstrom}},\ }\href@noop {} {\emph {\bibinfo {title} {User guide for MINPACK-1}}},\ \bibinfo {type} {Tech. Rep.}\ (\bibinfo  {institution} {CM-P00068642},\ \bibinfo {year} {1980})\BibitemShut {NoStop}%
\bibitem [{\citenamefont {Lepage}(1997)}]{lepage1997}%
  \BibitemOpen
  \bibfield  {author} {\bibinfo {author} {\bibfnamefont {G.~P.}\ \bibnamefont {Lepage}},\ }\bibfield  {title} {\bibinfo {title} {{How to renormalize the Schr{\"o}dinger equation}},\ }in\ \href {https://doi.org/10.48550/arXiv.nucl-th/9706029} {\emph {\bibinfo {booktitle} {{8th Jorge Andre Swieca Summer School on Nuclear Physics}}}}\ (\bibinfo {year} {1997})\ pp.\ \bibinfo {pages} {135--180},\ \Eprint {https://arxiv.org/abs/nucl-th/9706029} {arXiv:nucl-th/9706029} \BibitemShut {NoStop}%
\bibitem [{\citenamefont {Manohar}\ and\ \citenamefont {Georgi}(1984)}]{manohar1984}%
  \BibitemOpen
  \bibfield  {author} {\bibinfo {author} {\bibfnamefont {A.}~\bibnamefont {Manohar}}\ and\ \bibinfo {author} {\bibfnamefont {H.}~\bibnamefont {Georgi}},\ }\bibfield  {title} {\bibinfo {title} {Chiral quarks and the non-relativistic quark model},\ }\href {https://doi.org/10.1016/0550-3213(84)90231-1} {\bibfield  {journal} {\bibinfo  {journal} {Nuclear Physics B}\ }\textbf {\bibinfo {volume} {234}},\ \bibinfo {pages} {189} (\bibinfo {year} {1984})}\BibitemShut {NoStop}%
\bibitem [{\citenamefont {Georgi}(1993)}]{georgi1993}%
  \BibitemOpen
  \bibfield  {author} {\bibinfo {author} {\bibfnamefont {H.}~\bibnamefont {Georgi}},\ }\bibfield  {title} {\bibinfo {title} {Effective field theory},\ }\href {https://doi.org/10.1146/annurev.ns.43.120193.001233} {\bibfield  {journal} {\bibinfo  {journal} {Ann. Rev. Nucl. Part. Sci.}\ }\textbf {\bibinfo {volume} {43}},\ \bibinfo {pages} {209} (\bibinfo {year} {1993})}\BibitemShut {NoStop}%
\bibitem [{\citenamefont {Furnstahl}\ and\ \citenamefont {Hackworth}(1997)}]{Furnstahl1997}%
  \BibitemOpen
  \bibfield  {author} {\bibinfo {author} {\bibfnamefont {R.~J.}\ \bibnamefont {Furnstahl}}\ and\ \bibinfo {author} {\bibfnamefont {J.~C.}\ \bibnamefont {Hackworth}},\ }\bibfield  {title} {\bibinfo {title} {Skyrme energy functional and naturalness},\ }\href {https://doi.org/10.1103/PhysRevC.56.2875} {\bibfield  {journal} {\bibinfo  {journal} {Phys. Rev. C}\ }\textbf {\bibinfo {volume} {56}},\ \bibinfo {pages} {2875} (\bibinfo {year} {1997})}\BibitemShut {NoStop}%
\bibitem [{\citenamefont {{Coello P{\'e}rez}}\ and\ \citenamefont {{Papenbrock}}(2024)}]{coelloperez2024}%
  \BibitemOpen
  \bibfield  {author} {\bibinfo {author} {\bibfnamefont {E.~A.}\ \bibnamefont {{Coello P{\'e}rez}}}\ and\ \bibinfo {author} {\bibfnamefont {T.}~\bibnamefont {{Papenbrock}}},\ }\bibfield  {title} {\bibinfo {title} {{Effective field theories for collective excitations of atomic nuclei}},\ }\href {https://doi.org/10.48550/arXiv.2411.04895} {\bibfield  {journal} {\bibinfo  {journal} {arXiv e-prints}\ ,\ \bibinfo {eid} {arXiv:2411.04895}} (\bibinfo {year} {2024})}\BibitemShut {NoStop}%
\bibitem [{\citenamefont {{Papenbrock}}(2024)}]{papenbrock2024}%
  \BibitemOpen
  \bibfield  {author} {\bibinfo {author} {\bibfnamefont {T.}~\bibnamefont {{Papenbrock}}},\ }\bibfield  {title} {\bibinfo {title} {{Ab initio computations of atomic nuclei}},\ }\href {https://doi.org/10.48550/arXiv.2410.00843} {\bibfield  {journal} {\bibinfo  {journal} {arXiv e-prints}\ ,\ \bibinfo {pages} {arXiv:2410.00843}} (\bibinfo {year} {2024})}\BibitemShut {NoStop}%
\bibitem [{\citenamefont {M\"oller}\ \emph {et~al.}(2006)\citenamefont {M\"oller}, \citenamefont {Bengtsson}, \citenamefont {Carlsson}, \citenamefont {Olivius},\ and\ \citenamefont {Ichikawa}}]{moller2006}%
  \BibitemOpen
  \bibfield  {author} {\bibinfo {author} {\bibfnamefont {P.}~\bibnamefont {M\"oller}}, \bibinfo {author} {\bibfnamefont {R.}~\bibnamefont {Bengtsson}}, \bibinfo {author} {\bibfnamefont {B.~G.}\ \bibnamefont {Carlsson}}, \bibinfo {author} {\bibfnamefont {P.}~\bibnamefont {Olivius}},\ and\ \bibinfo {author} {\bibfnamefont {T.}~\bibnamefont {Ichikawa}},\ }\bibfield  {title} {\bibinfo {title} {Global calculations of ground-state axial shape asymmetry of nuclei},\ }\href {https://doi.org/10.1103/PhysRevLett.97.162502} {\bibfield  {journal} {\bibinfo  {journal} {Phys. Rev. Lett.}\ }\textbf {\bibinfo {volume} {97}},\ \bibinfo {pages} {162502} (\bibinfo {year} {2006})}\BibitemShut {NoStop}%
\bibitem [{\citenamefont {Bazak}\ \emph {et~al.}(2019)\citenamefont {Bazak}, \citenamefont {Kirscher}, \citenamefont {K\"onig}, \citenamefont {Valderrama}, \citenamefont {Barnea},\ and\ \citenamefont {van Kolck}}]{bazak2019}%
  \BibitemOpen
  \bibfield  {author} {\bibinfo {author} {\bibfnamefont {B.}~\bibnamefont {Bazak}}, \bibinfo {author} {\bibfnamefont {J.}~\bibnamefont {Kirscher}}, \bibinfo {author} {\bibfnamefont {S.}~\bibnamefont {K\"onig}}, \bibinfo {author} {\bibfnamefont {M.~P.}\ \bibnamefont {Valderrama}}, \bibinfo {author} {\bibfnamefont {N.}~\bibnamefont {Barnea}},\ and\ \bibinfo {author} {\bibfnamefont {U.}~\bibnamefont {van Kolck}},\ }\bibfield  {title} {\bibinfo {title} {Four-body scale in universal few-boson systems},\ }\href {https://doi.org/10.1103/PhysRevLett.122.143001} {\bibfield  {journal} {\bibinfo  {journal} {Phys. Rev. Lett.}\ }\textbf {\bibinfo {volume} {122}},\ \bibinfo {pages} {143001} (\bibinfo {year} {2019})}\BibitemShut {NoStop}%
\end{thebibliography}%
\end{document}